%% file: LRDoF_v2.tex
\documentclass[12pt,a4paper,openright,oneside]{book} %% See: REGULATION-LENGTHS

\usepackage{graphicx}
\usepackage{hyperref}
\usepackage{amsmath} % Taking out the \usepackage{subeqn} in the original template code seems to prevent a compiler error.
\usepackage{amssymb}

\usepackage{fancyheadings}

\usepackage{thesis}
\usepackage{setspace}
\usepackage{epsfig}
\usepackage{graphics}
\usepackage{psfrag}
\usepackage{titlesec}
\titleformat{\chapter}[display]{\sffamily
\Huge}{\thechapter}{2ex}{}[\vspace{2ex} \titlerule]
\renewcommand{\chaptername}{Chapter}

\usepackage{caption}
\usepackage{makeidx}

\makeindex
\usepackage{eurosans} %\usepackage{jmacro} \figuresarein{fig/}

\usepackage{afterpage}

\reversemarginpar

\def\be{\begin{equation}}
\def\ee{\end{equation}}
\def\bea{\begin{eqnarray}}
\def\eea{\end{eqnarray}}
\def\>{\rangle}
\def\<{\langle}
\newcommand{\ket}[1]{|#1\rangle}
\newcommand{\bra}[1]{\langle#1|}
\newcommand{\eq}[1]{Eq.~(\ref{eqn:#1})}
\newcommand{\sect}[1]{Sec.~\ref{sec:#1}}
\newcommand{\figu}[1]{Fig.~\ref{fig:#1}}

\begin{document}
\ifx\href\undefined\else\hypersetup{linktocpage=true}\fi 

\include{LRDoF_v2_titlepage}
\include{LRDoF_v2_frontmatter}

% Format for most of  thesis
\pagestyle{fancyplain}

\renewcommand{\chaptername}{Chapter}

\include{LRDoF_v2_chapter_1}

\include{LRDoF_v2_chapter_2}

\include{LRDoF_v2_chapter_3}
\include{LRDoF_v2_chapter_4}

\include{LRDoF_v2_chapter_5}

\include{LRDoF_v2_chapter_6}

% Make the following chapters appear as appendices.
\appendix
\include{LRDoF_v2_appendix_opt_one}
\include{LRDoF_v2_appendix_GaussianVis}

\include{LRDoF_v2_backmatter}

\end{document}

%% file: LRDoF_v2_titlepage.tex
\pagestyle{empty}

\begin{center}
  {\sc{University of London}}\\[0.5cm]
  \centerline{\includegraphics[width=3.5cm]{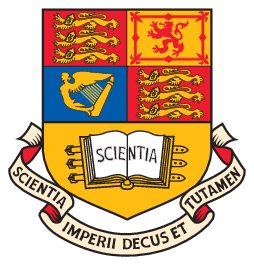}}
  Imperial College London \\ [0.2cm]
  Physics Department \\ [-0.15cm]
  Quantum Optics and Laser Science Group\\[1.5cm]

  {\Huge Localising Relational }\\[0.5cm]
  {\Huge Degrees of Freedom }\\[0.5cm]
  {\Huge in Quantum Mechanics}\\[1cm]

  {\Large by}\\[0.2cm]
  {\Large{Hugo Vaughan Cable}}\\[1.5cm]

  Thesis submitted in partial fulfilment of the \\
  requirements for the degree of\\
  Doctor of Philosophy\\
  of the University of London\\
  and the Diploma of Membership\\
  of Imperial College.\\[.8cm]
  \vfill
  October 2005
\end{center}

% Local Variables:
% TeX-master: "./thesis"
% End:

%% file: LRDoF_v2_frontmatter.tex
\clearpage \pagestyle{empty}
\begin{center}
  {\Huge Abstract}\\[1cm]
\end{center}

In this thesis I present a wide-ranging study of localising relational degrees of freedom, contributing to the wider debate on relationism
in quantum mechanics.  A set of analytical and numerical methods are developed and applied to a diverse range of physical systems.  Chapter \ref{chap:OpticalOne} looks at the interference of two optical modes with no prior phase correlation.  Cases of initial mixed states --- specifically Poissonian states and thermal states --- are investigated in addition to the well-known case of initial Fock states.  For the pure
state case, and assuming an ideal setup, a ``relational Schr{\"o}dinger cat'' state emerges localised at two values of the relative phase.
Circumstances under which this type of state is destroyed are explained.  When the apparatus is subject to instabilities, the states which emerge are sharply localised at one value.  Such states are predicted to be long lived.  It is shown that the localisation of the relative
phase can be as good, and as rapid, for initially mixed states as for the pure state case.  Chapter \ref{chap:OpticalTwo} extends the programme
of the previous chapter discussing a variety of topics --- the case of asymmetric initial states with intensity very much greater in one mode, the transitive properties of the localising process, some applications to quantum state engineering (in particular for creating large photon number states), and finally, a relational perspective on superselection rules.  Chapter \ref{chap:BEC} considers
the spatial interference of independently prepared Bose-Einstein condensates, an area which has attracted much attention since the work of Javanainen and Yoo.  The localisation of the relative atomic phase plays a key role here, and it is shown that the phase localises much faster than is intimated in earlier studies looking at the emergence of a well-defined pattern of interference.
A novel analytical method is used, and the predicted localisation is compared with the output of a full numerical simulation.
The chapter ends with a review of a related body of literature concerned with non-destructive measurement of relative atomic phases between
condensates.  Chapter \ref{chap:ParticlesScatteringLight} explores localising relative positions between mirrors or particles scattering light, addressing recent work by Rau, Dunningham and Burnett.  The analysis here retains the models of  scattering introduced by those authors but makes different assumptions.  Detailed results are presented for the case of free particles, initially in thermal states, scattering monochromatic light and thermal light.  It is assumed that an observer registers whether or not an incident light packet has been scattered into a large angle, but lacks access to more detailed information.  Under these conditions the localisation is found to be only partial, regardless of the number of observations, and at variance with the sharp localisation reported previously.

\clearpage \pagestyle{empty}
\begin{center}
  {\Huge Acknowledgements}\\[1cm]
\end{center}

Thanks, first and foremost, must go to Terry Rudolph who primarily supervised this project.  I am forever grateful to Terry
for his generosity, his sharing of deep insights and exciting ideas, his energy and humour, for helping me develop all aspects of my
life as a researcher, and for being such a good friend.  Many thanks also to Peter Knight, for guiding and supporting me throughout
my time at Imperial College, and for overseeing the joint programme on quantum optics, quantum computing and quantum information at
Imperial, which brings together so many very talented individuals.

I owe a debt of gratitude to many others at Imperial who have contributed in different ways during my postgraduate studies.   Thanks to
Jesus for sharing the template used to write this thesis.  Thanks to Almut for supervising a project on atom-cavity systems ---
during this time I learnt a tremendous amount on topics new to me at the start.  And thanks to Yuan Liang for being my ``best buddy''.
It has been great to have him to chat to about anything and everything, and I wish him and Puay-Sze the greatest happiness
on the arrival of Isaac, not too far away now.  I have made so many friends at Imperial, but will resist the tradition of
listing everybody, for fear of omitting some.  They know who they are.  Thanks to all for the comradely spirit I have enjoyed these
past three years.

I would also like to acknowledge all those with a common interest in ``all things relative in quantum mechanics''.  In particular, I have
fond memories of the workshop entitled ``Reference Frames and Superselection Rules in Quantum Information Theory'' and held in Waterloo, Canada in 2004.  Organised by Rob Spekkens and Stephen Bartlett, this workshop brought together for the first time people working in this budding area of research.  Thanks to Barry Sanders, whom I met for the first time at the workshop, for encouragement on my thesis topic.  Thanks also to Jacob
Dunningham and Ole Steuernagel for budding collaborations.

And finally, a big thanks to all my family.  Thanks particularly to Dad for helping me financially during my first year in the absence of
maintenance support, and to Dad and Aida for contributing towards a laptop which has transformed my working habits.

This work was supported in part by the UK EPSRC and by the European Union.

\vspace{1cm}
\textit{\center
I dedicate this thesis to Mum and Dad.  This work is my first major achievement since my mum's passing.  I am filled with the greatest
sadness that she cannot witness it.  Her love keeps me strong always.}

\tableofcontents

\listoffigures

% Local Variables:
% TeX-master: "./thesis"
% End:

%% file: LRDoF_v2_chapter_1.tex
\chapter{Introduction}
\label{chap:Introduction}

\section{Preface}

It is a widely accepted principle of modern physics that absolute physical quantities have no intrinsic usefulness or physical relevance.
Typically however, it is difficult to determine the extent to which some particular theory can be explicitly formulated in relational terms.
There is an ongoing debate, and a substantial literature, which explores different aspects of quantum mechanics from a relational point of
view, within diverse fields including quantum information, quantum gravity and foundational studies
of quantum mechanics.  This thesis contributes to this activity by presenting a comprehensive study of systems wherein some
relationally defined degree of freedom ``localises'' --- becoming well-defined, exhibiting strong correlation, and becoming
in some sense ``classical''\footnote{
The precise meaning of ``classical'' here depends on the physical system being studied.  Consider, for example, an optical system wherein a number of ``quantum'' light sources are phase-locked by processes of localisation, and are subsequently fed into a system of phase shifters, beam-splitters, and detectors.   The dynamical properties of the light are predicted to be the same as for ``classical light'' fields
with corresponding intensities and phase differences.  These fields are generically described by complex numbers and have absolute optical phases.}
--- under the action of some simple, well-characterised dynamical process.
Three physical systems are studied in depth, each built upon a simple measurement-based process.  The emphasis is then on the induced, post-measurement properties of states of these systems.  Chapters \ref{chap:OpticalOne} and \ref{chap:OpticalTwo} consider the localisation of {\it relative optical phase} in interference experiments wherein light from independent sources leaks onto a beam-splitter whose output ports are monitored by photodetectors.  Chapter \ref{chap:BEC} looks at the spatial interference of independently prepared Bose-Einstein condensates, a process wherein the localisation of the {\it relative atomic phase} plays a key role.  Finally, Chapter \ref{chap:ParticlesScatteringLight} discusses
the localisation of {\it relative positions} between massive particles as they scatter light.
This thesis contributes many new results using both analytical and numerical methods.  Up to now, there has been little attempt
to develop in detail the issues common to examples such as these.  This thesis lays out a ``modus operandi'' that can be applied widely.
Much of the content of this thesis has been published in \cite{Cable05}.

As an example of the relevance of this thesis topic, consider the highly involved controversy concerning the existence or otherwise of quantum coherence in several diverse contexts --- some key examples being the quantum states of laser light, Bose-Einstein condensates and Bardeen-Cooper-Schrieffer superconductors.  For a recent overview and fresh perspective on the controversy see \cite{Bartlett05}.
To take just one example, M{\o}lmer in his well-known publications \cite{Molmer97,MolmerB97} challenges the assumption, common in
quantum optics, that the state of the electromagnetic field generated by a laser is a Glauber coherent state,
with a fixed absolute phase and coherence between definite photon numbers.  By treating carefully the gain mechanism in a typical laser,
and pointing to the absence of further mechanisms to generate optical coherence in standard experiments,
he concludes that the correct form for the state of a laser field is in fact
a Poissonian improper mixture\footnote{An improper mixture is a density operator which cannot be given an ignorance interpretation.}
of number states with no absolute phase and no optical coherence.
A key question is then why two independent laser sources can demonstrate interference, as has been observed experimentally.
M{\o}lmer argued that in fact there is no contradiction, by detailed numerical studies.
He showed that a suitable process of photodetection acting on optical modes which are initially in number states,
can cause the optical modes to evolve to a highly entangled state, for which a stable pattern of interference may be observed.
This corresponds to the evolution of a well defined correlation in the phase difference between the modes.
Chapter \ref{chap:OpticalOne} takes up this example.

In studying localising relative degrees of freedom a number of questions suggest themselves regardless of the specific physical
realisation that is being considered.  How fast are the relative correlations created, and is there a limit to the
degree localisation that can be achieved?  What are the most appropriate ways of quantifying the degree of localisation?
How stable are the relative correlations once formed, for example against further applications of the
localising process with additional systems, interaction with a reservoir, and the free dynamics?
Does the emergence of relative correlations require entanglement between the component systems?
Does a localised relative quantum degree of freedom behave like a classical degree of freedom,
particularly in its transitive properties?  What role does an observer play in the process of localisation?

In addressing questions such as these, this thesis has several key objectives.
In addition to revisiting the more commonly considered examples of pure initial states,\footnote{
The studies of localising relative optical phase in \cite{Molmer97,MolmerB97,Sanders03}
consider specifically the case of initial photon number states, which are challenging to produce experimentally
(in fact M{\o}lmer's numerical simulations assume initial Fock states with order $10^4$ or $10^5$ photons).
Most of the existing literature concerning the spatial interference of independently prepared Bose-Einstein
condensates assumes initial atom number states, for example \cite{Javanainen96,Yoo97}.
The analysis of localising relative positions between initially delocalised mirrors or particles
in \cite{Rau03,Dunningham04} assumes initial momentum eigenstates.}
the focus is on initial states which can readily be prepared in the laboratory or are relevant
to processes happening in nature, and in particular on examples of initial states which are mixed.
Working with mixed states it is
important to be wary of common conceptual errors and in particular of committing the preferred ensemble fallacy.
The fallacy is to attribute special significance to a particular convex decomposition of a mixed state where it is wrong to do so.
Another key goal is to simplify the analyses as much as possible.  The emphasis is on deriving analytical results
rather than relying exclusively on stochastic numerical simulations.  A further goal is to identify descriptions of the various measurement
processes as positive operator-valued measures (POVM's), so as to separate out the characteristic localisation of the relevant relative
degree of freedom from the technical aspects of a particular physical system.  Identifying the relevant POVM's can also
facilitate analogy between localisation in different physical systems.  And finally
there is a preference for specifying in operational terms preparation procedures,
and measures of the speed and extent of localisation, rather than relying on abstract definitions with no clear physical meaning.

At this point mention should be made of some further topics with close connections to this thesis.  The first is that of
quantum reference frames.  Put simply, a frame of reference is a mechanism for breaking some symmetry.  To be consistent, the entities which act
as references should be treated using the same physical laws as the objects which the reference frame is used to describe.
However in pursuing this course in quantum mechanics there are a number of immediate difficulties to address.  There is the question
of the extent to which classical objects and fields are acceptable in the analysis.  Furthermore, in quantum mechanics establishing references
between the objects being referenced and the elements of the reference frame causes unavoidable physical disturbances.  Careful consideration
of the dynamical couplings between them is necessary, and in particular the effects of back-action due to measurement.  Finally,
translation to the reference frame of another observer itself requires further correlations to be established by some dynamical process
(in contrast to the simpler kinematical translations between frames possible in classical theories).

Another recurring topic is that of superselection rules.  In the traditional approach a superselection rule specifies that superpositions of the eigenstates of some conserved quantity cannot be prepared.  This thesis involves several examples where modes originally
prepared in states (pure or mixed) with a definite value of some conserved quantity express interference, but without violating
the corresponding superselection rule globally.  Following the more relational approach to superselection rules of Aharonov and Susskind \cite{Aharonov67} states thus prepared, with a well-defined value for the relative variable canonically conjugate to the conserved quantity, may be used in an operational sense to prepare and observe superpositions that would traditionally be considered ``forbidden''.  There is also the question as to whether superselection rules contribute to the robustness of the states with a well-defined relative correlation.  In particular, if typical dynamical processes obey the relevant  superselection rule then averaging over the ``absolute'' variables does not affect the longevity of these states.

\newpage
\section{Executive summary}

\begin{center}
\textbf{Chapter \ref{chap:OpticalOne}: Localising Relative Optical Phase}
\end{center}

Chapter \ref{chap:OpticalOne} looks at the interference of two optical modes with no prior phase correlation.
A simple setup is considered wherein light leaks out of two separate cavities onto a beam splitter whose output ports are monitored
by photocounters.  It is well known that when the cavities are initially in photon number states a pattern of interference is observed at the detectors \cite{Molmer97,MolmerB97}.  This is explored in depth in \sect{OptOneFockstates} following an analytic approach
first set out in \cite{Sanders03}.  The approach exploits the properties of Glauber coherent states which provide a mathematically
convenient basis for analysing the process of localisation of the relative optical phase which plays a central role in
the emerging interference phenomena.  For every run of the procedure the relative phase localises rapidly with successive photon detections.
A scalar function is identified for the relative phase distribution, and its asymptotic behaviour is explained.
The probabilities for all possible measurement outcomes a given time after the start are computed, and it is found that
no particular value of the localised relative phase is strongly preferred.  In the case of an ideal apparatus,
the symmetries of the setup lead to the evolution of what is termed here a ``relational Schr\"{o}dinger cat'' state,
which has components localised at two values of the relative phase.  However, if there are instabilities or asymmetries in the system,
such as a small frequency difference between the cavity modes, the modes always localise to a single value of the relative phase.
The robustness of these states localised at one value under processes which obey the photon number superselection rule is explained.

In \sect{OptOneQuantifyingLocalisation} a visibility is introduced so as to provide a rigorous and operational definition of
the degree of localisation of the relative phase.   The case of initial mixed states is treated in \sect{OptOneMixed}, looking specifically at the examples of Poissonian initial states and thermal initial states.  Surprisingly it is found that the localisation can
be as sharp as for the previous pure state example, and proceeds on the same rapid time scale
(though the localisation for initial thermal states is slower than for initial Poissonian states).  Differently from the pure state case
where the optical modes evolve to highly entangled states, in these mixed state examples the state of the optical modes remains separable
throughout the interference procedure.  Localisation at two values of the relative phase confuses the interpretation of the visibility,
and a mathematically precise solution is given for the Poissonian case.

\vspace{0.5cm}
\begin{center}
\textbf{Chapter \ref{chap:OpticalTwo}: Advanced Topics on Localising Relative Optical Phase}
\end{center}

Chapter \ref{chap:OpticalTwo} extends the programme of Chapter \ref{chap:OpticalOne} in a variety of directions.
The evolution of the cavity modes under the canonical interference procedure can be expressed simply in terms of
Kraus operators $a\pm b$ (where $a$ and $b$ are the annihilation operators for the two modes)\footnote{
If there is an additional fixed phase shift $\xi$ in the apparatus, the measurement operators take
the form $a\pm e^{i\xi} b$.  The characteristic localisation is the same.}
corresponding to photodetection at each of the photocounters which monitor the output ports of the beam splitter.
A formal derivation of these Kraus operators is presented in \sect{OPT2Krausopderivation}.
The situation of initial states with very different intensities in each mode is addressed in
\sect{OPT2AsymInitialStates}.  A key motivation for considering these asymmetric initial states is to shed
light on the situation when a microscopic system is probed by a macroscopic apparatus.
The discussion here takes the example of initial optical Poissonian states.
The relative phase distributions and the probabilities for different measurement outcomes are found to differ
substantially compared to the case of initial Poissonian states with equal intensities for the modes.
Special attention is paid to the questions of whether there are preferred values for the localised relative phase
and the speed of the localisation.
It is shown that the relative phase localises more slowly when the initial states are highly asymmetric.
The transitive properties of the localisation process are clarified in \sect{OPT2transitivity}, again
taking the example of initial Poissonian states.
The localisation process acts largely independently of prior phase correlations with external systems
(although the asymmetric depletion of population with respect to external modes has some effect).
The localised quantum relative phases have the same transitive properties as classical relative phases.
Loss of a mode does not alter the phase correlations between the systems which remain.

\sect{OPT2FockAddition} explores how the canonical interference procedure
could be used to engineer large photon number states using linear optics, classical
feed forward, and a source of single photons.  Single photons can be ``added'' probabilistically, by combining them at
a beam splitter and measuring one of the output ports, yielding a $2$-photon Fock state with probability $0.5$
(as suggested by the well-known Hong, Ou and Mandel dip experiment \cite{Hong87}).  In principle this procedure
could be iterated to yield progressively larger number states.  However, this is highly inefficient and
a protocol is presented here which greatly improves the success probabilities.  Simply stated, the idea is to
sacrifice a small number of photons from the input states before each ``addition'', in order to (partially) localise the relative
phase, and then to adjust the phase difference to $0$ or $\pi$.  Subsequent combination at a beam splitter, and measurement of
the output port with least intensity, yields a large Fock state at the other output port with much improved
probability.  Proposals for Heisenberg-limited interferometry provide one direct application of large photon number states.
In addition, when used as the initial states for the canonical interference procedure, large photon number states
can be used to make relational Sch\"{o}dinger cat states, and these also have potential applications.  For example,
when the components of a relational Schr\"{o}dinger cat state have relative phases different by approximately $\pi$,
the cat state can easily be converted into a ``NOON'' state\footnote{More specifically, a phase shifter and a beam splitter
can be used to convert the relational cat state into an ``approximate NOON state'' --- for which the amplitude of the
NOON state component is very much larger than other contributions.} (a superposition of Fock states of the form
$\ket{N}\!\ket{0}+\ket{0}\!\ket{N}$).  NOON states are currently attracting considerable research interest.

\sect{OPT2superselection} discussions a relational perspective on the topic of superselection rules.
Aharonov and Susskind suggest in \cite{Aharonov67} that superpositions forbidden in the conventional approach of
algebraic quantum field theory can, in fact, be observed in a fully operational sense, by preparing the apparatus
in certain special states.  However, the states suggested by Aharonov and Susskind are not easy to prepare.
Here optical mixed states with well localised relative phases and large intensities are presented as alternatives.
These can be readily prepared and serve the same purpose with no loss due to the lack of purity.
Finally \sect{OPT2opticalextensions} suggests possible future extensions of the work in Chapters \ref{chap:OpticalOne} and \ref{chap:OpticalTwo},
for example clarifying the consequences for the canonical interference procedure of detector inefficiencies typical
for experiments in the optical regime.

\vspace{0.5cm}
\begin{center}
\textbf{Chapter \ref{chap:BEC}: Interfering Independently Prepared Bose-Einstein Condensates and Localisation of the Relative Atomic Phase}
\end{center}

Chapter \ref{chap:BEC} looks at the interference of two, independently prepared, Bose-Einstein condensates which are released from their traps and imaged while falling, as they expand and overlap.  High contrast patterns of interference have been observed experimentally \cite{Andrews97}.
In many  theoretical treatments of Bose-Einstein condensation every condensate is assigned a macroscopic wavefunction.
This presumes an a priori symmetry breaking that endows a condensate with a definite absolute phase.
Interference is trivially predicted on this basis.  However this description poses various conceptual difficulties.
In particular, it implies coherences between different atom numbers at odds with conservation of atom number,
is commonly justified in terms of symmetry breaking fields with no clear physical relevance,
and invokes absolute phases with values which cannot be measured even in principle.
This is discussed more in \sect{BECstandardstory}. Such assumptions are, however,
not necessary to predict the spatial interference of independently prepared condensates,
as was first demonstrated in detail by Javanainen and Yoo \cite{Javanainen96}.
They studied the problem numerically for the case when the condensates are initially in number states.

A new analysis of the interference process is presented in \sect{BECspatialinterference}, based on the same measurement model as used by Javanainen and Yoo.  Localisation of the relative atomic phase plays a key role.  The process of localisation is the same as discussed in Chapter \ref{chap:OpticalOne} for localising relative phase between optical modes, in the case of an asymmetry or instability in the apparatus causing random phase shifts between photon detections.  The visibility, defined in \sect{OptOneQuantifyingLocalisation} of Chapter \ref{chap:OpticalOne}
to quantify the localisation of relative optical phase, can be translated into the current context.
It has a simple interpretation in terms of the probability distribution for single atom detection.  The case of initial Poissonian states is analysed in detail.  The localisation is messy and a novel method is presented to characterise it.  It is predicted that the relative phase distribution
after the first few atom detections, defined in terms of a basis of coherent states, takes the form of a Gaussian with width between $2/\sqrt{D}$ and $2\sqrt{2}/\sqrt{D}$ (where $D$ denotes the total number of such detections).  In particular, the relative phase is predicted to localise rapidly to one value, and very much faster than the emergence of the clearly defined patterns of interference simulated by Javanainen and Yoo, and others.  Numerical simulations produce results sustantially consistent with this analysis, although the analytically derived rate of localisation
is found to represent a slight underestimate.  The discussion proceeds to open questions asking what, in principle,
is lost when the spatial interference is analysed on the basis of a naive prior symmetry breaking.
Finally, \sect{BECOpticalRelPhaseMeasurement} reviews a recent experiment and several different theoretical proposals, concerning non-destructive measurements of the relative phases between condensates by optical means.

\vspace{0.5cm}
\begin{center}
\textbf{Chapter \ref{chap:ParticlesScatteringLight}: Joint Scattering off Delocalised Particles and Localising Relative Positions}
\end{center}

Chapter \ref{chap:ParticlesScatteringLight} looks at localising relative positions between massive particles scattering light.  The starting point is a recent article \cite{Rau03} which examines two simple models of scattering, which are reviewed at length in \sect{PSLtwoscatteringmodels}.
In the first ``rubber cavity'' model, a succession of photons pass through a Mach-Zehnder interferometer and are detected at photocounters monitoring the output ports, localising the relative position between two delocalised mirrors in the interferometer.  In the second ``free particle'' model, plane wave photons are scattered off two particles delocalised in a one dimensional region and are detected in the far field.  The photons are either deflected at a definite angle or continue in the forward direction.  The localisation in the ``rubber cavity'' model resembles that discussed in Chapter \ref{chap:OpticalOne} concerning relative optical phase.  Differently, when the light source is monochromatic the localisation
of the relative position is periodic on the order of the wavelength of the light.
For the ``free particle'' model the greatest difference is that the momentum kick imparted for each photodetection is variable, leading to localisation at a single value.

The localisation in the free particle model of scattering is explored in detail in \sect{PSLActionOnGaussians} and  \sect{PSLlocalisingthermalparticles}, making different assumptions.  The initial states in \cite{Rau03} are momentum eigenstates which are not particularly realistic.  Instead the initial state of the particles is taken to be thermal.  The situation considered is that of an observer viewing a distant light source.  The incident light is either forward scattered by the particles into the field of view of the observer or deflected, in which case the light source is observed to dim.  Results are presented for the cases of the incident light being monochromatic and thermal.  In both cases the localisation is only partial even after many detections, in contrast to the sharp localisation reported in \cite{Rau03}.  Possible future calculations are suggested in \sect{PSLfutureextensions}.

\vspace{0.5cm}
\begin{center}
\textbf{Chapter \ref{chap:Outlook}: Outlook}
\end{center}

The Outlook suggests several possible directions for future research on this thesis topic.
The ``modus operandi'' developed in Chapters \ref{chap:OpticalOne} through to \ref{chap:ParticlesScatteringLight}
can easily be adapted to answer a range of further questions, and can also be applied to other physical systems.
For example, it should be possible to analyse processes localising the relative angle between two spin systems
along the lines of calculations in this thesis.  In another direction, it is expected that the discussion in Chapter
\ref{chap:BEC}, concerning the interference of atomic Bose-Einstein condensates, is relevant to systems of superconductors.
This is suggested by the fact that Bose condensation of Cooper pairs --- weakly bound electron pairs ---
plays a central role in the Bardeen-Cooper-Schrieffer theory of superconductivity.
One concrete system where localisation of a relative superconducting order parameter
could be investigated is that of bulk superconductors placed close together, and coherently coupled by a mechanically
oscillating superconducting grain.  Theoretical studies have been published which provide a well characterised model for
the dynamical evolution of such a setup.

Chapters \ref{chap:OpticalOne}, \ref{chap:OpticalTwo} and \ref{chap:BEC}, which focus in different ways on the
localisation of relative quantum phases, are relevant to the debate concerning different mathematical
characterisation of phase measurements in quantum mechanics.  One possible future project might study the
expected decorrelation of some relative number variable, that would be expected to accompany the localisation of a given relative
phase variable.  For example, what happens when the respective modes have different intensities?
This might shed some light on the dictum ``number and phase are canonically conjugate quantum variables''.
In another direction, processes of localisation are relevant to synchronising ``quantum clocks'',
the subject of an extensive literature concerned with problems of time in
quantum mechanics.  Finally, states with a well defined relative correlation, of the type whose preparation is discussed in
this thesis, have potential application as pointer states in the theory of decoherence.  These states
exhibit quasi-classical properties and, in many cases are predicted to be long lived with respect
to coupling to an environment.

%% file: LRDoF_v2_chapter_2.tex
\chapter{Localising Relative Optical Phase}
\label{chap:OpticalOne}

This chapter looks at the interference of two, fixed frequency, optical modes which have no prior phase correlation,
presenting and extending work published in \cite{Cable05}.  A simple setup is considered
wherein light leaks out of two separate cavities onto a beam splitter whose output ports are monitored by
photodetectors.  It is well known that when the cavities are initially in photon number states
a pattern of interference is observed at the detectors.
The dynamical localisation of the relative optical phase plays a key role in this process.
Key studies are presented in \cite{Molmer97,MolmerB97,Sanders03} and some results are also reported in \cite{Chough97},
all focusing on the case of initial number states with the same photon number in each mode.
Some results for the first and second photodetections are also presented in \cite{Pegg05} for the general
case of mixed initial states with zero optical coherences.
The work here adopts the analytic approach introduced in \cite{Sanders03} and represents a substantial
development of the topic.

\sect{OptOneFockstates} looks in detail at the case when the cavity modes are initially in photon number states.
It first explains the basic setup and provides some intuition as to why interference phenomenon are observed.
M{\o}lmer's numerical treatment of the problem in \cite{Molmer97,MolmerB97} is also summarised.
Adopting a basis of Glauber coherent states,
\sect{OptOnelocalisingscalarfunction} investigates the evolution of the relative phase distribution for all possible
measurement sequences.  Expressions for the asymptotic forms of the relative phase distributions are also given.
The probabilities for all possible measurement outcomes
a given time after the start are evaluated in \sect{OptOneprobabilities},
and it is found that no particular value of the localised relative phase is preferred in this
example.\footnote{Different outcomes at the detectors are associated with localisation at different values of the relative phase.
Some measurement outcomes are more likely than others.  However the least likely measurement outcomes cause
localisation at values which are closer together.
Overall the localisation process does not ``prefer'' any
particular range for the localised relative phases for this choice of initial states.}
In addition it is seen that an additional fixed phase shift in one of the arms of the apparatus does not alter the experiment.
The emergence in an ideal experiment (one without phase instabilities) of what is termed a relational
Schr\"{o}dinger cat state is discussed in \sect{OptOnesymmetries}.  \sect{OptOnerobustness} discusses the
robustness of states sharply localised at one value of the relative phase,
including a brief summary of the relevant simulations in \cite{MolmerB97}.

A visibility is introduced in \sect{OptOneQuantifyingLocalisation}
as a means to rigorously quantify the degree of localisation of the relative phase.
It ranges from $0$ (no phase correlation) to $1$ (perfect phase correlation).
In \sect{OptOneMixed} the analysis of \sect{OptOneFockstates} is extended to the case of mixed initial states,
which is more realistic experimentally.
Specifically the case of Poissonian initial states is treated in \sect{OptOnePoissonian},
and of thermal initial states in \sect{OptOnethermal}.  Differently from the pure state case where
the localised state of the two cavity modes is highly entangled, in these mixed state examples
the state of the cavity modes remains separable throughout the interference procedure.
The visibilities for the final states are computed for all possible measurement sequences.
Surprisingly it is found for both examples that the localisation can be as sharp as for initial pure states,
and proceeds on the same rapid time scale.  In fact the localisation turns out to be slightly slower for the thermal case.
For Poissonian initial states the visibility jumps from $0$ to $1/2$ after just one measurement,
whereas for thermal initial states is jumps to $1/3$.
When the interference procedure involves the detection of more than one photons, and photons are registered at both detectors,
the relative phase localises at two values.  The visibilities for these cases
underestimate of the true degree of localisation, and a solution is presented for the Poissonian case.
Finally \ref{sec:OptOneNonIdealPhotodetectors} discusses the consequences of errors from the photodetectors,
which is a feature of any real experiment.

\section{Analysis of the canonical interference procedure for pure initial states}
\label{sec:OptOneFockstates}

\begin{figure}[h]
\begin{center}
\includegraphics[height=7cm]{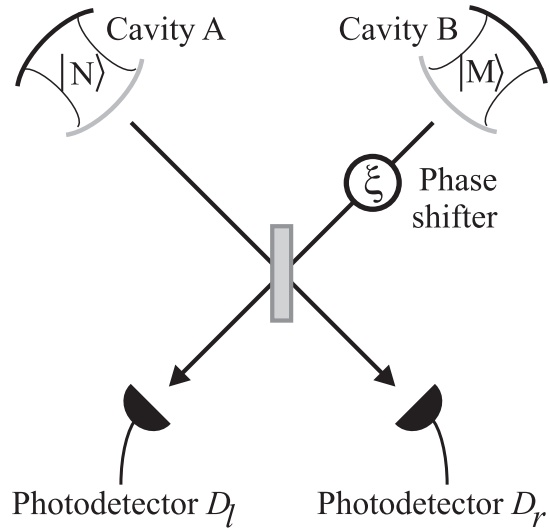}
\caption{\label{fig:OptOnetwocavities} Photon number states leak out of their cavities and are combined
on a 50:50 beam splitter. The two output ports are monitored by photodetectors. In the first instance the variable phase
shift $\xi$ is fixed at $0$ for the duration of the procedure.}
\end{center}
\end{figure}

A simple operational procedure for both causing and probing the localisation
between two optical modes
is depicted in \figu{OptOnetwocavities}. Two cavities
containing $N$ and $M$ photons respectively at the start (and thus described by pure initial
states $|N\>$ and $|M\>$) both leak out one end mirror (via linear mode coupling). Their outputs
are combined on a $50:50$ beam splitter, after which they are detected at two photocounters.

Despite the cavities initially being in Fock states with no well-defined relative phase it is well
known that an interference pattern is observed at the two detectors. The interference
pattern can be observed in time if the two cavities are populated by photons of slightly
differing frequencies or, as in standard interferometry, by varying a phase shifter
placed in one of the beam splitter ports.  Despite the evolution for the system taking place
under an effective superselection rule for photon number, coherence phenomena depending on the conjugate phases
are thus observed.  The reason for this contradiction with the
dictum ``number and phase are conjugate quantities'' may be understood as follows.

Consider the case after a single photon has been detected at one of the detectors. Then
the new state of the two cavities is $\sqrt{\tfrac{N}{N+M}}|N\!-\!1\>|M\>\pm\sqrt{\tfrac{M}{N+M}}|N\>|M\!-\!1\>$ i.e. it is entangled.
It is simple to show that the second photon is much more likely to be registered at
the same detector. The exact ratio of the probabilities of being counted at the same detector and
at the other is $N^2+M^2-N-M+4NM$ to $N^2+M^2-N-M$.
When $N=M$ this ratio is strictly greater than $3$, and tends sharply to infinity as $N$ and $M$ approach $1$.  This is in
agreement with the phenomenon demonstrated by the well-known Hong, Ou and Mandel dip experiment \cite{Hong87},
whereby two uncorrelated and identical photons, simultaneously incident on the input ports of a $50:50$ beam splitter must
both be registered at the same output port.  Further detections lead to a more and more entangled
state. It is not so surprising then that detections on an entangled state lead to some form of
interference pattern. In essence after a small number of detections the relative number
of photons in each cavity is no longer well defined, and so a well defined relative phase
can emerge. Note that this is only possible if the beam splitter, detectors and cavities all
have well-defined relative positions.

One method for confirming this intuition is to use  a quantum jumps approach
(for a review of quantum jump methods see \cite{Plenio98} and references therein), and
numerically simulate such a system through a number of detection procedures.
Such an approach was taken in Sec.~III of \cite{Molmer97} and Sec.~4 of \cite{MolmerB97},
and some important features of those studies are highlighted here.
M{\o}lmer assumes a frequency difference $\omega_b\!-\!\omega_a$  between the two cavities,
which gives rise to an interference pattern in time.
The cavities are assigned equal decay rates $\Gamma$ for leakage onto the beam splitter,
leading to exponential decay laws for the cavity intensities and the photon count rates.
M{\o}lmer looks at the case that both cavities start with the same number of photons, $N$ say,
and the chosen values of $N$ are large.
In \cite{Molmer97} the choice of parameters is $N\!=\!10^5$ and
$\omega_b\!-\!\omega_a\!=\!1000\Gamma$, and in \cite{MolmerB97} $N\!=\!5000$ and $\omega_b\!-\!\omega_a\!=\!30\Gamma$.\footnote{
In contrast the analysis in this chapter for pure initial states applies for arbitrary initial cavity photon number $N$.}
Sharp localisation of the relative phase occurs as the first few photons are detected, which
manifests itself as continuous oscillation
in the count rates at the two photodetectors as further photons are registered.
This oscillation takes place on a time scale $1/(\omega_b-\omega_a)$,
with a high count rate alternating between the detectors.
M{\o}lmer also looks at the evolution of the quantity $Q=\bra{\psi} a^\dagger b \ket{\psi}$ (where $\ket{\psi}$
is the state of the two cavities and, $a^{\dagger}$ and $b$ are creation and annihilation operators for cavities $A$ and $B$ respectively).\footnote{
More specifically he considers $2\text{Re}(Q)/(2N-D)$ (where $\text{Re}(\cdot)$ denotes the real part of some complex number
and $D$ denotes the total number of detected photons).} This quantity determines the count rates at the detectors and
makes a rapid transition from a random to a harmonic evolution as the first few photons are registered.
However simulations such as these yield little in the way of physical insight.
The following discussion is based instead on an analysis introduced in \cite{Sanders03}.

\subsection{Evolution of the localising scalar function}
\label{sec:OptOnelocalisingscalarfunction}

To begin,
the initial state $|\psi_I\>=|N\>|M\>$ of the cavities
is expanded in terms of coherent states
$|\alpha\>,|\beta\>$:
\be
\label{eqn:OptOneISpure}
|\psi_I\>=\mathcal{N}\int_0^{2\pi} \!\!\!\int_0^{2\pi}
\!\!\!d\theta d\phi e^{-i(N\theta+M\phi)} |\alpha\>|\beta\>
\ee
with $\alpha=\sqrt{N}e^{i\theta}$, $\beta=\sqrt{M}e^{i\phi}$, and the normalisation
$\mathcal{N}=1/\sqrt{\Pi_N(N) \Pi_M(M)}{4\pi^2}$  where $\Pi_n(\mu)=\mu^n e^{-\mu}/n!$ is the
Poissonian distribution.  For the properties of coherent states refer \cite{Glauber63}
(for pedagogical reviews see \cite{Klauder85,Gerry04}).
At first the normalisation in \eq{OptOneISpure} will be ignored for simplicity.

Consider now the case that a single photon is detected at either the left detector $D_L$
or the right one $D_R$. Since only the change of the state of the cavity modes is of interest
the exterior modes may be treated as ancillae, and corresponding Kraus operators $K_L$ and $K_R$
describing the effect of the detection on the cavity modes only may be determined
(for an explanation of the basic properties of Kraus operators see for example
Chapter $8$ of \cite{NielsenChuang00}).  Treating in a basis of coherent states
the leakage of the cavity populations into the ancillae, and subsequent
combination at a beam splitter and photodetection, as is done in
\cite{Sanders03}, immediately suggests the form of $K_L$ and $K_R$ as
proportional to $a\pm b$ (where $a$ and $b$ are annihilation operators for the modes in cavity
$A$ and $B$ respectively).  The constant of proportionality depends on the transmittivity of the cavity
end mirrors.  For the purposes here of examining the main features of the
localising relative phase, this result is assumed to be correct, and a careful verification
is delayed to \sect{OPT2Krausopderivation} of Chapter \ref{chap:OpticalTwo}.\footnote{
The operators $a\pm b$ are also derived in \cite{Molmer97,MolmerB97} and \cite{Pegg05} using different arguments.
Detailed calculations later in this chapter do not assume this result but treat
the evolution of the cavity and ancilla modes in full.}
It may be observed immediately that these simple expressions for $K_L$ and $K_R$ constitute a complete
measurement process.  This reflects the assumptions that the beam splitter is lossless and acts
as a unitary process on the ancillae.  The equal weighting of $a$ and $b$ reflects the assumptions
that the beam splitter is $50:50$ splitting, and that rates of the leakage from both cavities onto the beam splitter
are the same.

In the event that some number $l$ of photons are registered at $D_L$ and $r$ at
$D_R$, the state of the two cavities evolves as follows:
\begin{eqnarray*}
|\psi_I\>&\rightarrow& K_L^l K_R^r |\psi_I\>\\
 &\propto& K_L^l K_R^r \int \!\!\!\int\!\!\!d\theta d\phi e^{-i(N\theta+M\phi)} |\alpha\>|\beta\> \\
 &\propto& \int \!\!\!\int\!\!\!d\theta d\phi e^{-i(N\theta+M\phi)} (\alpha-\beta)^l (\alpha+\beta)^r  |\alpha\>|\beta\>
\end{eqnarray*}
The scalar function,
\be
C_{l,r}\equiv (\alpha-\beta)^l (\alpha+\beta)^r,
\ee
encodes information about the localisation in the relative phase which occurs between the two cavities.
It should be pointed out that in this example there is some ambiguity over the definition of
$C_{l,r}$.  In expanding a Fock state in terms of coherent states,
$\ket{N}=\frac{1}{\sqrt{\Pi_n(m)}} \int_0^{2\pi} \frac{d\varphi}{2\pi} e^{-in\varphi} \ket{\sqrt{m} e^{i\varphi}}$,
the number $m$ is free to take any positive value --- the key feature of the expansion for a state with $N$ photons
is that the integral must encircle the origin in phase space $N$ times (where the phase space is the complex plane for the
variable labelling the basis of coherent states).
Altering the relative amplitudes of
$\alpha$ and $\beta$ in \eq{OptOneISpure} will alter the relative phase distribution contained in
$C_{l,r}$.
When the cavities begin in the same photon number state the
natural choice is to set the amplitudes of the coherent states the same for both modes.

\begin{figure}[h]
\begin{center}
\includegraphics[height=5.6cm]{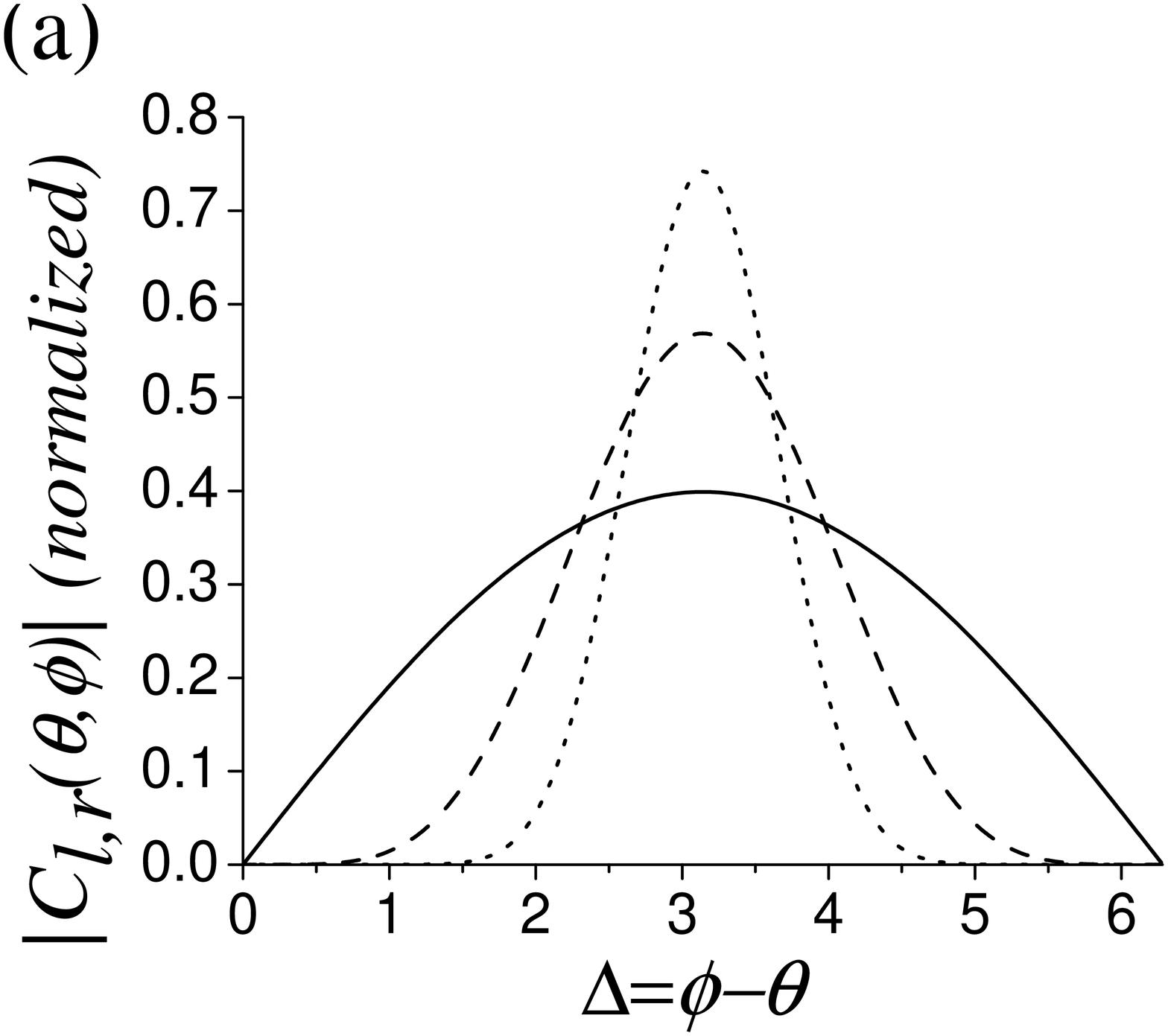}
\includegraphics[height=5.6cm]{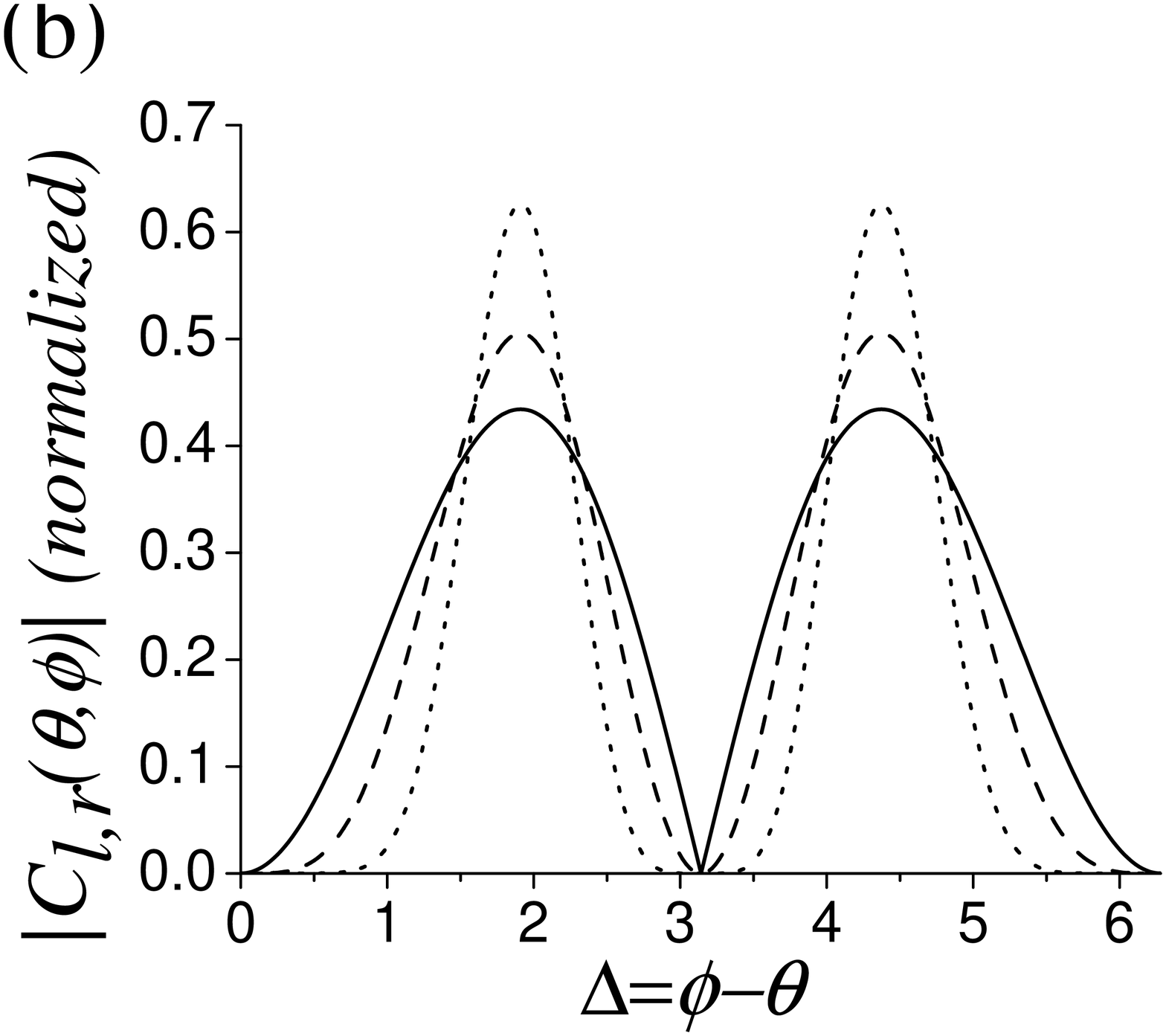}
\caption{\label{fig:OptOneRLoptical} The evolution of $C_{l,r}(\theta,\phi)$.
In (a) localisation about $\Delta_0=\pi$ after 1, 5 and 15 counts when photons are recorded in the left photodetector only.
(b) localisation about $\Delta_0 = \pm 2\arccos \left( 1/\sqrt{3} \right) \sim 1.9$ after 3, 6 and 15 counts when twice as many photons are recorded in the left detector as the right one. The symmetry properties of the Kraus operators $K_L$ and $K_R$ cause
$C_{l,r}$ to have multiple peaks (either one or two for $\Delta$ ranging on an interval of $2\pi$).}
\end{center}
\end{figure}

For the purposes here it is sufficient to focus on the symmetric case $N=M$
when both cavities begin in the same state.
It should be noted however that the physics of the highly asymmetric
case is somewhat different (for further discussion refer \sect{OPT2AsymInitialStates} of Chapter \ref{chap:OpticalTwo}).
Setting $|\alpha|=|\beta|$, the ``localising scalar function'' $C_{l,r}$ takes the form,
\begin{eqnarray}
\label{eqn:OptOneClrcavitiesequal}
C_{l,r}(\theta,\phi)
\!\!\! &=& \!\!\! N^{(l+r)/2} (e^{i\theta}-e^{i\phi})^l (e^{i\theta}+e^{i\phi})^r \\
\!\!\! &=& \!\!\! (4N)^{(l+r)/2} (-i)^l e^{i(l+r)(\theta+\phi)/2}  \sin^l {\! \tfrac{\Delta}{2}} \cos^r \! \tfrac{\Delta}{2} \nonumber
\end{eqnarray}
where $\Delta\equiv \phi-\theta$. Factors that do not
depend on $\theta,\phi$ will be ignored for the moment, since they will be taken care of by normalisation.

Of particular interest is the behaviour of $C_{l,r}(\theta,\phi)$ as the total number
of detections $l+r$ gets larger.
Asymptotic expansions \cite{Rowe01} for $C_{l,r}(\theta,\phi)$
can be used to examine this limit.
When photons are detected at both detectors,
\begin{eqnarray}
\label{eqn:OptOnebothportsasympt}
\left| \sin^{l} \tfrac{\Delta}{2} \cos^{r} \tfrac{\Delta}{2} \right|
\! &\approx& \!
\sqrt{\frac{l^{l}r^{r}}{ l\!+\!r \,^{l+r}}}\,{\exp} \Big[ -  \tfrac{l\!+\!r}{4} \left( \Delta \!-\! \Delta_0 \right) ^{2} \Big]
\end{eqnarray}
where $\Delta_0 \equiv 2 \arccos\sqrt{r/(r+l)}$ when $\Delta$ takes values between $0$ and $\pi$,
and $\Delta_0 \equiv 2\pi \! - \!  2 \arccos\sqrt{r/(r+l)}$ between $\pi$ and $2\pi$.
$\Delta_0$ denotes the values of the relative phase around which the localisation occurs.
When all the photons are detected at one detector the appropriate expressions are
\begin{eqnarray}
\label{eqn:OptOneoneportasympt}
\left\vert \cos ^{r}\tfrac{\Delta }{2}\right\vert
\! &\approx& \!
{\exp} \Big[ - \tfrac{r}{8}\Delta ^{2} \Big] \;\;\;{\rm for}\; \Delta \in [-\pi,\pi], \nonumber \\
\left\vert \sin ^{l}\tfrac{\Delta }{2}\right\vert
\! &\approx& \!
{\exp} \Big[ - \tfrac{l}{8}\! \left( \Delta \!-\! \pi \right) ^{2} \Big] \;\;\;{\rm for}\; \Delta \in [0,2\pi].
\end{eqnarray}
In every case asymptotically $\vert C_{l,r} \vert$ takes the form of a Gaussian distribution with
width (two standard deviations), $4/\sqrt{r}$ (or $4/\sqrt{l}$) when the counts are all at one detector
or $2\sqrt{2}/\sqrt{l+r}$ otherwise, decreasing with the total number of detections.

As $l+r$ grows the state of the two cavities evolves into a
superposition (over global phase) of coherent states with an increasingly sharply defined
relative phase. A plot showing the evolution of $C_{l,r}(\theta,\phi)$ is shown in
\figu{OptOneRLoptical}. This localisation in the relative phase is
responsible for the interference phenomena seen at the two detectors, as was examined
numerically in \cite{Molmer97,MolmerB97}.

\subsection{The probabilities for different measurement outcomes}
\label{sec:OptOneprobabilities}

The value $\Delta_0$ at which the relative phase
localisation occurs depends on (the ratio of) the specific number of photons $l$ and $r$
detected at each detector. This is of course probabilistic. Letting $P_{l,r}$ denote the
probability of detecting $l$ and $r$ photons at the left and right detectors respectively,
a complete expression for $P_{l,r}$ may be obtained by a simple heuristic treatment of the dynamics as in \cite{Sanders03}.
It is supposed that population leaks out of each cavity modes into an ancilla
according to a linear coupling with parameter $\epsilon$, where $\epsilon$ is small.
The ancillae evolve under the action of a $50:50$ beam splitter.  Subsequent action of
the projection operators $\ket{l}\bra{l}$ and $\ket{r}\bra{r}$ on the ancillae corresponds to the
detection of $l$ photons at the left detector and $r$ photons at the right detector.

It is helpful here to take a brief diversion to clarify the mathematics of a general dynamical
evolution under some linear mode coupling, which is supposed here to govern the leakage from each cavity,
and the action of the beam splitter.  Denoting the creation and annihilation operators for some pair of modes
$\hat{c}^{\dagger}, \hat{c}$ and $\hat{d}^{\dagger},\hat{d}$,
a linear mode coupling Hamiltonian is a linear combination of energy conserving terms of the form,
\[
H_{\rm lmc}=i\left( \vartheta e^{-i\xi }\hat{c}^{\dagger }\hat{d}-\vartheta e^{i\xi }\hat{c}\hat{d}^{\dagger }\right) \text{,}
\]
where $\vartheta$ and $\xi$ are real parameters.
Evolution over some time $t$ is then given by the unitary operator,
\bea
U_{\rm lmc}\left( \varphi ,\xi \right)
&=&\exp \left( -i\hat{H}_{\rm lmc}t\right) \nonumber \\
&=&\exp \left( \varphi e^{-i\xi }\hat{c}^{\dagger }\hat{d}-\varphi e^{i\xi }\hat{c}\hat{d}^{\dagger }\right), \nonumber
\eea
where $\varphi \equiv \vartheta t$.   One particularly important property of linear mode couplings is that they evolve products
of coherent states to products of coherent states.  Specifically for coherent states with complex parameters $\alpha$ and $\beta$,
\[
U_{\rm lmc}\left( \vartheta ,\xi \right) \left\vert \alpha \right\rangle \left\vert \beta \right\rangle
=\left\vert \alpha \cos \vartheta +\beta e^{-i\xi }\sin \vartheta \right\rangle
\left\vert -\alpha e^{i\xi }\sin \vartheta +\beta \cos \vartheta \right\rangle.
\]
Leakage into the vacuum is described by $\beta=0$ and a small value for $\vartheta$ so that
$\cos\vartheta=\sqrt{1\!-\!\epsilon}$ and $\sin\vartheta=\sqrt{\epsilon}$ where $\epsilon$ is small
(and say $\xi=\pi$).  The linear mode coupling then acts as,
\[
\ket{\alpha}\ket{0}\longrightarrow\ket{\sqrt{1\!-\!\epsilon}\alpha}\ket{\sqrt{\epsilon}\alpha}.
\]
For a $50:50$ beam splitter $\vartheta=\pi/4$ so that (with the additional phase shift $\xi=0$ say),
\[
\left\vert \alpha \right\rangle \left\vert \beta \right\rangle
\rightarrow \left\vert \frac{\alpha +\beta }{\sqrt{2}}\right\rangle \left\vert \frac{-\alpha +\beta }{\sqrt{2}}\right\rangle.
\]
Many calculations are facilitated by considering how a linear mode coupling transforms the creation and annihiliation operators.
This is given by,\footnote{Sometimes $\hat{c}$ and $\hat{d}$ are referred as ``input'' field operators and denoted $\hat{c}_{\rm in}$ and
$\hat{d}_{\rm in}$ say, while after unitary transformation the operators are called ``output'' field operators so that
$\hat{c}_{\rm out}=U^{\dagger} \hat{c}_{\rm in} U$ and $\hat{d}_{\rm out}=U^{\dagger} \hat{d}_{\rm in} U$.  This notation is not used here.}
\bea
U^{\dagger }(\theta ,\phi )\hat{c}U\left( \theta ,\phi \right) &=& \cos \theta \hat{c}+e^{-i\phi }\sin \theta \hat{d} \nonumber \\
U^{\dagger }(\theta ,\phi )\hat{d}U\left( \theta ,\phi \right) &=& -e^{i\phi }\sin \theta \hat{c}+\cos \theta \hat{d} \,.\nonumber
\eea

Returning to the interference procedure
the full expression for the cavity modes after the measurement process has acted is,
\[
 \mathcal{N} \!\! \int \!\! d\theta d\phi e^{-iN(\theta +\phi )}C_{l,r}(\epsilon ,\theta ,\phi )\left\vert \sqrt{1-\epsilon }\alpha \right\rangle \left\vert \sqrt{1-\epsilon }\beta \right\rangle,
\]
where the normalisation factor $\mathcal{N} = (\Pi_{N}(N)4\pi ^2)^{-1}$ and,
\be
\label{eqn:OptOneClrepsilon}
 C_{l,r}(\epsilon ,\theta ,\phi )
 =
 \left\<r \, \Big{\vert} \sqrt{\epsilon }\frac{\alpha +\beta }{\sqrt{2}}\right\rangle
 \left\<l \, \Big{\vert} \sqrt{\epsilon }\frac{-\alpha +\beta }{\sqrt{2}}\right\rangle,
\ee
where $\bra{r}$ and $\bra{l}$ denote photon number states.
The probability $P_{l,r}$ is given by,
\[
\mathcal{N}^{2} \!\! \int \!\! d\theta d\theta ^{\prime }d\phi d\phi ^{\prime }e^{-\frac{i}{2}\left( 2N-r-l\right) (\theta +\phi )}e^{\frac{i}{2}\left( 2N-r-l\right) (\theta ^{\prime }+\phi ^{\prime })}
\]
\[
\times \,\, C_{l,r}(\epsilon ,\theta ,\phi )C_{l,r}(\epsilon ,\theta ^{\prime },\phi ^{\prime })^{\ast }
\]
\[
\times \,\, \left\langle \sqrt{1-\epsilon }\alpha ^{\prime }\right\vert \left\vert \sqrt{1-\epsilon }\alpha \right\rangle \left\langle \sqrt{1-\epsilon }\beta ^{\prime }\right\vert \left\vert \sqrt{1-\epsilon }\beta \right\rangle.
\]

To simplify this expression one can at first proceed na\"{i}vely, taking the overlaps of the basis
states to be zero, to derive an approximate expression for $P_{l,r}$ which can be compared numerically to the real values.
It might be guessed that the approximate expression thus derived will hold good when several,
but not too many, photons have been recorded so that
$C_{l,r}$ is narrow while the amplitudes $\sqrt{1-\epsilon}\alpha, \sqrt{1-\epsilon}\beta$ are still large.
Hence assuming,
\begin{eqnarray*}
\left \langle \alpha' \vert \alpha \right \rangle &=& \exp\left(-|\alpha-\alpha'|^2\right) \sim \delta (\phi-\phi') \\
\left \langle \beta' \vert \beta \right \rangle &=& \exp\left(-|\beta-\beta'|^2\right) \sim \delta (\theta-\theta'),
\end{eqnarray*}
and using the relation for the gamma function $\Gamma(.)$,
\[
\int_{0}^{2\pi } \!\!\! \int_{0}^{2\pi }\tfrac{d\phi }{2\pi }\tfrac{d\theta }{2\pi }\cos ^{2r} \tfrac{\Delta }{2} \sin ^{2l} \tfrac{\Delta}{2} = \frac{\Gamma (r+0.5)\Gamma (l+0.5)}{\pi \Gamma (r+l+1)}
\]
the following approximation for $P_{l,r}$ is obtained,
\be
\label{eqn:OptOneFockProbApprox}
P_{l,r} \! \approx \!\!
\left[ \frac{(2\epsilon N)^{r+l}}{\left( r+l\right) !}e^{-2\epsilon N} \! \right] \!\!
\frac{\left( r+l\right) !}{r!l!}\frac{\Gamma (r+0.5)\Gamma (l+0.5)}{\pi \Gamma (r+l+1)}.
\ee
It is found numerically that this approximation is surprisingly good,
and applies quite generally whenever $\epsilon$ is small.
The fractional error goes roughly as $0.6\epsilon$, growing linearly with the leakage parameter.
As for its general features $P_{l,r}$ is seen to be a product
of a global Poissonian distribution in the total number of detected photons $l+r$
and a second function depending on the precise ratio of counts at $D_l$ and $D_r$.

\begin{figure}
\begin{center}
\includegraphics[height=9cm]{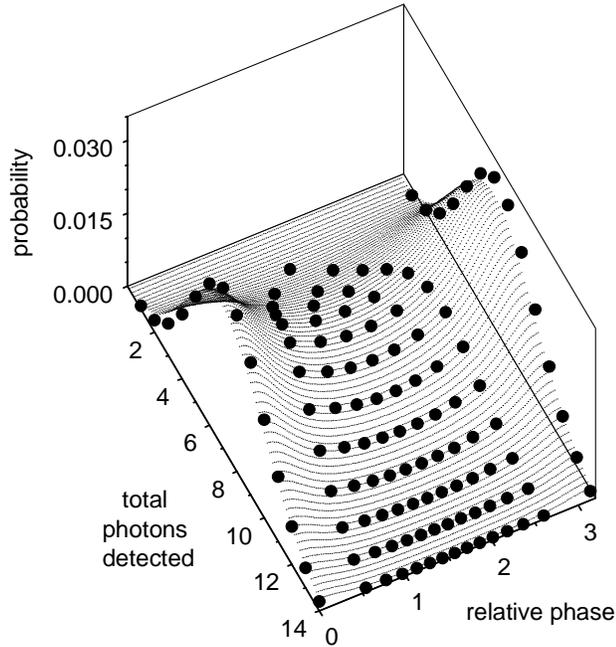}
\caption{\label{fig:OptOneFockPlr}
A plot of the exact values of the probabilities $P_{l,r}$ for all the possible measurement outcomes to the procedure
a finite time after the start, against the absolute value of the relative phase which is evolved.
The initial state is $\left\vert 20 \right\rangle \left\vert 20 \right\rangle$ and the leakage parameter $\epsilon$,
corresponding roughly to the time, has a value of $0.2$.  Each spot corresponds to
a different measurement outcome with $l$ and $r$ counts at detectors $D_l$ and $D_r$ respectively.
The value $\Delta_0$ of the relative phase which evolves in each case is given by $2 \arccos \left( \sqrt{r/(r+l)} \right)$.}
\end{center}
\end{figure}

A plot of the exact values for the probabilities $P_{l,r}$ for different measurement outcomes is plotted
in \figu{OptOneFockPlr} for typical parameter values $\epsilon=0.2$ and initial state
$\left\vert 20 \right\rangle \left\vert 20 \right\rangle$, with each spot corresponding to a possible outcome.
$\epsilon$ corresponds to a time parameter, an approximation which holds good provided $\epsilon$ is not too large.
The distribution $P_{l,r}$ gives the likely degree of localisation of the relative phase $\Delta$
a finite time after the start of the procedure, and the values $\Delta_0$ which are picked out.
Looking at the precise distribution in \figu{OptOneFockPlr},
it is seen that given $\epsilon=0.2$ it is most likely that 7 photons (approximately $2 \epsilon N$) have been counted
(corresponding to the ridge).  For the most probable outcomes all the photons are counted at one detector,
in which case the relative phase localises at $0$ or $\pi$.
However the density of points is greatest about $\Delta_0=\frac{\pi}{2}$, and for these outcomes there are approximately equal counts
at both detectors. Overall no particular value of the localised relative phase is preferred in this example.

Finally it is interesting to ask how the probabilities for different sequences
of detections are altered if the phase shift $\xi$
between cavity B and the beam splitter (refer \figu{OptOnetwocavities}) is a fixed at a value not equal to
$0$ for the duration of the procedure.
The phase shift alters the function $C_{l,r}(\epsilon,\theta,\phi)$,
\eq{OptOneClrepsilon}, according to a translation in the relative phase i.e.
$\Delta \equiv \phi-\theta\rightarrow\Delta+\xi$.  The Kraus operators $K_l$ and $K_r$
for the altered apparatus are $a\pm e^{i\xi}b$.
Looking now at the effect of r ``right''
detections and l ``left'' detections on the initial state $\ket{N}\ket{N}$
in a basis of Fock states,
\bea
\ket{N}\ket{N} &\rightarrow& (a+e^{i\xi}b)^r (a-e^{i\xi}b)^l \ket{N}\ket{N} \nonumber \\
&=& \Sigma_{j=0}^{l+r} d_j e^{i(l+r-j)\xi} \ket{N-j}\ket{N-l-r+j},
\eea
where the $d_j$ are some coefficients independent of $\xi$, reveals that
the dependence on $\xi$ drops out on taking the norm of the final state.
Hence the probabilities $P_{l,r}$
are not affected by the additional phase shift despite the asymmetry in the apparatus.
The phase shift $\xi$ is seen to be relevant operationally only if altered at different stages during the procedure.

\subsection{Symmetries of the localising procedure and relational Schr\"{o}dinger cat states}
\label{sec:OptOnesymmetries}

Once a given measurement outcome has occurred with $l$ and $r$ counts at the left and right detectors
respectively the resultant state of the two cavities has two symmetries.
This is demonstrated by the explicit form of $C_{l,r}$, \eq{OptOneClrcavitiesequal},
and in \figu{OptOneRLoptical}. A $2\pi$ translational symmetry
identifies physically identical phases. In addition there is symmetry in $C_{l,r}$ about $\Delta=0$. This exists
because the procedure as described so far localises the absolute value of the relative phase.

When photons are detected at both ports $C_{l,r}$ is peaked at two different values $\pm \Delta_0$.
Looking at the asymptotic form of $C_{l,r}$ as $l$ and $r$ tend to large values it is observed that
the state that emerges, $|\psi_\infty\>$, takes the following form:
\begin{eqnarray}
\label{eqn:OptOnepsiinfinitydouble}
|\psi_\infty\>
\!\! &\propto& \!\!
\int \! d\theta e^{-i2|\gamma|^2\theta}
|\gamma\> \nonumber \\
\!\! && \!\!
\otimes \left [
e^{\!-i|\gamma|^2 \! {\Delta_0}} |\gamma e^{i{\Delta_0}}\>
\!+\!
 e^{i|\gamma|^2 \! {\Delta_0}}|\gamma e^{\!-i{\Delta_0}}\>
\right ],
\end{eqnarray}
where $\vert\gamma\>=\vert|\gamma|e^{i\theta}\>$ and $|\gamma|=\sqrt{N-\left(\frac{l+r}{2}\right)}$.
The relative component of the two mode state, contained in the square brackets,
is a superposition of two coherent states with the same amplitude
but different phases $\pm\Delta_0$ --- ordinarily called a Schr\"{o}dinger cat state. $|\psi_\infty\>$ has
in addition a sum over all values of the global phase $\theta$. A state of the form \eq{OptOnepsiinfinitydouble}
could be termed a \textit{relational Schr\"{o}dinger cat state}.
It may be asked why values of the relative phase with the same magnitude but opposite sign may
not be identified as equivalent,
in particular since
either one of the two components
of \eq{OptOnepsiinfinitydouble} separately gives rise to the same probabilities for further detections at the left
and right photocounters.  As an example of why the two values of the relative phase
are not equivalent consider the case that a relational Schr\"{o}dinger cat state is formed with $\vert \Delta_0 \vert \simeq \pi/2$,
and the phase shifter (refer back to \figu{OptOnetwocavities}, initially fixed at $\xi=0$) is subsequently adjusted by $\pi/2$.
With high probability later photons
will all be detected at one detector, which is randomly the ``left'' or ``right'' one
 varying from run to run of the experiment.
This behaviour is not consistent with a state sharply localised at one value of the relative phase.

Creating the superposition \eq{OptOnepsiinfinitydouble} would however be experimentally challenging
as it requires perfect phase stability. In practise it is found that the relational Schr\"{o}dinger cat is sensitive
to any asymmetry or instability in the system.
The effect of a randomly varying phase is to cause localisation about
\textit{one} particular value of the relative phase. This phenomenon is evident in the numerical studies of M{\o}lmer \cite{Molmer97,MolmerB97}.
These incorporate a slight frequency difference between the two cavity modes causing the free evolution to have
an additional detuning term $\exp i(\omega_b - \omega_a) b^{\dagger}b t$.  Combined with the random intervals
between detections, this means that the process can be described by Kraus operators $a \pm e^{i\sigma}b$
where the phase $\sigma$ takes random values for each photodetection. The relative phase then takes a unique value
varying randomly for each run. A dynamically equivalent process occurs when atoms
from two overlapping Bose Einstein condensates drop onto an array of detectors and are detected at random positions;
a detailed discussion of this point is given in  \sect{BECdetectionmodelvisiblity} of Chapter \ref{chap:BEC}.

In the case of an idealised setup for which the phase shifts throughout the apparatus remain fixed,
one component of the relational Schr\"{o}dinger cat state can be removed manually.
Suppose that after $l$ and $r$ photons have been detected at $D_l$ and $D_r$ in the usual way,
the phase shifter is adjusted by $\pm \Delta_0$, and then the experiment is continued until a
small number of additional photons have been detected. The phase shift translates the interference pattern
in such a way that the additional counts will occur at one detector (with high probability).
The additional measurements eliminate the unwanted component of the cat state and confirm
a well-defined relative phase.

\subsection{Robustness of the localised states and operational equivalence to tensor products of coherent states}
\label{sec:OptOnerobustness}

The next important feature of the canonical interference process concerns the robustness of the localisation.
In the limit of a large number of detections, the state of the two cavities becomes equivalent to
\be
\label{eqn:OptOnepsiinfinity}
|\psi_\infty\>=\int \!\! d\theta e^{-2i|\gamma|^2\theta} |\gamma\>|\gamma e^{i{\Delta_0}}\>,
\ee
with $|\gamma\>=||\gamma|e^{i\theta}\>$ some coherent state. The coherent states, being
minimum uncertainty gaussian states, are the most classical of any quantum states
(see for example Chapter $2$ of \cite{Scully97}). Thus
states of the form $|\gamma\>|\gamma e^{i{\Delta_0}}\>$ are expected to be robust.
However $|\psi_{\infty}\>$ is a superposition over such states, and this could
potentially affect the robustness. That this is not the case can be understood by noting
that the superposition in \eq{OptOnepsiinfinity} is summed over the global
phase\footnote{``global phase'' here is not
referring to the always insignificant total phase of a wavefunction, but rather the phase
generated by translations in photon number: $e^{ia^\dagger a}$. This is still a relative
phase between different states in the Fock state expansion of a coherent state.}
$\theta$ of the
coherent states. Under evolutions obeying an additive conservation of energy rule
(photon-number superselection), which is essentially the extremely good rotating-wave approximation of
quantum optics, this global phase becomes operationally insignificant (refer \cite{Sanders03}).
This is discussed further in \sect{OPT2superselection}, Chapter \ref{chap:OpticalTwo}.

Evidence of the robustness of the localised state \eq{OptOnepsiinfinity} has also been provided
by numerical studies reported in Sec.~5 of \cite{MolmerB97}, and a summary of
M{\o}lmer's findings is as follows.  The setup considered was similar to that
depicted in \figu{OptOnetwocavities} except the cavities had a slight frequency difference,
and each cavity was assumed to be coupled to an additional,
independent, reservoir. M{\o}lmer first looked at the scenario in which
the reservoirs acted as additional decay channels and had decay rates equal
to that which determined the leakage onto the beam splitter.
The characteristic oscillation of the counts rates at the two measured ports of the beam splitter
was found to be unchanged by this additional leakage, although the intensities of the cavity fields
and the counts rates decayed twice as fast.  In the second scenario considered the reservoirs
were thermal and photons could feed into the cavity as well as out.  The photodetector count rates were
more difficult to interpret in this example.  The diagnostic variable
$2\text{Re}(\bra{\psi} a^\dagger b \ket{\psi})$, where $a$ and $b$ denote annihilation operators
for the two cavity modes and $\ket{\psi}$ the state of the cavities,  was observed to
oscillate consistent with localisation of the relative phase within the first few detections.
However the oscillations were not as smooth as in the absense of the reservoirs.
Furthermore the oscillations were disrupted
and partially reestablished, a phenomenon which was found to coincide with the cavities being nearly empty.
It was concluded that the relative phase correlation of the field modes
is not robust when the cavities are populated by only a few photons.

Finally a state of the form \eq{OptOnepsiinfinity} is, for any processes
involving relative phases between the cavities, operationally equivalent to a tensor
product of pure coherent states for each cavity $|\gamma\>|\gamma e^{i\Delta_0}\>$.
However, because of the phase factor $e^{-2i|\gamma|^2\theta}$, the state is in
fact highly entangled. Expanding in the (orthogonal) Fock bases (as opposed to the
non-orthogonal coherent states) the state is seen to be of the form:
\begin{eqnarray}
\left\vert \psi _{\infty }\right\rangle \!\!\!\!
&=& \!\!\!\!
\int \tfrac{d\theta }{2\pi }e^{-2i\left\vert \gamma \right\vert ^{2}\theta }\left\vert \gamma \right\rangle \left\vert \gamma e^{i{\Delta_0}}\right\rangle \nonumber \\
&=& \!\!\!\!
\!\!\! \sum^{\infty}_{n,m=0} \!\!\! \!\! \sqrt{\Pi_{n} \! \left( |\gamma |^{2}\!\right) \! \Pi_{m} \! \left( |\gamma |^{2}\!\right) }
\!\!\int\!\! \tfrac{d\theta}{2\pi} e^{i\left(\! n+m-2\left\vert \gamma \right\vert ^{2}\!\right) \theta }e^{im{\Delta_0}} \! \left\vert n,\!m \!\right\rangle \nonumber \\
&=& \!\!\!\!
\! \sum^{2|\gamma|^2}_{m=0} \!\! \sqrt{\Pi_{2\left\vert \gamma \right\vert^{2}\!-\!m} \! \left( |\gamma |^{2}\! \right) \! \Pi_{m} \! \left( |\gamma |^{2} \! \right) } \: e^{im{\Delta_0}} \! \left\vert 2 \! \left\vert \gamma \right\vert ^{2}\!\!-\!m,\!m\!\right\rangle
\end{eqnarray}
where $\left\vert n,\!m \! \right\rangle$ denotes a product of photon number states, $\Pi_.(.)$
denotes a Poissonian factor and $2|\gamma|^2$ is a whole number of photons.
Note that every term in the superposition has the same total photon number, consistent with a
photon number superselection rule.

\section{Quantifying the degree of localisation of the relative phase}
\label{sec:OptOneQuantifyingLocalisation}

\begin{figure}[h]
\begin{center}
\includegraphics[height=5.8cm]{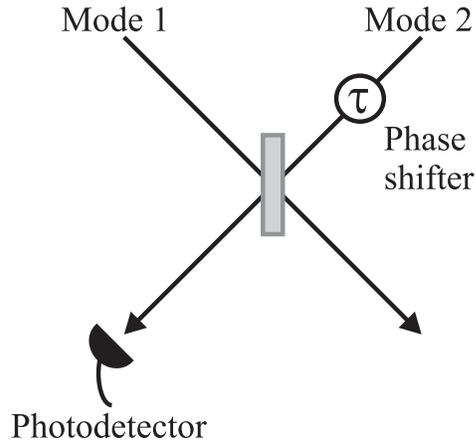}
\caption{\label{fig:OptOnevisibilitydef}
$I(\tau)$ is the intensity at the left output port after the second
mode undergoes a phase shift of $\tau$ and is combined with the first at a $50:50$ beam splitter.
This intensity is evaluated for all possible settings of the phase shifter.
Extremising over $\tau$, the visibility for the two mode state is defined as
$V=(I_{\rm max}-I_{\rm min}) / (I_{\rm max}+I_{\rm min})$.}
\end{center}
\end{figure}

A definition of the visibility suitable for rigorously quantifying the degree of localisation of
the relative phase for a prepared two mode state is illustrated by \figu{OptOnevisibilitydef}.
It is supposed that the second mode undergoes a phase shift $\tau$ before being completely
combined with the first at a 50:50 beam splitter. The expected photon number at the left port
is then denoted $I(\tau)$.  This intensity is evaluated for all possible phase shifts $\tau$,
allowing a visibility for the two mode optical state to be defined in terms of the difference
between the maximum and minimum values as follows,
\be
\label{eqn:OptOneVisFromInt}
V=( I_{\rm max} - I_{\rm min} ) \, / \, ( I_{\rm max} + I_{\rm min} ).
\ee
By definition the visibility takes values between $0$ and $1$.

For a product of photon number states $|N\rangle|M\rangle$ the action of a phase shifter on the second mode
merely introduces an irrelevant factor of $e^{iM}$, and hence the intensity $I(\tau)$ is constant for different phase shifts $\tau$,
and the visibility $V$ is $0$.  In a similar way the visibility is $0$ for any product of mixed states
diagonal in the photon number basis, such as the product of Poissonian states in \eq{OptOnePoissonianstart} (discussed later in \sect{OptOnePoissonian}).
On the other hand, for a product of coherent states $|\sqrt{\bar{N}}e^{i\theta}\rangle|\sqrt{\bar{N}}e^{i(\theta+\Delta_0)}\rangle$
and, $|\psi_\infty\>$ \eq{OptOnepsiinfinity} and $\rho_\infty$ \eq{OptOnerhoinfinity} (discussed later in \sect{OptOnePoissonian})
summed over the global phase,
and all three with exactly one value $\Delta_0$ for the localised relative phase, it is easily shown that
$I(\tau)$ is proportional to $\cos^2 \left( \frac{\Delta_0+\tau}{2} \right)$.  $I(\tau)$ is then maximized if the phase shifter is set to
$\tau=-\Delta_0$ and $0$ for $\tau=-\Delta_0+\pi$.  Therefore the visibility is $1$ for these three examples for which
the relative phase is perfectly correlated.

\section{Mixed initial states}
\label{sec:OptOneMixed}

\subsection{Poissonian initial states}
\label{sec:OptOnePoissonian}

The example of \sect{OptOneFockstates}, while usefully illustrating many features of
relative localisation, is not experimentally accessible due to the assumption of the
availability of large photon number, initially pure, Fock states populating the cavities. In particular, if
looking for a mechanism by which relative localisation occurs naturally in our
interactions with surrounding objects, the previous example is somewhat implausible as it
stands, in as much as it would suggest that macroscopic levels of entanglement are
necessary to localise relative degrees of freedom.

With this in mind a more realistic scenario is considered in this section. While it is implausible that the
cavities are populated by large Fock states, it is \textit{not} implausible that they are
populated by a large number of photons, and that only the mean number $\bar{N}$
of photons is known. In such a situation the quantum state of the cavity would be
assigned a Poissonian distribution over photon number under a maximum entropy
principle (refer \cite{Jaynes03}). Alternatively, the cavities may be populated by light from
independent lasers, for which standard laser theory leads to the photon number distribution being
Poissonian (refer Chapter $11$ of \cite{Scully97}).

The canonical localisation procedure discussed in \sect{OptOneFockstates} is now reconsidered
assuming that the initial state of the cavities is,
\begin{eqnarray}
\label{eqn:OptOnePoissonianstart}
\rho_I&=&\sum_n \Pi_n(\bar{N})|n\>\<n| \otimes \sum_m \Pi_m(\bar{N})|m\>\<m| \nonumber \\
&=& \frac{1}{4\pi^2}\int\!\!\!\int \!\!d\theta d\phi \;|\alpha\>\<\alpha|\otimes
|\beta\>\<\beta|,
\end{eqnarray}
where $\alpha=\sqrt{\bar{N}} \exp i\theta$ and $\beta=\sqrt{\bar{N}} \exp i\phi$.
The evolution of $\rho_I$ is as follows, given that $l$ and $r$
photons are detected at the left and right detectors respectively:
\begin{eqnarray*}
\rho_I&\longrightarrow& K_L^l K_R^r \rho_I K_L^{\dagger l} K_R^{\dagger r} \\
 &\propto& K_L^l K_R^r \int\!\!\!\int \!\!d\theta d\phi \;|\alpha\>\<\alpha|\otimes
|\beta\>\<\beta| K_L^{\dagger l} K_R^{\dagger r} \\
 &\propto& \int \!\!\!\int\!\!\!d\theta d\phi \; |\alpha+\beta|^{2r} |\alpha-\beta|^{2l} |\alpha\>\<\alpha|\otimes
|\beta\>\<\beta| \\
&\propto& \int \!\!\!\int\!\!\!d\theta d\phi \; |C_{l,r}(\theta,\phi)|^2
|\alpha\>\<\alpha|\otimes |\beta\>\<\beta|,
\end{eqnarray*}
with $C_{l,r}(\theta,\phi)$ as in \eq{OptOneClrcavitiesequal}. Clearly the discussion about the
localising nature of $C_{l,r}(\theta,\phi)$ applies
equally well in this case. The expression \eq{OptOneFockProbApprox} approximating the probabilities
for different measurement records when the initial state is a product of Fock states is exact
for a product of Poissonian states. In the limit of a large number of detections,
\be
\label{eqn:OptOnerhoinfinity} \rho_\infty =  \int_0^{2\pi}\!\! \frac{d\theta}{2\pi} \;
|\alpha\>\<\alpha|\otimes |\alpha e^{i\Delta_0}\>\<\alpha e^{i\Delta_0}|.
\ee
Quite remarkably \emph{the relative phase localisation of the mixed states is
just as sharp and just as rapid as that of the pure states}!

However, in striking contrast to the large entanglement observed for the pure state case considered in \sect{OptOneFockstates},
the state of the two cavities in this example remains manifestly separable (unentangled) throughout.
This fact might seem surprising at first sight.
If the calculation had been done in terms of photon number states,
products of photon number states $|N\rangle\otimes |M\rangle$ would evolve to highly entangled states.
However the photon number variables are summed according to Poissonian probability distributions and this changes things.
As an example of the fact that a mixture of entangled states can be separable consider the case of
an incoherent mixture of all $4$ maximally entangled (Bell) states of $2$ qubits
(refer \cite{NielsenChuang00}).  This
is equivalent to the completely mixed state $\mathbb{I}/4$ which is certainly separable.
Any two-party mixed state of the form $\rho_a\otimes\rho_b$ is certainly unentangled.
Furthermore, any ``convex sum'' $\Sigma_{a,b}C_{a,b}\rho_a\otimes\rho_b$,
where all the $\rho_a$ and $\rho_b$ are bona fide density operators, has only classical correlation
and by the standard definition is separable.  This is exactly the case for the mixed state calculation
considered in this section both before and after the measurement process has acted.
A calculation in terms of mixtures of photon number states
would likely be much messier, but ultimately must be consistent with this.
It should be noted further here that
$\rho_{\infty}$ in \eq{OptOnerhoinfinity} has the interesting feature of being formally
separable but \emph{not} locally preparable under
a superselection rule (or equivalently lack of a suitable reference frame), a feature
first noted in \cite{Rudolph01}.

After involved calculation, extending methods and results developed earlier in this chapter,
a simple expression for the intensity and the visibility can be found for the case of
two initial Poissonian states and an idealised experiment for which the phase shifts
throughout the apparatus remain fixed
(for the derivation refer appendix \ref{appendix:PoissonianVisibilityDerivation} and
specifically Sec.~\ref{appendix:PoissonianvisibilitiesSCS}):
\be
\label{eqn:OptOnePoissonianintensity}
I(\tau )\propto r\cos ^{2}\tfrac{\tau }{2}+l\sin ^{2}\tfrac{\tau }{2}+\tfrac{1}{2},
\ee
\be
\label{eqn:OptOnePoissonianvisibility}
V_{l,r} = \frac{|r-l|}{r+l+1}.
\ee
If the detections are all at one detector, the right one say,
\eq{OptOnePoissonianvisibility}
simplifies to $r/\left(r+1\right)$ which tends rapidly $1$, and is $1/2$ after just one detection.
However, it is also seen that the expression decreases to $0$ as the proportion of counts at the two detectors becomes equal.
This does not reflect less localisation in these cases but is an artefact of the definition of the visibility.
It is easy to see that if the state of the two cavities is localised at two values of the relative
phase these will both contribute to the intensity at one port in the definition of the visibility; changing the phase shift
$\tau$ will tend to reduce the contribution of one while increasing that of the other so that overall the variation in
the intensity is reduced.  In more detail, the visibility is underestimated when $r\approx l$ because of the
occurrence of two peaks in $C_{l,r}$ approximately $\pi$ apart. However, in realistic situations such
localisation at multiple values
is expected to be killed by small instabilities in the apparatus (as in the case of
relational Schr\"{o}dinger cat states in pure state case) and thus
the visibility is expected to tend to $1$ in all cases.

The definition of the visibility can in fact be modified so as to provide a mathematically rigorous measure of
the degree of localisation when the relative phase parameter is peaked at more than one value.
The localising scalar $C_{l,r}(\theta ,\phi )$ for the case of initial Poissonian states
(with the same average photon number) is the same as for the case of initial Fock states (with the same number),
and is illustrated in \figu{OptOneRLoptical}.  In particular it should be observed that
$C_{l,r}(\theta ,\phi )=0$ when $\phi=\theta$.  It is clear then that the final state of the two cavity modes
may be considered a sum of two separate components, one
localised at $+\Delta_0=2\arccos\sqrt{r/(r+l)}$ with the relative phase parameter
$\frac{\Delta}{2}\equiv\frac{\phi-\theta}{2}$ varying on $0$ to $\pi/2$,
and another at $-\Delta_0$ with $\frac{\Delta}{2}$ restricted to $-\pi/2$ to $0$.
Hence an improved measure of the degree of localisation is achieved by computing the visibility
for one of these components taken separately.
This is done in Sec.~\ref{appendix:SplitPoissonianVisibilities} of the appendix
\ref{appendix:PoissonianVisibilityDerivation}
 where full analytic expressions are provided.
Given a total number $D$ of detections at both detectors, the visibility $V$ is found to be smallest
when all the photons are detected at one photocounter, with $V=D/(D+1)$.
The same expression is obtained from \eq{OptOnePoissonianvisibility} when $r=0$ or $l=0$
as would be expected as the localisation process picks out a unique value for the relative phase
in these two cases.
$V$ is found to be greater when there are counts
at both photocounters, and in the
``fastest case'' where there are equal counts at both photocounters,
\bea
V&=&\frac{2\left[\, \Gamma \! \left( D/2\!+\!1\right) \,\right] ^{2}}{\left( D\!+\!1\right)
\left[\, \Gamma \! \left( D/2\!+\!1/2\right)\, \right] ^{2}} \nonumber \\
&=&\lambda \frac{D}{\left( D\!+\!1\right) }\,, \nonumber
\eea
where $\Gamma \left(\cdot\right)$ denotes the gamma function,
$\lambda$ is $1.27$ when $D=2$, $1.13$ when $D=4$ and $1.05$ when $D=10$, decreasing rapidly to $1$.

\subsection{Thermal initial states}
\label{sec:OptOnethermal}

In the examples presented above there was no limit to how sharp the localisation could
be. The case that both cavities are initially populated by
thermal states with the same mean photon numbers $\bar{N}$ is now examined along the same lines.
A thermal probability distribution in the photon number $n$ with mean $\bar{N}$ is given by
$\bar{N}^n/(1+\bar{N})^{n+1}$,
decreasing monotonically with $n$, and as such the initial state of the two cavities is,\footnote{
Note that in the analysis here the thermal light is assumed to be single mode.
A valuable extension would be to consider the interference of multimode thermal light,
characteristic of unfiltered chaotic light from natural sources.}
\begin{eqnarray*}
\rho_I
\!\!&=&\!\!
\sum_n \frac{\bar{N}^n}{(1+\bar{N})^{n+1}}|n\>\<n|
\otimes
\sum_m \frac{\bar{N}^m}{(1+\bar{N})^{m+1}}|m\>\<m| \\
\!\!&=&\!\!
\frac{1}{4\pi^2\bar{N}^2} \! \int \!\! \int \!\! d\bar{n} d\bar{m} d\theta d\phi
e^{-(|\alpha|^2+|\beta|^2)/\!\bar{N}} |\alpha\>\<\alpha| \! \otimes \! |\beta\>\<\beta|,
\end{eqnarray*}
where $\alpha\!=\!\sqrt{\bar{n}}\exp i\theta$ and $\beta\!=\!\sqrt{\bar{m}}\exp i\phi$.
Under the measurement of $l$ and $r$ photons at the left and right detectors respectively,
\begin{eqnarray*}
\rho_I \!\!\! &\Rightarrow& \!\!\! K_L^l K_R^r \rho_I K_L^{\dagger l} K_R^{\dagger r} \\
\!\!\! &\propto& \!\!\!\! \int\!\! d^2\!\alpha d^2\!\beta \, e^{\!-\left(\!|\alpha|^2+\!|\beta|^2\!\right)/\!\bar{N}} K_L^l K_R^r |\alpha\>\!\<\alpha|
 \!\otimes\!
|\beta\>\!\<\beta| K_L^{\dagger l} K_R^{\dagger r} \\
\!\!\! &\propto& \!\!\!\! \int \!\! d^2\!\alpha d^2\!\beta \, e^{\!-\left(\!|\alpha|^2+\!|\beta|^2\!\right)/\!\bar{N}} |\alpha\!+\!\beta|^{2r} |\alpha\!-\!\beta|^{2l} |\alpha\>\!\<\alpha| \!\otimes\! |\beta\>\!\<\beta| \\
\!\!\! &\propto& \!\!\!\! \int \!\! d^2\!\alpha d^2\!\beta \, e^{\!-\left(\!|\alpha|^2+\!|\beta|^2\!\right)/\!\bar{N}} |C_{l,r}(\bar{n},\bar{m},\theta,\phi)|^2
|\alpha\>\!\<\alpha| \!\otimes\! |\beta\>\!\<\beta|.
\end{eqnarray*}
As in the previous example of initial Poissonian states the state of the cavities remains separable throughout the localising
process.

Unlike previous examples $|C_{l,r}\left(\bar{n},\bar{m},\theta,\phi\right)|^2$ does not provide a simple picture of the localisation of the relative
phase due to the additional dependence on the mean photon number variables.  One possibility would be to average
the scalar function over the two photon number variables leading to a function,
\be
\int_{0}^{\infty}\!\!\! \int_{0}^{\infty}
\! \frac{d\bar{n}}{\bar{N}} \! \frac{d\bar{m}}{\bar{N}}\,e^{-\bar{n}/\bar{N}}e^{-\bar{m}/\bar{N} }
|C_{l,r}(\bar{n},\bar{m},\theta,\phi)|^2.
\ee
After normalising this can be shown to be,
\be
\left\vert C_{l,r}\left( \Delta \right) \right\vert ^{2}=\frac{1}{2^{r+l}r!l!}
\int_0^\infty \!\!\!
\int_0^\infty \!\!\!
d\bar{n}d\bar{m} \, e^{-\bar{n}-\bar{m} }\left\vert \sqrt{\bar{n}}+\sqrt{\bar{m}}e^{i\Delta }\right\vert ^{2r}\left\vert -\sqrt{\bar{n}}+\sqrt{\bar{m}}e^{i\Delta }\right\vert ^{2l}\!\!\!,
\nonumber
\ee
where $\Delta \equiv \phi-\theta$. $\left\vert C_{l,r}\left( \Delta \right) \right\vert ^{2}$ is independent of $\bar{N}$.
However it is not necessary to consider such a function in detail to
quantify the rate and extent of the localisation of the relative phase parameter.

An intensity and a visibility
can be computed as in the Poissonian case above (see appendix \ref{appendix:ThermalVisibilityDerivation}).
For an arbitrary measurement record the results are,
\be
I(\tau ) \propto l\cos ^{2}\tfrac{\tau }{2}+r\sin ^{2}\tfrac{\tau }{2}+1
\ee
\be
V_{l,r}=\frac{|r-l|}{r+l+2}.
\ee
If all the measurements occur in one detector, the right one say, the visibility is  $r/\left(r+2\right)$
which is $1/3$ after just one detection and which tends to $1$ rapidly
- but slower than in the Poissonian case.
Again it is found that
\emph{the localisation of the relative phase parameter can be as sharp
for initial mixed states as for initial pure states}.  In addition,
\emph{the localisation for initial thermal states
is rapid, but slower than for initial photon number states or Poissonian states}.

\begin{figure}[h]
\begin{center}
\includegraphics[height=7.5cm]{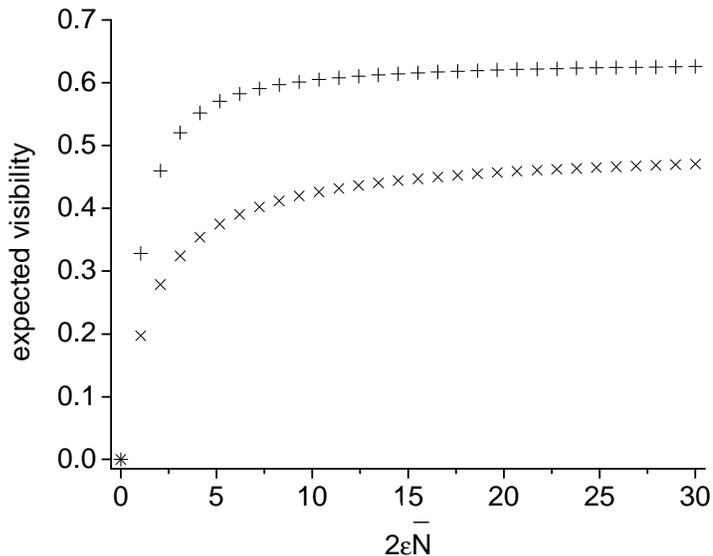}
\caption{\label{fig:OptOnemixedstatevisibilities} Expected visibilites for (a) an initial product of two Poissonian states (plusses) and
(b) an initial product of two thermal states (crosses), with average photon number $\bar{N}$ for both cavities.
}
\end{center}
\end{figure}
Averaging over all detection outcomes, an expected visibility
a finite time after the start of the procedure
$\Sigma_{l,r}P_{l,r}V_{l,r}$ can be computed.
An exact expression for
the probabilities of different measurement sequences
derived in (see appendix \ref{appendix:ThermalVisibilityDerivation}) is,
\be
P_{l,r}= \frac{(\bar{N}\epsilon )^{r+l}}{\left( 1+\epsilon \bar{N}\right) ^{r+l+2}}\,.
\ee
This notably has the form of the probabilities for two independent sources of thermal light with mean photon number
$\epsilon \bar{N}$.
The expected visibilities
for the thermal and Poissonian cases are compared in \figu{OptOnemixedstatevisibilities}.
These averages do not tend to one - as the visibility underestimates the degree of localisation
when the prepared states are localised at two values of the relative phase,
as discussed previously in \sect{OptOnePoissonian}.
However the general trend is clear.
\emph{The cavity modes initially in thermal states tend, as in the Poissonian case,
to a state which is perfectly correlated in relative phase while remaining unentangled.}
Although more photons must be detected to achieve the same
degree of localisation when the initial states are thermal, the localisation
proceeds very rapidly in both cases.

\subsection{The consequences of non-ideal photodetection}
\label{sec:OptOneNonIdealPhotodetectors}

An important issue regarding the experimental feasibility of the interference procedure for preparing states
with a well-defined relative phase is the problem of errors in the photodetection process.
There are two types of errors --- detector inefficiency (or photon loss) and dark counts
(for a recent review of the current state of technology see \cite{JMOReview04}).
In general, dark counts are more serious if they occur unpredictably in the experiment, and to incorporate them in calculations
it is necessary to mix over the outcomes in the ideal case consistent with what is actually registered.
However detector inefficiency presents the greater problem technologically (typically $20\%$ at optical frequencies).
An inefficiency detector can be modelled as in \cite{Yuen83}.  For the analysis of the interference procedure here
where one of the detectors has efficiency $\eta<1$,
the ideal detector assumed previously is replaced by a beam splitter with transmittivity
$\sqrt{\eta}$ and reflectivity $\sqrt{1-\eta}$, which couples the ancilla to an additional vacuum mode, and
with the primary output port now monitored by a perfect detector.  The dynamical evolution during the procedure causes the additional
mode to accumulate some population which is not measured, and the mode is traced over at the end.

In fact photons lost from the apparatus due to detector inefficiency cannot affect the process of localisation directly.
Instead, the beam splitter which combines the ancillae ensures that the origin of the lost photons from either cavity is uncertain,
and the situation is equivalent to one where both cavities leak separately into external modes
not involved in the measurement process.  The localisation of the relative phase depends on
those photons which are registered by the detectors in the same way as for the ideal case,
while the photon loss reduces the total number of photons in the final state.
For the case of large initial number states this situation has been studied numerically
by M{\o}lmer, as discussed in \sect{OptOnerobustness}.  The relevant simulation assumes loss rates out of the cavities
equal to the leakage onto the beam splitter, a situation M{\o}lmer identifies as equivalent to $50\%$ efficiencies for both
detectors.  As another example, the calculations in \sect{OptOnePoissonian} for
the case of Poissonian initial states can readily be extended to incorporate detector inefficiencies
and, as is straightforward to see, the relative phase localisation conditioned
on a particular measurement outcome is as for the ideal case,
although the probabilities for specific outcomes a fixed time after the start are
altered.  Finally it should be pointed out that postselection provides one potential aide when the photodetectors are subject to errors.
For example, one can reject all runs of the experiment where the total number of detections is significantly
different from what would be expected in the ideal case.

%% file: LRDoF_v2_chapter_3.tex
\chapter{Advanced Topics on Localising Relative Optical Phase}
\label{chap:OpticalTwo}

This chapter extends the programme of Chapter \ref{chap:OpticalOne}
on localising relative optical phase and the canonical
interference procedure explained in \sect{OptOneFockstates},
and presents original work on several topics.
\sect{OPT2Krausopderivation} formally proves
the assertion that the action of photodetections during the interference procedure
can be described by Kraus operators $a\pm b$ (where $a$ and $b$
are the annihiliation operators for the two modes).
The full form of the measurement operators incorporates also the leakage from
the cavity modes.  The identification of the leakage parameter
$\epsilon$ with time is explained.

\sect{OPT2AsymInitialStates} looks at
the localisation of the relative phase for
initial states with very different intensities at each mode, focusing on the example
of initial Poissonian states.  This example is treated along similar lines
to the symmetric case discussed in \sect{OptOnePoissonian} of Chapter \ref{chap:OpticalOne}.
The evolution of the localising scalar function and the probability
distribution for different measurement outcomes are found to be very different
in the highly asymmetric case compared to the symmetric one.  The
issue of whether there are preferred values for the localised relative phase
in the asymmetric case is discussed.
Intuitively one might expect the
localisation of the relative phase to be slower
for increasingly asymmetric initial states, and this is verified here
by looking at the evolution of the average visibility.  In fact the definition
of the visibility given in \sect{OptOneQuantifyingLocalisation}
of Chapter \ref{chap:OpticalOne} must be modified for this case.
It is observed that the limit of sharp localisation becomes harder to attain
for increasing asymmetric initial
states.

The transitive properties of the localisation process are considered in
\sect{OPT2transitivity}.  Specifically the localisation process is assumed to act pairwise
on three modes initially in Poissonian states which can have different intensities.
The conclusions are expected to hold generally.  It is concluded that
the localisation process acts independently of prior phase correlations with further systems,
and that the localised quantum phases have the same transitive properties as classical phases.
However the localisation process does affect
phase correlations with external systems through (asymmetric) photon depletion.
It is pointed out that for three modes with well localised relative phases
loss of one of the systems does not disrupt the phase correlation between the remaining two.

The possibility of using the canonical interference procedure for
linear optical and classical feedforward based state engineering
is suggested in \sect{OPT2FockAddition}, studying in detail
the ``addition'' of photon number states.
First the situation is analysed wherein two Fock states are combined at a beam splitter
and one of the output ports is measured by a photodetector, to yield a larger Fock state
at the free output port.
An improved method of addition is then suggested,
motivated by the fact that if the input states are phase locked, all the photons can
be directed to one output port of the beam splitter using only an additional phase shifter.
The probability for detecting the vacuum at the free out port, given input Fock states
which are identical, can be doubled compared to the first method
at the cost of one photon used to partially localise the relative phase,
and almost tripled at the cost of two. As an aside
it is pointed out that relational Sch\"{o}dinger cat states produced
from large number states as in \sect{OptOnesymmetries} of Chapter \ref{chap:OpticalOne},
can easily be converted to ``NOON'' states whenever the relative phases of the cat components are
different by approximately $\pi$.  Hence the analysis in this section points
towards a possible new route to generating NOON states based on procedures
which establish relative phase correlations.

\sect{OPT2superselection} discusses the application of states prepared
from initial Poissonian or thermal states as in Chapter \ref{chap:OpticalOne},
with well localised relative phase and a fixed total energy,
for fundamental tests of superselection rules (as published in \cite{Cable05}).
In algebraic quantum field theory the existence of absolute conservation laws, and
associated superselection rules forbidding the creation
of superpositions of states with different values of the conserved
quantity, is taken to be true axiomatically. A less absolutist and more operational
approach was initiated by Aharonov and Susskind \cite{Aharonov67},
who suggested that the forbidden
superpositions can in fact be observed provided that the apparatus
used by an observer are prepared in certain special
states. The states suggested by Aharanov and Susskind are
not particularly realistic.  Here
mixed states with well localised relative phase, which are
much more experimentally feasible, are presented as alternatives
which can reproduce the desired effects with no loss due to the lack of purity.
To end the chapter, \sect{OPT2opticalextensions}
suggests possible future calculations following the
programme set out in Chapters \ref{chap:OpticalOne} and \ref{chap:OpticalTwo}.

\section{Derivation of the measurement operators for the canonical interference procedure}
\label{sec:OPT2Krausopderivation}

In the canonical interference procedure the effect of photodetections on the two cavity modes
is given by Kraus operators $K_L$ and $K_R$
proportional to $a \pm b$, where $a$ and $b$ are annihilation operators for the cavity fields,
as assumed by fiat in \sect{OptOnelocalisingscalarfunction} of chapter \ref{chap:OpticalOne}.
This result assumes ancillae which are initially in the vacuum state, and for simplicity the
phase shifts in the apparatus are taken to be $0$.
In this section a mathematical derivation of these measurement operators is presented,
and in addition the effect of leakage from the cavities is fully accounted for.
The action of the interference procedure on an arbitrary tensor product of Glauber
coherent states is evaluated, and corresponding operators for the full evolution of the
system are deduced.  In fact this is all that is required.
The same measurement operators must also be valid for arbitrary initial states for the cavities, pure or mixed,
since coherent states form a complete basis and all the operators involved are linear.\footnote{
The general form of Kraus operators for a system which is
coupled to an ancilla is derived in Section $8.2$ of \cite{NielsenChuang00}.
It is assumed there that the system and ancilla are initially uncorrelated
with the ancilla in some state $\ket{e_0}$ at the start.
The system and ancilla then couple under some unitary process which denoted $U$.
At the end the ancilla is projected onto one of the members of a complete orthonormal basis $\left\{\ket{e_k}\right\}$ for the ancilla.
The Kraus operators for this entire process are shown to be of the form
$\bra{e_k} U \ket{e_0}$.  The derivation in this section
identifies the operators $\bra{e_k} U \ket{e_0}$ relevant to the canonical interference procedure.
Note that strictly speaking the discussion in \cite{NielsenChuang00} concerns a general system described by a finite dimensional
Hilbert space, whereas in the current problem the state space of each mode is infinite dimensional.
}

The canonical localisation process is described in detail in \sect{OptOneFockstates} of
Chapter \ref{chap:OpticalOne}, and is summarised in what follows.
Each cavity mode couples to an ancilla mode, initially the vacuum,
according to a linear mode coupling with small parameter $\epsilon$.  The ancillae are combined at a $50:50$ beam splitter,
and the output channels are measured by photodetectors with $r$ photons detected at the ``right'' detector
and l at the ``left'' detector.  Mathematically the steps of this process are as follows.\footnote{
It should be emphasized that this heuristic treatment of the dynamics is expected to be valid only
when $\epsilon$, also the fraction of the cavity populations which has leaked into the ancillae, is small.}
An initial product of coherent states
$\ket{\alpha}_A \otimes \ket{\beta}_B$ for the cavity modes
and ancilla modes $\ket{0}_{a} \otimes \ket{0}_b$ external to the cavities
are first transformed as,
\bea
&& \left\vert 0\right\rangle _{a}\otimes \left\vert 0\right\rangle _{b}\otimes \left\vert \alpha \right\rangle _{A}
\otimes \left\vert \beta \right\rangle _{B} \nonumber \\
&\longrightarrow&
\left\vert \sqrt{\epsilon} \alpha \right\rangle _{a}
\otimes \left\vert \sqrt{\epsilon} \beta \right\rangle _{b}
\otimes \left\vert \sqrt{1-\epsilon} \alpha \right\rangle _{A}
\otimes \left\vert \sqrt{1-\epsilon} \beta \right\rangle _{B}, \nonumber
\eea
due to the leaking of a small fraction of the cavity populations as determined by $\epsilon$.
The $50:50$ beam-splitter acts on the ancillae and they evolve to,
\bea
&&
\left\vert \sqrt{\epsilon} \left(\frac{\alpha+\beta}{\sqrt{2}} \right) \right\rangle _{a}
\otimes \left\vert \sqrt{\epsilon} \left(\frac{-\alpha+\beta}{\sqrt{2}} \right) \right\rangle _{b}
\otimes \left\vert \sqrt{1-\epsilon} \alpha \right\rangle _{A}
\otimes \left\vert \sqrt{1-\epsilon} \beta \right\rangle _{B}. \nonumber
\eea
Finally $r$ photons are measured in the ancilla labeled $a$, and $l$ photons are measured in the ancilla labelled $b$,
corresponding to action of the operator $\ket{r}\!\bra{r}_a \otimes \ket{l}\!\bra{l}_b$, leading to the final state,
\bea
&&
e^{-\left( \epsilon /2\right) \left( \left\vert \alpha \right\vert ^{2}+\left\vert \beta \right\vert ^{2}\right) }\frac{\sqrt{\epsilon }^{l+r}}{\sqrt{l!r!}}
\left( \frac{\alpha \!+\! \beta }{\sqrt{2}}\right) ^{r}
\left( \frac{-\alpha \!+\! \beta }{\sqrt{2}}\right) ^{l}\left\vert r\right\rangle _{a}
\! \otimes \! \left\vert l\right\rangle _{b}
\! \otimes \! \left\vert \sqrt{1\!-\!\epsilon }\alpha \right\rangle _{A}
\! \otimes \! \left\vert \sqrt{1\!-\!\epsilon }\beta \right\rangle _{B}. \nonumber
\eea
Overall the measurement process involves decay of the cavity field amplitudes
in addition to the action of the operators $a\pm b$ on the cavity modes,
and there are coefficients relating to normalisation.

An obvious candidate for an operator causing the decay of the amplitude of a coherent state
$\ket{\alpha}_A$ would take the form $\exp \left[-\left(\kappa/2\right) \hat{N}_A t\right]$,
where $\hat{N}_A$ is the number operator and $\kappa$ is a decay rate and $t$ a time parameter.
In a continuous measurement model of photon detection such an operator would generically correspond to
a temporal evolution conditioned on no photon
detection (for a review of quantum jump methods refer \cite{Plenio98} and references therein).
Expanding the coherent state $\ket{\alpha}_A$ in a basis of Fock states,
it may be deduced that a suitable
operator taking $\ket{\alpha}_A$ to $\ket{\sqrt{1\!-\!\epsilon}\alpha}_A$
is of the form $e^{\hat{N}_{A}\ln \sqrt{1-\epsilon }}$.  In detail,
\be
e^{\hat{N}\ln \sqrt{1-\epsilon }}\left\vert \alpha \right\rangle =e^{-\frac{1}{2}\epsilon \left\vert \alpha \right\vert ^{2}}\left\vert \sqrt{1-\epsilon }\alpha \right\rangle. \nonumber
\ee
For two modes with annihilation operators $a$ and $b$ the appropriate operator would be
$e^{\left( a^{\dagger }a+b^{\dagger }b\right) \ln \sqrt{1-\epsilon }}$.
Identifying a potential time parameter,
$\ln \sqrt{1-\epsilon }= -\frac{1}{2}\left( \epsilon +O\left( \epsilon ^{2}\right) \right)$,
and hence $\epsilon \simeq\kappa t$, provided $\vert\epsilon\vert \ll 1.$\footnote{
An identification of $\epsilon$ and time $t$ could be deduced just by considering dimensions.
However the full expression $t\propto-\,\ln \sqrt{1\!-\!\epsilon}$ is not so obvious.}

A complete measurement operator can now be proposed which has the correct action on
$\ket{0}_{a} \otimes \ket{0}_b \otimes \ket{\alpha}_A \otimes \ket{\beta}_B$,
\be
\left\vert r\right\rangle \!\! \left\langle 0\right\vert _{a}
\left\vert l\right\rangle \!\! \left\langle 0\right\vert _{b}
\exp \left\{ \! -\frac{1}{2}\left( a^{\dagger }a\!+\!b^{\dagger }b\right) \left[\! - \!\ln \left( 1\!-\!\epsilon \right) \right] \! \right\} \sqrt{\frac{\left( l\!+\!r\right) !}{l!r!}}\sqrt{\frac{\epsilon ^{l+r}}{\left( l\!+\!r\right) !}}
\left( \frac{a\!+\!b}{\sqrt{2}}\right)^{r}
\!\! \left( \frac{-a\!+\!b}{\sqrt{2}}\right)^{l}.
\ee
Only the action of the procedure on the cavity fields is of interest so the initial component
$
\left\vert r\right\rangle \!\! \left\langle 0\right\vert _{a}
\left\vert l\right\rangle \!\! \left\langle 0\right\vert _{b}
$
may be dropped.  As for the additional coefficients: the $\sqrt{\left( l\!+\!r\right)! / \left(l!r!\right)}$
is the root of the number of sequences of detections with $l$ ``left'' counts and $r$ ``right'' counts;
the $\sqrt{\epsilon ^{l+r}/\left( l\!+\!r\right) !}$ can be understood as the root
of a factor relating to the probability of having a total of $l+r$ detections.

It is worth here emphasizing that this heuristic treatment of the dynamics
leads to measurement operators with components, which can be identified as ``jump operators'' and a ``decay operator'',
which are also the principle elements of a quantum jump type analysis.
However the latter analysis is based on many additional approximations.  Specifically
the reservoir modes are assumed  to be multimode
(associated for example with a continuous spread of wavelengths and
different directions of propagation).  In addition they are assumed to be strictly Markovian ---
the reservoir correlation time is taken to be very much smaller than the typical relaxation time of the ``principle system''.
The interaction of the principle system with the reservoir is supposed ``weak'' over short times ---
so that all but the leading order interaction terms can be discarded (the Born approximation).
One assumption which is shared with the heuristic analysis is that
the reservoir and principle system are uncorrelated initially.
Furthermore in both cases it is assumed that the governing
Hamiltonians are consistent with the rotating wave approximation
(for which components of the interaction Hamiltonian which do not
explicitly conserve energy are discarded).  In conclusion, the treatment of the dynamics
in this section is seen to be reliable for short times despite its simplicity.

\section{Analysis of the canonical interference procedure for asymmetric initial states}
\label{sec:OPT2AsymInitialStates}

This section considers key features of the localisation of the relative optical phase between two cavity modes
which undergo the canonical interference procedure
when the initial intensity of one of the modes is very much greater than the other.
Specifically the methods employed in Chapter \ref{chap:OpticalOne} are extended for the case of
initial asymmetric Poissonian states, an important case since
--- is as much as the state of laser light may be taken to be Poissonian ---
lasers with a wide range of intensities is readily available in the laboratory.
More generally this discussion is relevant to a wider set of issues ---
understanding the situation when a microscopic system is probed by a macroscopic
apparatus; analysing reference systems composed of quantum resources; and explaining the
emergence of classical behaviour in closed quantum systems.

Suppose that two cavity modes undergo the canonical interference procedure described in \sect{OptOneFockstates}
of Chapter \ref{chap:OpticalOne}, but now the cavities are populated initially by
Poissonian states with different mean photon numbers $\bar{N}$ and $\bar{M}$ respectively.
Following the analysis in \sect{OptOnePoissonian} of Chapter \ref{chap:OpticalOne},
the initial states of the two cavities may be expanded in terms of coherent states,
\bea
\rho _{I}
&=&\sum_{n=0}^{\infty} \Pi _{n}(\bar{N})\left\vert n\right\rangle \! \left\langle n\right\vert \otimes
\sum_{m=0}^{\infty} \Pi _{m}(\bar{M})\left\vert m\right\rangle \! \left\langle m\right\vert \nonumber \\
&=&\frac{1}{4\pi ^{2}}\int_{0}^{2\pi } \!\!\! \int_{0}^{2\pi } \!\! d\theta d\phi
\left\vert \alpha \right\rangle \! \left\langle \alpha \right\vert
\otimes \left\vert \beta \right\rangle \! \left\langle \beta \right\vert, \nonumber
\eea
where $\alpha =\sqrt{\bar{N}}\exp i\theta $ and $\beta =\sqrt{\bar{M}}\exp i\phi $,
and $\Pi _{n}\left( \mu \right) =\mu ^{n}e^{-\mu }/n!$ is a Poissonian probability factor with mean $\mu$.
After $l$ detections at the left detector and $r$ at the right detector the state of the two cavities has evolved to,
\bea
\label{eqn:OPTTWOasympartiallyloc}
\rho^{\prime}
&=&\int_{0}^{2\pi} \!\!\! \int_{0}^{2\pi} \!\!
\frac{d\theta }{2\pi }\frac{d\phi }{2\pi }e^{-\epsilon \left( \bar{N}+\bar{M}\right) }\frac{\epsilon ^{l+r}}{l!r!}
\left( \frac{\alpha \!+\! \beta }{\sqrt{2}}\right) ^{r}\!\left( \frac{\alpha ^{\ast } \!+\! \beta ^{\ast }}{\sqrt{2}}\right) ^{r} \!
\left( \frac{-\alpha \!+\! \beta }{\sqrt{2}}\right) ^{l}\!\left( \frac{-\alpha ^{\ast } \!+\! \beta ^{\ast }}{\sqrt{2}}\right) ^{l} \nonumber \\
&\times&
\left\vert \sqrt{1\!-\!\epsilon }\,\alpha \right\rangle \! \left\langle \sqrt{1\!-\!\epsilon }\,\alpha \right\vert \otimes
\left\vert \sqrt{1\!-\!\epsilon }\,\beta \right\rangle \! \left\langle \sqrt{1\!-\!\epsilon }\,\beta \right\vert,
\eea
where $\epsilon$ denotes the leakage parameter for both cavities and takes a value very much less than $1$.
$\rho^\prime$ here is subnormalised and the value of the trace is the probability for
the specific measurement outcome at the two detectors.
The scalar function
$\left\vert C_{l,r}\left( \Delta \right) \right\vert ^{2}$,
where $\Delta\equiv\phi-\theta$, is defined by,
\[
\rho^{\prime }
=\int_{0}^{2\pi} \!\!\! \int_{0}^{2\pi} \!\!
\frac{d\theta}{2\pi}\frac{d\phi}{2\pi}\left\vert C_{l,r}\left( \Delta \right) \right\vert ^{2}
\left\vert \sqrt{1\!-\!\epsilon }\,\alpha \right\rangle \! \left\langle \sqrt{1\!-\!\epsilon }\,\alpha \right\vert \otimes
\left\vert \sqrt{1\!-\!\epsilon }\,\beta \right\rangle \! \left\langle \sqrt{1\!-\!\epsilon }\,\beta \right\vert,
\]
and takes the form,
\be
\left\vert C_{l,r}\left( \Delta \right) \right\vert ^{2} \propto \left( 1+R\cos \Delta \right) ^{r}\left( 1-R\cos \Delta \right) ^{l},
\ee
where $R$ is defined as the ratio $2\sqrt{\bar{N}\bar{M}}/\left(\bar{N}+\bar{M}\right)$, which must take values between $0$ and $1$.
$R$ is $1$ when $\bar{N}=\bar{M}$,
$0.94$ when $\bar{N}$ is twice $\bar{M}$ (or vica-versa),
0.57 when the factor difference is 10, and 0.20 when the factor is 100.
Noting that $\left\vert C_{l,r}\left( \Delta \right) \right\vert ^{2}=\left\vert C_{r,l}\left( -\Delta \right) \right\vert ^{2}$,
it is seen that the probability of $l$ and $r$ counts at the left and right detectors, and the degree of localisation
of the relative phase parameter for the final state, are identical if
the outcomes at the two detectors are swapped around.

\begin{figure}[h]
\begin{center}
\includegraphics[height=8cm]{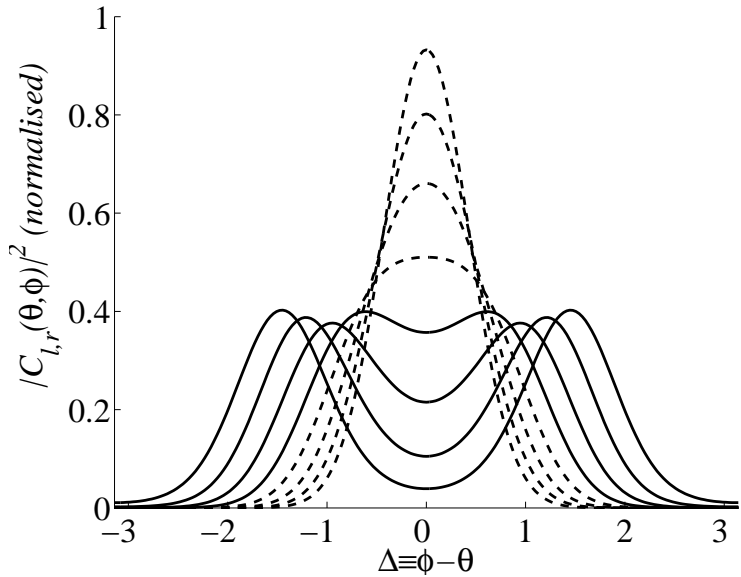}
\caption{\label{fig:OptTwoLSF}
The normalised scalar function $|C_{l,r}(\Delta)|^2$ for a total of $15$ photocounts
and the number $l$ of ``left'' detections going
from $0$ to $7$.  Initial Poissonian states are assumed with the intensity
in one mode $10$ times the other (parameter $R=0.57$).
The dashed curves peaked at $\Delta_0=0$ correspond to $l=0,\ldots,3$
and have progressively larger spreads.
The solid curves are for $l=4,\ldots,7$ and are
peaked progressively further from $\Delta=0$.
The curves for the case of more left then right detections would be centered about
$\Delta=\pi$ and related symmetrically to the ones shown.}
\end{center}
\end{figure}
For the highly asymmetric case when $R$ takes smaller values,
the $\left\vert C_{l,r}\left( \Delta \right) \right\vert ^{2}$
are qualitatively different to the case of symmetric case $R=1$
and are as follows.
If the counts are predominately at the right detector,
$\vert C_{l,r}(\Delta) \vert^2$
is peaked around the value $\Delta_0=0$ for the relative phase.
Given a fixed total number of detections $l+r$,
the localising function is progressively less well localised as the number of left counts increases,
leading to
a qualitative change when $l$ and $r$ are close in value.
Specifically, for the cases when
$r-R(l+r) <l\leq r$,
$\vert C_{l,r}(\Delta)\vert^2$ is peaked at two values,
$\Delta_0=\pm\arccos\left[\frac{1}{R}\frac{(r-l)}{(r+l)}\right]$.
Differently here from the cases of initial symmetric Fock or Poissonian states
(discussed in \sect{OptOnesymmetries} and \sect{OptOnePoissonian} of Chapter \ref{chap:OpticalOne}),
$|C_{l,r}(\Delta)|^2$ is non-zero at $\Delta=0$ and a partition of the final state into two
components localised at $+\Delta_0$ and $-\Delta_0$ is harder to justify.
When the $l\geq r$ the pattern of the $\vert C_{l,r}(\Delta)\vert^2$ is reversed, and when the detections
are predominately at the left detector $\vert C_{l,r}(\Delta)\vert^2$ is peaked at $\Delta_0=\pi$.
The situation is illustrated in \figu{OptTwoLSF},
for the case that $r\geq l$, and a total of $15$ detections at both detectors.
The initial state is chosen with $R=0.57$ corresponding to the case that one cavity begins with 10 times
the intensity of the other.  Dashed curves are for $l=0,\ldots,3$ and solid ones are for $l=4,\ldots,7$.

For a fraction of approximately (1-R) of the measurement outcomes,
the relative phase is localised at
$0$ or $\pi$, and the question arises as to whether these are ``preferred values'' when the initial states
are highly asymmetric.  To answer such questions it is
necessary to investigate the behaviour of $P_{l,r}$, the probability distribution
for different detection outcomes, for small values of the parameter $R$.
A full expression for the probability $P_{l,r}$
of $l$ ``left'' and $r$ ``right'' detections is given by,
\bea
P_{l,r}&=&\textrm{tr}\left( \rho ^{\prime }\right) \nonumber \\
&=&\Pi _{l+r}\left( \epsilon \bar{N}+\epsilon \bar{M}\right)
\left(
\begin{array}{c}
l+r \\
l
\end{array}
\right)
\left( \frac{\sqrt{1-R^{2}}}{2}\right)^{r+l}
\left( -1\right)^{l} \nonumber \\
&\times&
\sum _{j=0}^{l}
\left(
\begin{array}{c}
l \\
j
\end{array}
\right)
\left( \frac{-2}{\sqrt{1-R^{2}}}\right) ^{j}P^{\text{Legendre}}_{r+l-j}\left( \frac{1}{\sqrt{1-R^{2}}}\right),
\eea
where
$\left(^t_s\right)$
abbreviates the combinatorial factor $\left( s+t\right)!\,\big{/}\left(s!t!\right)$, and
$P^{\text{Legendre}}_n$ denotes the Legendre polynomial of degree $i$ given by,
\[
P^{\text{Legendre}}_{n}(z)=\frac{1}{2^{n}n!}\frac{d^{n}}{dz^{n}}\left( z^{2}-1\right) ^{n}.
\]
The Poissonian factor $\Pi_{l+r}\left(\epsilon \bar{N} + \epsilon \bar{M} \right)$
depends on the leakage parameter and the sum of the initial cavity intensities, and
controls the probability for the total number of detections $l+r$.
Evaluating $P_{l,r}$ numerically reveals the following.
The distribution changes qualitatively as $R$ is reduced in value.  For small values of
R, The situation is essentially reversed from the
symmetric case $R=1$ (refer \figu{OptOneFockPlr} of Chapter \ref{chap:OpticalOne}).
Given a fixed total number of detections, outcomes with detections at one detector are found to be least likely.
Instead, the probabilities increase to a maximum at $l=r$, (or this peak is split
and the most likely events are at two points for which $r\approx l).$

Turning attention now to the total probability of localisation of the relative phase
at $\Delta_0=0,\pi$, let $P_{0,\pi}(R,D)$ denote the total probability of such events
for a fixed total number of detections $D=l+r$, and where the the Poissonian factor
$\Pi_{l+r}(\epsilon \bar{N} + \epsilon {M})$ is divided out.
Events with localisation at $0,\pi$ are determined by the inequality,
\be
\label{eqn:OPTTWOsinglevaluedcondition}
\vert l -r \vert \geq R \left( l+r \right).
\ee
For the case of symmetric initial states,
$P_{0,\pi}(1,D)=2\Gamma(D+0.5)\Gamma(0.5)/\pi\Gamma(D+1)$,
which is well approximated by $2/\sqrt{\pi D}$.
Evaluating $P_{0,\pi}(R,D)$ numerically for other values of $R$ reveals the following.
For cases when the initial state is asymmetric but $R$ is not too small,
$P_{0,\pi}(R,D)$ follows the same general trend as in the symmetric case
although it is weakly oscillatory and can be bigger or smaller than $P_{0,\pi}(1,D)$.
In fact, even for $R$ as small $0.57$,
when the intensities of the cavity modes are initially different
by a factor of $10$,
$P_{0,\pi}(0.57,D)$ and $P_{0,\pi}(1,D)$ take similar values for $D$ up to $100$.
However, for more extreme values of $R$, $P_{0,\pi}(R,D)$ is found to be
consistently greater than $P_{0,\pi}(1,D)$.
For example, for $R=0.2$, $P_{0,\pi}(0.2,D)$ is very roughly double the probability for symmetric case
for $D$ up to $50$.  Going to still smaller values of $R$,
$P_{0,\pi}(R,D)$/$P_{0,\pi}(1,D)$ is found to increase both with increasing detections $D$ and
$R$ tending to $0$.  In conclusion, in the limit of highly asymmetric initial states,
the increasing probabilities for events with $ l\approx r$
is less significant compared to the increase in the proportion of measurement outcomes
satisfying relation (\ref{eqn:OPTTWOsinglevaluedcondition}).

Thinking now about applications to understanding the transition between quantum and classical behaviour in the natural world
(or even decoherence in controlled laboratory experiments), the canonical interference procedure could be viewed as a simple prototype
for a natural process involving essential elements of indistinguishability (provided by coherent
combination under a linear mode coupling) and measurement.  It is of interest then to consider the localisation
of the relative phase after mixing over all possible measurement outcomes.
For the case of highly asymmetric Poissonian initial states it might be expected on the basis of the discussion above
that the averaged mixed state describes localisation of the relative phase at $0$ and $\pi$.
In fact this is not so, which can easily be demonstrated for the general case
of initial states of the form
$\rho _{A}=\int \frac{d^{2}\alpha }{2\pi }P_{A}(\alpha )\left\vert \alpha \right\rangle \!\! \left\langle \alpha \right\vert $
and
$\rho _{B}=\int \frac{d^{2}\beta }{2\pi }P_{B}(\beta )\left\vert \beta \right\rangle \!\! \left\langle \beta \right\vert $.
Here the modes are labelled $A$ and $B$,
$\left\vert \alpha \right\rangle $
and $\left\vert \beta \right\rangle$
are coherent states and $P_{A}(\alpha )$ and $P_{B}(\beta)$ are
the Glauber-Sudarshan P representation \cite{Glauber63,Sudarshan63}) of the states.
Under the action of leakage with parameter $\epsilon$ into the external modes labelled $ExtA$ and $ExtB$,
\bea
\rho _{I}
\! &=& \! \rho _{A}\otimes \rho _{B} \nonumber \\
\! &\rightarrow& \!
\int \frac{d^{2}\alpha }{2\pi }\frac{d^{2}\beta }{2\pi }P_{A}(\alpha )P_{B}(\beta )
\left\vert \sqrt{\epsilon }\alpha \right\rangle \!\! \left\langle \sqrt{\epsilon }\alpha \right\vert _{ExtA} \! \otimes \!
\left\vert \sqrt{\epsilon }\beta \right\rangle \!\! \left\langle \sqrt{\epsilon }\beta \right\vert _{ExtB} \nonumber \\
\! &&
\,\,\,\,\,\,\, \otimes \!
\left\vert \sqrt{1\!-\!\epsilon }\alpha \right\rangle \!\! \left\langle \sqrt{1\!-\!\epsilon }\alpha \right\vert _{A} \! \otimes \!
\left\vert \sqrt{1\!-\!\epsilon }\beta \right\rangle \!\! \left\langle \sqrt{1\!-\!\epsilon }\beta \right\vert _{B}. \nonumber
\eea
Letting $U_{comb}$ denote the combined unitary action of fixed phase shifts and
a linear mode coupling of arbitrary strength and duration on the external modes,
and supposing a subsequent measurement of $l$ and $r$ photons in modes $ExtA$ and $ExtB$,
the final state of modes $A$ and $B$ is,
\bea
\rho _{l,r} \! &=& \!
\int \frac{d^{2}\alpha }{2\pi }\frac{d^{2}\beta }{2\pi }P_{A}(\alpha )P_{B}(\beta )
\Big\vert \left\langle l\right\vert \! \left\langle r\right\vert U_{comb}
\left\vert \sqrt{\epsilon }\alpha \right\rangle \!\! \left\vert \sqrt{\epsilon }\beta \right\rangle _{Ext}\Big\vert ^{2} \nonumber \\
\! &\times&
\! \left\vert \sqrt{1\!-\!\epsilon }\alpha \right\rangle \!\! \left\langle
\sqrt{1\!-\!\epsilon }\alpha \right\vert _{A}\otimes \left\vert
\sqrt{1\!-\!\epsilon }\beta \right\rangle \!\!  \left\langle
\sqrt{1\!-\!\epsilon }\beta \right\vert _{B}\text{,} \nonumber
\eea
where  $\rho_{l,r}$ is subnormalised and $\textrm{tr} \left( \rho_{l,r} \right)$ is the associated probability.
If the record of the photodetections is unavailable, the final state of modes $A$ and $B$ is given by
mixing over all measurement outcomes and,
\bea
\label{eqn:OPTTWOmixingprocess}
\rho _{f}&=&\sum_{l,r=0}^{\infty}\textrm{tr}\left(\rho _{l,r}\right) \frac{\rho _{l,r}}{\textrm{tr} \left(\rho _{l,r}\right) } \\
&=&\sum_{l,r=0}^{\infty }\rho _{l,r}\text{.} \nonumber
\eea
However summing over the localising factors,
\be
\sum_{l,r=0}^{\infty }\left\vert \left\langle l\right\vert \! \left\langle r\right\vert
U_{comb}\left\vert \sqrt{\epsilon }\alpha \right\rangle \!\! \left\vert
\sqrt{\epsilon }\beta \right\rangle _{Ext}\right\vert^2 =1. \nonumber
\ee
Hence this complete mixing washes out the measurement-induced
localisation.
This result is in contrast to the situation when a state such as a Poissonian state
is beat against the vacuum at a beam splitter.
In this case a relative phase is acquired
depending on fixed physical properties of the beam splitter, and something
similar would happen for highly asymmetric states for which both modes
are populated.
A different point of view on the mixing process \eq{OPTTWOmixingprocess} is provided by
the theory of decoherence as proposed by Zurek et al \cite{Zurek03},
and is discussed in the Outlook, Chapter \ref{chap:Outlook}.

Returning to the example of initial Poissonian states, another question concerns the
speed of localisation for the symmetric versus asymmetric case.  This issue is complicated
by the fact that even a perfectly phase locked state,
\be
\label{eqn:OPTTWOasymrhoinfty}
\rho_{locked}
=
\int_{0}^{2\pi }\!\frac{d\theta}{2\pi}\,\left\vert \sqrt{\bar{N}}e^{i\theta }\right\rangle \!\! \left\langle \sqrt{\bar{N}}e^{i\theta }\right\vert
\otimes \left\vert \sqrt{\bar{M}}e^{i(\theta +\Delta _{0})}\right\rangle \!\! \left\langle \sqrt{\bar{M}}e^{i(\theta +\Delta _{0})}\right\vert,
\ee
(where the basis states are coherent states)
has a visibility less than $1$ whenever the amplitudes $\sqrt{\bar{N}}$ and $\sqrt{\bar{M}}$ are different from each other
(for the definition of the visibility refer \sect{OptOneQuantifyingLocalisation} of Chapter \ref{chap:OpticalOne}),
and will lead to an imperfect pattern of interference.
In fact, the visibility for $\rho_{locked}$ can easily be shown to be
$\frac{2\sqrt{\bar{M}\bar{N}}}{\bar{N}+\bar{M}}$, equal to the parameter $R$ assigned above
to products of Poissonian states with corresponding amplitudes.
However,
a combination of a single mode phase shifter and a beam splitter with reflection and transmission
coefficients determined by the ratio $\sqrt{\bar{N}/\bar{M}}$ is sufficient to
arrange deterministically for one or other of the modes of $\rho_{locked}$ to be
in the vacuum state, provided the phase parameter $\Delta_0$ is known accurately.
Hence, it seems reasonable to attribute perfect phase localisation to $\rho_{locked}$
and to define a modified visibility taking values between $0$ and $1$.
In what follows, visibilities are computed for the partially localised states \eq{OPTTWOasympartiallyloc}
generated from initial products of Poissonian states
by the canonical interference procedure,
and are adjusted
by a factor $1/R$.  It should be noted that in this example
the leakage and measurement processes do not
alter the relative intensities of the two modes, and that in
the limit of sharp localisation, the final state will be of the form of
$\rho_{locked}$.

\begin{figure}[h]
\begin{center}
\psfrag{PSFRAGINSERTION}{$\,\,\,\,\,\,\,\,\,\,\,\,\,\,\,\,\,\,\,\,
\epsilon (\bar{N} + \bar{M})$}
\includegraphics[height=6.3cm]{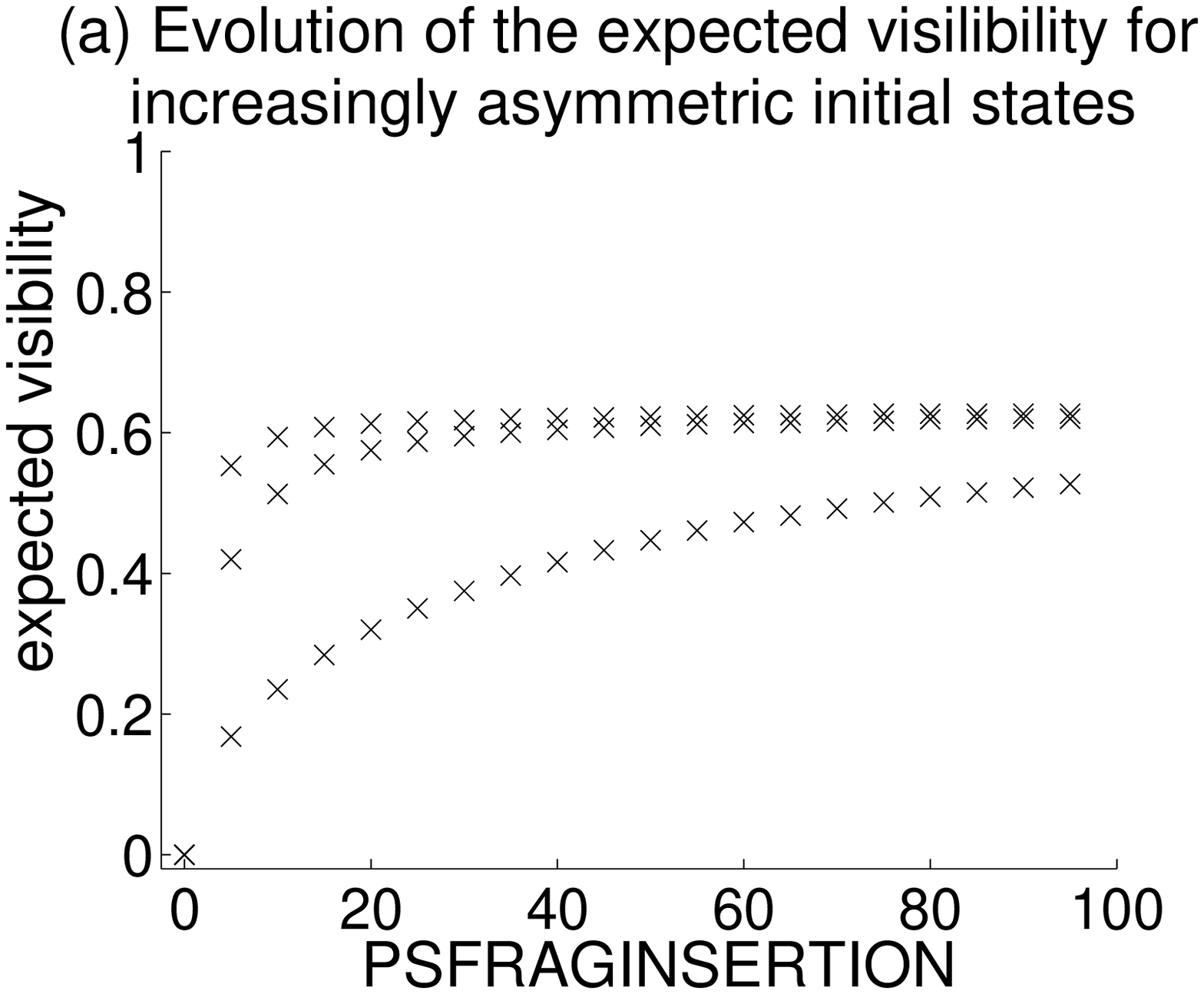}
\includegraphics[height=6.3cm]{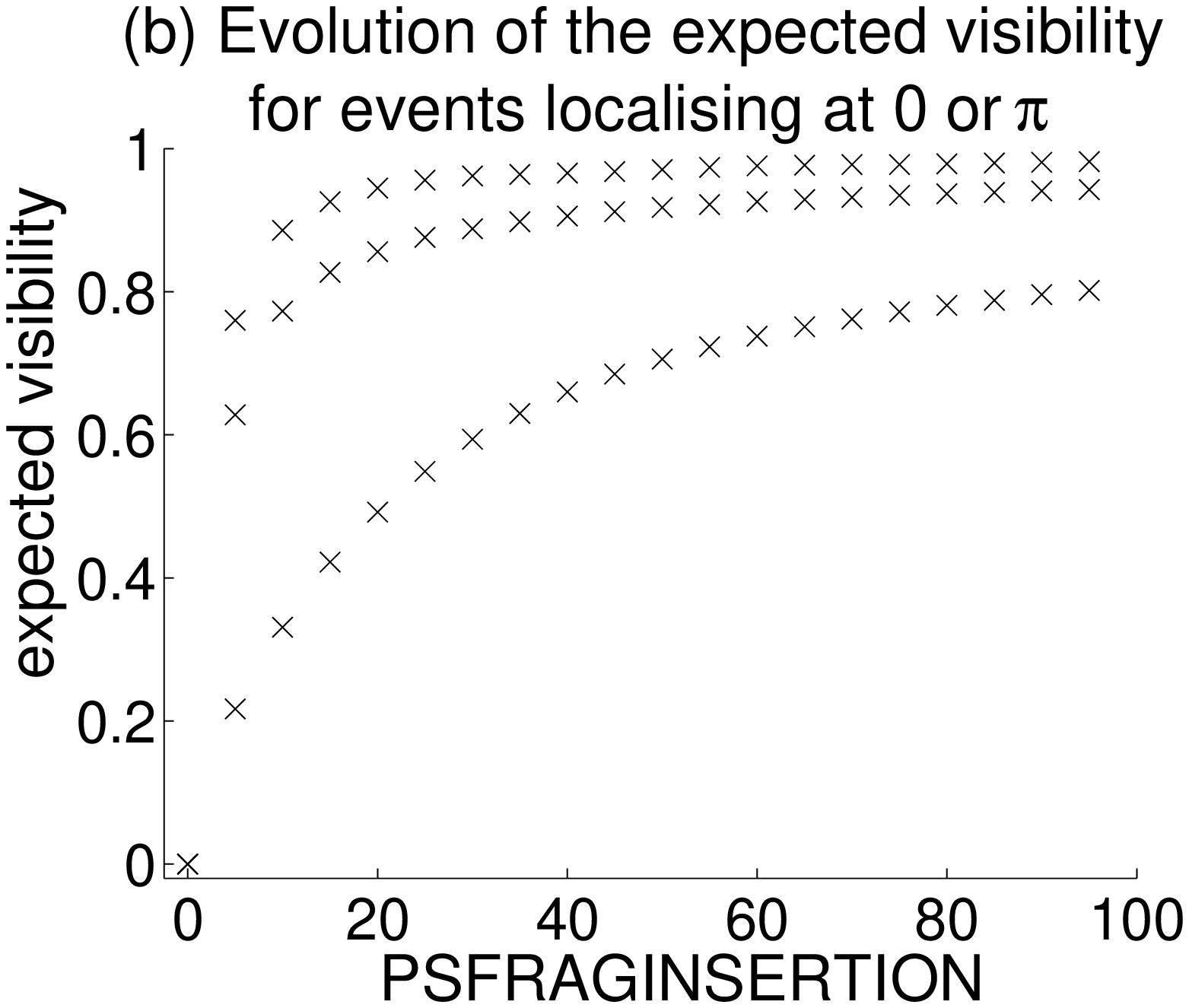}
\caption{\label{fig:OptTwoAsymAveragedVisibilities}
Evolution of the expected visibility $\sum_{l,r} P_{l,r} \widetilde{V}_{l,r}$
with increasing values of $\epsilon (\bar{N}+\bar{M})$
for initial asymmetric Poissonian states with intensities $\bar{N}$
and $\bar{M}$ in each mode.  Curves are for $R=0.94$, $0.57$ and $0.2$
and increase more slowly for smaller values of $R$.
In (a) all possible detections outcomes are included in the averaged visibility.
In (b) only outcomes with localisation at a single value of the relative phase
are included in the average.}
\end{center}
\end{figure}
The symmetry of the localising scalar functions $\vert C_{l,r}(\theta,\phi)\vert^2$,
as illustrated in \figu{OptTwoLSF}, allows a simple formal expression for the visibilities
to be written down in terms of the probability distribution $P_{l,r}$ for corresponding detection outcomes.
When $l=r$
the visibility can be shown to be $0$.  In
the case of more right than left detections, $r>l$,
the intensity $I(\tau)$ in the definition of the visibility
(refer \sect{OptOneQuantifyingLocalisation} of Chapter \ref{chap:OpticalOne})
is maximised and minimised for phase shifts of $\tau=0,\pi$
for which it is proportional to $(r+1)P_{l,r+1}$ and $(l+1)P_{l+1,r}$ respectively.
The situation when $l>r$ is analogous.  The following expression for the visibility,
including the additional rescaling factor $1/R$,
\be
\widetilde{V}_{l,r}=
\frac{1}{R}
\left\vert \frac{(r+1)P_{l,r+1}-\left( l+1\right) P_{l+1,r}}{(r+1)P_{l,r+1}+\left( l+1\right) P_{l+1,r}}\right\vert,
\ee
is valid for any outcome at the detectors.

The evolution of the expected visibility $\sum_{l,r} P_{l,r} \widetilde{V}_{l,r}$
as a function of $\epsilon (\bar{N}+\bar{M})$ is plotted in \figu{OptTwoAsymAveragedVisibilities} (a)
for the cases $R=0.94$, $0.57$ and $0.2$.  The average rate of localisation of the
relative parameter is seen to be progressively slower for more asymmetric initial states.
For the slowest case $R=0.2$ the expected visibility attains $85\%$ the value
for the fastest case $R=0.94$ by
$\epsilon (\bar{N}+\bar{M})=100$, while the cases of $R=0.57$ and $R=0.94$ converge
over the range shown.  More analysis would be required to clarify whether
the localisation can be as sharp (as in the symmetric case)
for small values of $R$ given enough detections.
However, the slower rate of localisation for highly asymmetric initial states
would in any case necessitate a large total intensity for the initial states if the limit is to be attained.
Another feature of \figu{OptTwoAsymAveragedVisibilities} (a) is that the
expected visibilities increase to a value very much less than one
--- in fact the maximum is $0.63$ on the range shown.  This reflects
localisation at two values of the relative phase.  In
\figu{OptTwoAsymAveragedVisibilities} (b) the averaging for the visibility
is restricted to outcomes for which there is localisation at single values, which can be
$0$ or $\pi$ (in other words outcomes satisfying \eq{OPTTWOsinglevaluedcondition}),
and account is taken of the total probability of such events.  The expected
visibilities for these events is $0.98$, $0.94$ and $0.81$ by $\epsilon (\bar{N}+\bar{M})=100$
for $R=0.94$, $0.57$ and $0.2$ respectively.

\section{Transitivity of the canonical interference procedure}
\label{sec:OPT2transitivity}

This section considers the transitive properties of the process of localisation
underlying the canonical interference procedure,
introduced in \sect{OptOneFockstates} of Chapter \ref{chap:OpticalOne}.
The following examples assume that the states of the optical modes
prior to the localisation processes are Poissonian, which can readily be prepared in the laboratory,
but the conclusions are expected to apply generally.

Suppose that two systems, modes 1 and 2, have been prepared in a phase locked state
by an interference experiment in which several photons were detected,
and are brought together with a third system, mode 3, in a Poissonian state with no
phase correlation with either mode 1 or mode 2.
The state of the first two modes is,
\be
\rho _{I}^{(1,2)}=\int_{0}^{2\pi }\frac{d\theta }{2\pi }\left\vert \sqrt{\bar{N}_{1}}e^{i\theta }\right\rangle \left\langle \sqrt{\bar{N}_{1}}e^{i\theta }\right\vert _{1}\otimes \left\vert \sqrt{\bar{N}_{2}}e^{i\left( \theta +\Delta _{0}\right) }\right\rangle \left\langle \sqrt{\bar{N}_{2}}e^{i\left( \theta +\Delta _{0}\right) }\right\vert _{2},
\ee
and of the third mode is,
\be
\rho _{I}^{(3)}=\int_{0}^{2\pi }\frac{d\phi }{2\pi }\left\vert \sqrt{\bar{N}_{3}}e^{i\phi }\right\rangle \left\langle \sqrt{\bar{N}_{3}}e^{i\phi }\right\vert _{3},
\ee
where the basis states are coherent states with amplitudes $\sqrt{\bar{N}_j}$ and $j=1,2,3$ is the mode label.
Suppose now that the interference procedure is performed on modes 2 and 3, with $l$ and $r$
photons detected at the ``left'' and ``right'' photocounters, as described in \sect{OptOneFockstates}.
The state of the three
systems is transformed as \linebreak $\rho _{I}^{(1,2)}\otimes \rho _{I}^{(3)} \rightarrow \rho ^{\prime}$ as follows,
\bea
\rho ^{\prime}
\!\!\! &\propto& \!\!\! \left( a_{2}+a_{3}\right) ^{r}\left( a_{2}-a_{3}\right) ^{l}\rho _{I}^{(1,2)}\otimes \rho _{I}^{(3)}\left( a_{2}^{\dagger }-a_{3}^{\dagger }\right) ^{l} \left( a_{2}^{\dagger }+a_{3}^{\dagger }\right) ^{r} \nonumber \\
\rho ^{\prime}
\!\!\! &=& \!\!\! \int_{0}^{2\pi } \!\!\! \int_{0}^{2\pi }
\frac{d\theta }{2\pi } \! \frac{d\phi }{2\pi }
\left\vert C_{l,r}\left( \theta ,\phi \right) \right\vert ^{2}
\left\vert \sqrt{\bar{N}_{1}}e^{i\theta }\right\rangle \!\! \left\langle \sqrt{\bar{N}_{1}}e^{i\theta }\right\vert _{1}\otimes \nonumber \\
\!\!\! && \,\,\,\,\,\,\,\,\,\,\,\,\,\,\,\,\,\,
\left\vert \sqrt{\bar{N}_{2}}e^{i\left( \theta +\Delta _{0}\right) }\right\rangle \!\! \left\langle \sqrt{\bar{N}_{2}}e^{i\left( \theta +\Delta _{0}\right) }\right\vert _{2}\otimes
\left\vert \sqrt{\bar{N}_{3}}e^{i\phi }\right\rangle  \!\! \left\langle \sqrt{\bar{N}_{3}}e^{i\phi }\right\vert _{3} \! ,
\eea
where $\rho'$ here is assumed to be subnormalised so that $\textrm{tr}(\rho^{\prime})$ gives the probability of the specific
outcome at the detectors.
The scalar function $\vert C_{l,r}\vert^2$ encoding information about the localisation of the relative phases takes the form,
\be
\left\vert C_{l,r}\left( \theta ,\phi \right) \right\vert ^{2}\propto\left\vert \sqrt{\bar{N}_{2}}e^{-i\Delta /2}+\sqrt{\bar{N}_{3}}e^{i\Delta /2}\right\vert ^{2r}\left\vert \sqrt{\bar{N}_{2}}e^{-i\Delta /2}-\sqrt{\bar{N}_{3}}e^{i\Delta /2}\right\vert ^{2l},
\ee
where $\Delta \equiv \phi -\theta -\Delta _{0}$.
For example if $\vert C_{l,r}\vert^2$ is sharply peaked around the value $\Delta_1$ for $\Delta$,
the final state of the three systems is,
\bea
\rho ^{\prime }
\!\!\! &\propto& \!\!\! \int_{0}^{2\pi } \! \frac{d\theta }{2\pi }
\left\vert \sqrt{\bar{N}_{1}}e^{i\theta }\right\rangle \!\! \left\langle \sqrt{\bar{N}_{1}}e^{i\theta }\right\vert _{1}
\otimes \nonumber \\
\!\!\! && \,\,\,\,\,\,\,\, \left\vert \sqrt{\bar{N}_{2}}e^{i\left( \theta +\Delta _{0}\right) }\right\rangle \!\! \left\langle \sqrt{\bar{N}_{2}}e^{i\left( \theta +\Delta _{0}\right) }\right\vert _{2}
\otimes \left\vert \sqrt{\bar{N}_{3}}e^{i\left( \theta +\Delta _{0}+\Delta _{1}\right) }\right\rangle \!\! \left\langle \sqrt{\bar{N}_{3}}e^{i\left( \theta +\Delta _{0}+\Delta _{1}\right) }\right\vert _{3} \! . \nonumber
\eea
The relative phase of mode 3 relative to mode 1 has evolved to $\Delta_0+\Delta_1$.
If the second system  is lost --- i.e. mode 2 is traced out --- the final state of modes 1 and 3 is,
\[
\int_{0}^{2\pi } \! \frac{d\theta }{2\pi }
\left\vert \sqrt{\bar{N}_{1}}e^{i\theta }\right\rangle \!\! \left\langle \sqrt{\bar{N}_{1}}e^{i\theta }\right\vert _{1}
\otimes \left\vert \sqrt{\bar{N}_{3}}e^{i\left( \theta +\Delta _{0}+\Delta _{1}\right) }\right\rangle \!\! \left\langle \sqrt{\bar{N}_{3}}e^{i\left( \theta +\Delta _{0}+\Delta _{1}\right) }\right\vert _{3},
\]
and the phase correlation remains unaffected by the loss.

The previous example demonstrates that the localised relative quantum phases have the same additive properties as classical phases between
different systems.  Furthermore, the interference procedure on the second and third systems is independent of the prior
phase correlation with the first system.  This is clear from
the dependence of $\vert C_{l,r} \vert^2$ on $\Delta =\phi -\theta -\Delta _{0}$
and the fact that,
\be
\textrm{tr}\left(\rho ^{\prime}\right)=\int_{0}^{2\pi }\int_{0}^{2\pi }\frac{d\theta }{2\pi }\frac{d\phi }{2\pi }\left\vert C_{l,r}\left( \phi -\theta -\Delta _{0}\right) \right\vert ^{2}=\int_{0}^{2\pi }\int_{0}^{2\pi }\frac{d\theta }{2\pi }\frac{d\phi }{2\pi }\left\vert C_{l,r}\left( \phi -\theta \right) \right\vert ^{2}.
\ee
$\textrm{tr}\left(\rho^{\prime}\right)$, determining the probability for different measurement outcomes of the interference procedure,
is the same here as
for an initial product of Poissonian states with corresponding mean photon numbers $N_2$ and $N_3$
uncorrelated with a further system.

In practice however establishing a relative phase between two systems will have some effect on phase correlations with external systems
due to the effects of photon depletion. With reference to the previous example,
if the amplitudes of modes 2 and 3 are reduced by a factor
$\sqrt{1-\epsilon }$
during the interference procedure, the relative amplitude of modes 1 and 2 changes as
$\sqrt{\bar{N}_{1}/\bar{N}_{2}}\longrightarrow \left( 1/\sqrt{1-\epsilon }\right) \sqrt{\bar{N}_{1}/\bar{N}_{2}}$.
This does not affect the value $\Delta _{0}$ of the relative phase between modes 1 and 2,
but does change, for example, the behaviour of the two modes if combined on a beam splitter.

The independence of the canonical interference procedure of phase correlations with
external systems makes it interesting to consider what happens
in a three mode system where, as previously, mode 3 is initially Poissonian,
and an interference procedure
is performed on modes 2 and 3, but where differently, modes 1 and 2 begin in a
state where the relative phase is localised at two values
$\pm \Delta_0$.  Such states are produced
in an ideal interference experiment (refer \sect{OptOnesymmetries} of Chapter \ref{chap:OpticalOne}).
It is easy to see that after the second interference procedure on systems 2 and 3,
which for simplicity is assumed to determine the value 0 for the relative phase,
the final state of the three mode system is,
\bea
\int_{0}^{2\pi} \!\! \frac{d\theta }{2\pi }
\left\vert \sqrt{\!\! \bar{N}}e^{i\theta }\right\rangle
\!\! \left\langle \!\! \sqrt{\bar{N}}e^{i\theta }\right\vert _{1}
\!\!\! \otimes \!\!
\left\vert \! \sqrt{\!\! \bar{N}}e^{i\left(\theta +\Delta _{0}\right)}\right\rangle
\!\! \left\langle \!\! \sqrt{\bar{N}}e^{i\left(\theta +\Delta _{0}\right)}\right\vert _{2}
\!\!\! \otimes \!\! \left\vert \! \sqrt{\bar{N}}_{3}e^{i\left(\theta +\Delta _{0}\right)}\right\rangle
\!\! \left\langle \!\! \sqrt{\bar{N}}_{3}e^{i\left(\theta +\Delta _{0}\right)}\right\vert _{3}
\,\nonumber \\
+ \int_{0}^{2\pi} \!\! \frac{d\theta }{2\pi }
\left\vert \sqrt{\!\! \bar{N}}e^{i\theta }\right\rangle
\!\! \left\langle \!\! \sqrt{\bar{N}}e^{i\theta }\right\vert _{1}
\!\!\! \otimes \!\!
\left\vert \! \sqrt{\!\! \bar{N}}e^{i\left(\theta -\Delta _{0}\right)}\right\rangle
\!\! \left\langle \!\! \sqrt{\bar{N}}e^{i\left(\theta -\Delta _{0}\right)}\right\vert _{2}
\!\!\! \otimes \!\! \left\vert \! \sqrt{\bar{N}}_{3}e^{i\left(\theta -\Delta _{0}\right)}\right\rangle
\!\! \left\langle \!\! \sqrt{\bar{N}}_{3}e^{i\left(\theta -\Delta _{0}\right)}\right\vert _{3}.
\nonumber
\eea
Interestingly
if the second system is lost (i.e. if mode 2 is traced over), modes 1 and 3 are
left in a state localised at two values, but if the first system is lost (i.e. if mode 1 is traced over),
modes 2 and 3 are left localised at one value of the relative phase 0.
It should also be pointed out that for a well localised state
$\int \frac{d\theta}{2\pi} \ket{\alpha}\bra{\alpha} \otimes \ket{\beta}\bra{\beta}$,
where $\alpha=\sqrt{(1-\epsilon)N_1}e^{i\theta}$ and $\beta=\sqrt{(1-\epsilon)N_2}e^{i(\theta+\Delta_0)}$,
the expected numbers of photons in each mode --- $(1-\epsilon)N_1$ and $(1-\epsilon)N_2$  ---
are independent of $\Delta_0$, and hence the effects of photon depletion in an interference experiment
involving one mode of pair of systems in a state having two components
with different localised relative phases
will in general be the same on both components.

\section{Application to engineering large optical Fock states}
\label{sec:OPT2FockAddition}

This section presents an example of how the canonical interference procedure introduced in \sect{OptOneFockstates}
of Chapter \ref{chap:OpticalOne}
can be utilised in schemes for quantum state engineering based on linear optics and classical feedforward of measurement
results.
The example considered is the generation of large Fock
states from a source of single photons by a probabilistic process of ``addition''.
Large photon number states are of interest for several reasons.
They have applications to interferometry with phase uncertainty reduced to the Heisenberg limit, considered
theoretically in \cite{Holland93,Kim98}.
Second, the coherence properties of Fock states are of fundamental interest, particularly in relation to those of
Glauber coherent states.
Finally relational Schr\"{o}dinger cat states, generated by applying the canonical interference procedure to
pairs of Fock states, can easily be converted to ``High NOON states'' --- states of the form
$\frac{1}{\sqrt{2}}\left( \left\vert N\right\rangle _{1}\left\vert 0\right\rangle _{2}+
\left\vert 0\right\rangle _{1}\left\vert N\right\rangle _{2}\right)$ in a basis of Fock states where $N$ is assumed to be
large --- whenever the two components of the cat have relative phases approximately $\pi$ apart.
This conversion is explained at the end of the section.
For a recent discussion of some important applications of NOON states, in particular for interferometry
beyond the shot-noise limit, see \cite{Lee05}.

The problem considered here is how to create large Fock states given (idealised) linear optical components
--- beam splitters, photon detectors, phase shifters and a highly efficient source of single photons
(for recent progress on sources see \cite{Grangier04} and on detectors see \cite{Rosenberg05}).
One obvious and simple but inefficient method is as follows,
suggested by the well-known Hong, Ou and Mandel dip experiment \cite{Hong87},
whereby two uncorrelated and identical photons, simultaneously incident on the input ports of a $50:50$ beam splitter,
must both be registered at the same output port.
Starting with a product of one photon states $\ket{1}\ket{1}$, coherent combination of the optical modes at a $50:50$
beam splitter,\footnote{Fixing the phase convention for calculations in this section,
the beam splitter transformation is taken to be
$\exp \left( \frac{\pi }{4}ab^{\dagger }-\frac{\pi }{4}a^{\dagger }b\right)$,
$a$ and $b$ denoting the annihilation operators for the modes labelled 1 and 2.} yields a superposition
of states with two photons in a single mode,
\[
\sqrt{\frac{1}{2}}
\Big\{ -\left\vert 2\right\rangle _{1}\left\vert 0\right\rangle _{2}+\left\vert 0\right\rangle _{1}\left\vert 2\right\rangle _{2}
\Big\} . \nonumber
\]
Assuming the second output port is monitored by a (non-discriminating) photodetector, a two photon Fock state
is created at the first output port with probability $1/2$.
This process could be iterated to yield successively larger Fock states.
For example if two two-photon states are combined on the beam splitter,
\[
\left\vert 2\right\rangle _{1}\left\vert 2\right\rangle _{2}
\longrightarrow
\frac{1}{4}
\Big\{ \sqrt{3.2}\left\vert 4\right\rangle _{1}\left\vert 0\right\rangle _{2}-2\left\vert 2\right\rangle _{1}\left\vert 2\right\rangle _{2}+\sqrt{3.2}\left\vert 0\right\rangle _{1}\left\vert 4\right\rangle _{2}
\Big\},
\]
in a basis of Fock states,
and if the second mode is monitored by a photon-number discriminating detector, the output state in the first mode is $\ket{4}$
with probability $3/8$, and $\ket{2}$ with probability $1/4$.
For the case of an input state with the same number $N$ of photons in each mode
it is seen that in half, or a little under half, of such operations
the output is worse than the inputs.

Looking in more detail it is helpful to determine a general expression for the action of a $50:50$ beam-splitter on the state
$\ket{N}_1\ket{N}_2$ with $N$ photons in each mode, which can be done as follows.
Adopting a representation in terms of coherent states,
\[
\left\vert N\right\rangle_1 \left\vert N\right\rangle_2=\frac{1}{\Pi _{N}\left( N\right) }
\int_{0}^{2\pi } \!\!\!\! \int_{0}^{2\pi }\frac{d\theta }{2\pi }\frac{d\phi }{2\pi }e^{-iN\left( \theta +\phi \right) }
\left\vert \sqrt{N}e^{i\theta }\right\rangle _{1}\left\vert \sqrt{N}e^{i\phi }\right\rangle _{2},
\]
where $\Pi_N\left(N\right)$ denotes a Poissonian factor.
Under the action of the beam splitter this is transformed to,
\be
\left\vert \psi^f\right\rangle
=
\frac{1}{\Pi _{N}\left( N\right) }
\int_{0}^{2\pi } \!\!\!\! \int_{0}^{2\pi }\frac{d\theta }{2\pi }\frac{d\phi }{2\pi }e^{-iN\left( \theta +\phi \right) }\left\vert \frac{\sqrt{N}e^{i\theta }\!+\!\sqrt{N}e^{i\phi }}{\sqrt{2}}\right\rangle _{1}\left\vert \frac{-\sqrt{N}e^{i\theta }\!+\!\sqrt{N}e^{i\phi }}{\sqrt{2}}\right\rangle _{2}\!\!. \nonumber
\ee
Expanding the coherent states in the Fock basis
and making the replacement\break
$\int_{0}^{2\pi } \!\! \int_{0}^{2\pi }\frac{d\theta }{2\pi }\frac{d\phi }{2\pi }\left[ \cdot \right]
=\int_{0}^{2\pi } \!\! \int_{0}^{2\pi }
d\left( \frac{\phi -\theta }{2}\right) d\left( \frac{\theta +\phi }{2}\right) \left( \frac{1}{2\pi }\right) ^{2}\left[ \cdot \right] $
(where $\frac{\phi-\theta}{2}$ and $\frac{\theta+\phi}{2}$
are the variables of integration),
valid because of the $2\pi$-periodicity in $\theta$ and $\phi$,
\bea
\left\vert \psi^f\right\rangle &=& \frac{e^{-N}}{\Pi _{N}\left( N\right) }
\sum_{A,B=0}^{\infty }\frac{1}{\sqrt{A!B!}}\left( 2N\right) ^{\left( A+B\right) /2}i^{B} \nonumber \\
&\times&
\left[ \int_{0}^{2\pi} \!\! \frac{1}{2\pi} \, d\left( \frac{\theta +\phi }{2}\right)
e^{i\left( A+B-2N\right) \left( \theta +\phi \right) /2}\right] \nonumber \\
&\times&
\left[ \int_{0}^{2\pi} \!\! \frac{1}{2\pi} \, d\left( \frac{\Delta }{2}\right)
\cos ^{A}\left( \frac{\Delta}{2}\right) \sin ^{B}\left( \frac{\Delta}{2}\right) \right]
\left\vert A\right\rangle _{1}\left\vert B\right\rangle _{2}, \nonumber
\eea
where $\Delta \equiv \phi-\theta$.
The integral over the average of the phase variables enforces conservation of photon number $A+B=2N$.
As for the second integral the following result is required,
\[
\int_{0}^{\pi /2} \!\!  \frac{dz}{2\pi} \cos ^{A}z\sin ^{B}z
=\frac{1}{4\pi } \, \Gamma \left( \frac{A+1}{2}\right) \Gamma \left( \frac{B+1}{2}\right)
\Big{/} \, \Gamma \left( \frac{B+A}{2}+1\right),
\]
for an integral over the first quadrant, where $\Gamma(\cdot)$ denotes the usual gamma function.
The integral over the phase difference variable
$\Delta/2$ over all four quadrants is seen to be four times this when $A$ and $B$ are both even, and $0$ otherwise
since positive and negative valued contributions cancel.
Overall
$\ket{\psi^f}$
is seen to be a sum over contributions with even photon numbers for both modes. The final expression simplifies
to the following (using the standard identity
$\Gamma \left( M+0.5\right) =\left( 2M\right) !\sqrt{\pi }/\left( 2^{2M}M!\right) $ valid for $M\geq 0)$,
\be
\left\vert \psi^f\right\rangle =\frac{1}{2^{N}}\sum_{j=0}^{N}
\frac{\sqrt{\left( 2j\right)!\left( 2N-2j\right)!}}{j!\left( N-j\right) !}\,
e^{i\left( N-j\right) \pi }\left\vert 2j\right\rangle _{1}\left\vert 2N-2j\right\rangle _{2},
\ee
(in agreement with \cite{Campos89}).
When a photon number discriminating detector monitors the second mode
the most probable outcomes are measurements of
$0$ or $2N$ photons, and the least likely measurement is $N/2$ photons (or $\left(N\pm1\right)/2$ photons if $N$ odd).
If a non-discriminating detector monitors the second mode only
a measurement of $0$ photons would be useful, which happens with probability $P_0$ where,
\bea
\label{eqn:OPTTWOPzeroBASIC}
P_0&=&
\left\vert \left\langle 2N\right\vert_1 \left\langle 0\right\vert _2
\left\vert \psi ^f\right\rangle \right\vert ^{2} \nonumber \\
&=&\frac{\left( 2N\right)!}{ 2^{2N}N!^{2} }\,,
\eea
which well approximated by the expression
$1 \big{/} \sqrt{\pi N}$ ( obtained using Stirling's formula for large factorial numbers).

It is interesting to compare this result with the action of a lossless beam splitter
on a product of coherent states $\ket{\alpha}_1\ket{\beta}_2$
or a phase locked state\break
$\int_{0}^{2\pi }\frac{d\theta}{2\pi}e^{-i2N\theta }\left\vert \alpha \right\rangle_1\left\vert \beta \right\rangle_2$,
where
$\alpha =\sqrt{N_{1}}e^{i\theta}$
and
$\beta =\sqrt{N_{2}}e^{i\left( \theta +\Delta _{0}\right) }$
---
products of coherent states evolve deterministically to products of coherent states.
A combination of a single mode phase shifter and a beam splitter with reflection and transmission
coefficients determined by the ratio $\sqrt{N_1/N_2}$ of the amplitudes of the coherent states
suffices to get mode 2 into the vacuum state.  This
motivates an approach to ``adding'' Fock states based on the interference procedure
discussed in \sect{OptOneFockstates} of Chapter \ref{chap:OpticalOne}
which localises the relative phase degree of freedom.
Since the localisation in such a procedure is rapid,
it would be expected that a substantial improvement in the success probability of the previous method is possible
by sacrificing a small number of photons, and proportionally less for larger input states.
In fact when the input states are one-photon Fock states there can be no benefit
--- a single photon Fock state could be obtained with certainty but is the same as the inputs.

In detail a revised method of ``adding'' a product of Fock states $\ket{N}_1\ket{N}_2$ is as follows.
Refer back to \sect{OptOneFockstates} of Chapter \ref{chap:OpticalOne} for
an explanation of the canonical interference procedure; the notation and assumptions here are the same.
It is assumed now that: the canonical interference procedure is applied first and stopped after a
fixed total number of detections $W$ have been recorded; that the apparatus is ideal i.e.
phase stable; and that phase shifts throughout the apparatus are fully characterised so that full information
is available about the value $\Delta_0$ of the (partially) localised relative phase.
If the process is performed slowly then only non-disciminating photocounters would be needed.
If $W\geq2$ it is assumed that the process is
interrupted half way through and that a phase shifter in the second arm (refer \figu{OptOnetwocavities} of
\sect{OptOneFockstates}) is adjusted by $\pi/2$.  This enforces
localisation at one value of the relative phase.  Without this step a relational Schr\"{o}dinger cat state would be obtained
with high probability and it would not be possible to arrange for such large variation in the
intensity at one of the output ports of a beam splitter applied subsequently as can be achieved otherwise.
The state of the optical modes in a basis of Glauber coherent states is then,
\[
\left\vert \psi ^{loc}\right\rangle =
\int \frac{d\theta }{2\pi }\frac{d\phi }{2\pi }e
^{-i\left( 2N-W\right) \left( \theta +\phi \right) /2}
C^{0}\!\left(\frac{\Delta }{2}\right)\left\vert \sqrt{N^\prime}e^{i\theta }\right\rangle _{1}
\left\vert \sqrt{N^\prime}e^{i\phi }\right\rangle _{2},
\]
where $\Delta \equiv \phi - \theta$,
$N^\prime=\left( 1-\epsilon \right) N$ (where $\epsilon$ is a leakage parameter)
and the scalar function $C^{0}\!\left(\frac{\Delta }{2}\right)$ is peaked at $\Delta =\Delta _{0}$
and is normalised such that $\langle \psi ^{loc}|\psi ^{loc}\rangle =1$.

The next steps in the procedure are a phase shift of the second mode by $-\Delta_0+\pi$
and coherent combination of the two modes at a $50:50$ beam splitter, so that the final state is,
\bea
\left\vert \psi ^f\right\rangle
&=&\int \frac{d\theta}{2\pi}\frac{d\phi}{2\pi}
e^{-i\left( 2N-W\right)\left( \theta +\phi \right)/2}
C^{0}\!\left(\frac{\Delta}{2}\right) \nonumber \\
&\times&
\left\vert \frac{\sqrt{N^\prime}e^{i\theta }+\sqrt{N^\prime}e^{i\phi }e^{-i\Delta _{0}}}{\sqrt{2}}\right\rangle _{1}
\left\vert \frac{\sqrt{N^\prime}e^{i\theta }-\sqrt{N^\prime}e^{i\phi }e^{-i\Delta _{0}}}{\sqrt{2}}\right\rangle _{2} \!\!.
\nonumber
\eea
If a detector monitoring the second output port registers 0 photons,
$\left\vert \psi ^f\right\rangle $
is projected onto the Fock state
$\left\vert 2N-W\right\rangle _{1}$
as follows,
\bea
\left\langle 0\right\vert _{2}\left\vert \psi^f\right\rangle
&=& e^{-i\left( 2N-W\right) \Delta _{0}/2}\sqrt{\Pi _{2N-W}\left[ 2\left( 1-\epsilon \right) N\right]} \nonumber \\
&\times& \int_{0}^{2\pi } \!\! \frac{1}{2\pi } \, d\left( \frac{\Delta }{2}\right)
C^{0}\!\left( \frac{\Delta }{2} \right)
\cos ^{2N-W} \!
\left( \frac{\Delta -\Delta _{0}}{2}\right) \left\vert 2N-W\right\rangle _{1}, \nonumber \\
&&
\eea
where $\Pi _{2N-W}\left[ 2\left( 1-\epsilon \right) N\right] $
is a Poissonian factor with mean $2\left( 1-\epsilon \right) N$,
and the variable of integration $\Delta /2$ runs between $0$ and $2\pi$.  If a photon number
discriminating detector is available for monitoring the second output port then the output at the first
output port need not be discarded when the detector fails to measure a vacuum state;
in general the yield of Fock states with more photons than the input states is expected
to be considerably improved compared to the earlier method.

Letting $\left\vert \psi^{proc}\right\rangle $ denote the normalised state of the two optical modes
after the localisation procedure has been performed and after a phase shift of
$-\Delta _0+\pi $ has been applied to the second mode,
$\left\vert \psi^{proc}\right\rangle=\exp\left[i\left(-\Delta _{0}+\pi\right)b^\dagger b\right]\ket{\psi^{loc}}$
where
$b$ denotes the annihilation operator on mode 2,
and letting $U(50:50)$ denote the
beam splitter transformation, the amplitude for detecting the vacuum state is given in
a basis of orthogonal Fock states by,
\bea
&& \!\! \left\langle 2N-W\right\vert _{1}\left\langle 0\right\vert _{2}U(50:50)\left\vert \psi^{proc}\right\rangle \nonumber \\
&=& \!\!\!  \left( \frac{1}{2}\right) ^{\left( 2N-W\right) /2}
\left\{
\sum _{s=0}^{\infty}
\left(
\begin{array}{c}
2N-W \\
S
\end{array}
\right) ^{1/2}
\!\!\!\!
\left( -1\right) ^{\left( 2N-W-s\right)}
\left\langle s\right\vert_1 \left\langle 2N-W-s\right\vert_2 \right\}
\left\vert \psi ^{proc}\right\rangle. \nonumber \\
&&
\eea
Consider the case when $W=1$ and the interference procedure involves just one detection.
If the photon is detected at mode 1 then clearly no additional phase shift is required. After normalisation,
\[
\left\vert \psi ^{proc}\right\rangle
=\frac{\left\vert N-1\right\rangle \left\vert N\right\rangle
-\left\vert N\right\rangle \left\vert N-1\right\rangle }{\sqrt{2}}.
\]
$P_{0}$, denoting the success probability
$\left\vert \bra{2N-1}_1\bra{0}_2 U(50:50) \left\vert \psi ^{proc}\right\rangle \right\vert ^{2}$,
evaluates to $2 \left( 2N\right)! \big{/} 2^{2N}N!^2$,
exactly double the value for the first procedure
(refer \eq{OPTTWOPzeroBASIC}), and
consistent with the Hong, Ou and Mandel phenomenon when $N=1$.
If the localising detection occurs mode 1, an additional phase adjustment
of $\pi$ is required, and $P_{0}$ is the same.

Consider now the case when $W=2$, and the detections are both at mode 2.
The protocol demands a fixed phase shift of $\pi/2$ between two detections so that
$\left\vert \psi ^{loc}\right\rangle \propto \left( a+b\right) \left( a+ib\right)
\left\vert N\right\rangle_1 \left\vert N\right\rangle_2 $.
Other outcomes at the detectors work out analogously.
After normalising,
\[
\left\vert \psi ^{loc}\right\rangle
\!=\!
\sqrt{\frac{ N\!\!-\!1 }{4N\!\!-\!2}}\left\vert N\!\!-\!2\right\rangle_1 \left\vert N\right\rangle_2
+\frac{\left( 1\!+\!i\right) \sqrt{N}}{\sqrt{4N\!\!-\!2}}\left\vert N\!\!-\!1\right\rangle_1 \left\vert N\!\!-\!1\right\rangle_2
+i\sqrt{\frac{N\!\!-\!1}{4N\!\!-\!2}}\left\vert N\right\rangle_1 \left\vert N\!\!-\!2\right\rangle_2,
\]
peaked at a relative phase of
$\Delta_{0}=-\pi/4$.
Hence a phase shift on mode 2 of
$5\pi/4$ is required to maximise the success probability $P_{0}$,
after which the state of the optical modes is,
\[
\left\vert \psi ^{proc}\right\rangle
=
\sqrt{\frac{N\!\!-\!1}{4N\!\!-\!2}}\left\vert N\!\!-\!2\right\rangle \left\vert N\right\rangle
-\sqrt{\frac{2N}{4N\!\!-\!2}}\left\vert N\!\!-\!1 \right\rangle \left\vert N\!\!-\!1\right\rangle
+\sqrt{\frac{ N\!\!-\!1}{4N\!\!-\!2}}\left\vert N\right\rangle \left\vert N\!\!-\!2\right\rangle.
\]
Then the success probability $P_0$ is given by,
\[
\left\vert \left\langle 2N\!\!-\!2\right\vert _{1}\left\langle 0\right\vert _{2}
U(50:50)\left\vert \psi ^{proc}\right\rangle \right\vert ^{2}
=\frac{\left( 2N\!\!-\!2\right)!}{2^{2N\!\!-\!2}(N\!\!-\!1)!^2}
\left[ 2\sqrt{\frac{N\!\!-\!1}{N}}\sqrt{\frac{N\!\!-\!1}{4N\!\!-\!2}}+\sqrt{\frac{N}{2N\!\!-\!1}}\right] ^{2}.
\]
A simpler expression valid asymptotically is,
\[
P_{0}\simeq 4\frac{1}{2^{2N}}\frac{\left( 2N-2\right) !}{(N-1)!^{2}}\left[ 1+\sqrt{\frac{1}{2}}\right] ^{2}\text{,}
\]
having a fractional error of less than $5\%$ for $N\geq15$, but it is invalid for small values of $N$.
Taking the full expression for $P_0$, the ratio of the success probability here with
that of the method without the interference procedure
\eq{OPTTWOPzeroBASIC} is $2.6$
for the uninteresting case $N=2$ (when the input and output states are the same size), rising to $2.9$ for larger values for $N$.
These preliminary results for $W=1,2$
suggest that
sacrificing even more photons to achieve sharper localisation of the relative phase
may be helpful for optimising the addition process for larger Fock states.

One potential application of engineering large photon number states is generating relational
Schr\"{o}dinger cat states, using again the canonical interference process set out
in \sect{OptOneFockstates} of Chapter \ref{chap:OpticalOne},
and from these NOON states.
This could be done as follows, continuing with the same notation
and assumptions.  Consider the case after several detections
(a total number $D$) for an initial product of Fock states $\ket{N}_1\ket{N}_2$
when the counts at the ``left'' and ``right'' detectors are equal.
The final state of the two optical modes is a relational
Schr\"{o}dinger cat state with components having sharply defined relative phases of $\pm \pi/2$, and is of the form,
\[
\left\vert \psi _{r.s.c.}\right\rangle \propto
\int_{0}^{2\pi }\frac{d\theta }{2\pi }e^{-i\left\vert \gamma \right\vert ^{2}\left[ 2\theta +\left( \pi /2\right) \right] }
\left\vert \gamma \right\rangle_1 \left\vert \gamma e^{i\pi /2}\right\rangle_2
+\int_{0}^{2\pi }\frac{d\theta }{2\pi }e^{-i\left\vert \gamma \right\vert ^{2}\left[ 2\theta -\left( \pi /2\right) \right] }
\left\vert \gamma \right\rangle_1 \left\vert \gamma e^{-i\pi /2}\right\rangle_2,
\]
where $\ket{\gamma}$ denotes a coherent state,
$\gamma =\left\vert \gamma \right\vert e^{i\theta }$
and $\left\vert \gamma \right\vert =\sqrt{N-\left(D/2\right)}$.
Application of a phase shift of $\pi/2$ to the second mode and combination of the two modes at a $50:50$ beam splitter yields a NOON state.
Under the phase shift the relational cat state is transformed as,
\be
\left\vert \psi _{r.s.c.}\right\rangle
\longrightarrow
\int_{0}^{2\pi }\frac{d\theta }{2\pi }e^{-i\left\vert \gamma \right\vert ^{2}\left[ 2\theta +\left( \pi /2\right) \right] }
\left\vert \gamma \right\rangle_1 \left\vert -\gamma \right\rangle_2
+\int_{0}^{2\pi }\frac{d\theta }{2\pi }e^{-i\left\vert \gamma \right\vert ^{2}\left[ 2\theta -\left( \pi /2\right) \right] }
\left\vert \gamma \right\rangle_1 \left\vert \gamma \right\rangle_2, \nonumber
\ee
and the beam splitter transforms the state to,
\bea
&&
\int_{0}^{2\pi }\frac{d\theta }{2\pi }e^{-i\left\vert \gamma \right\vert ^{2}\left[ 2\theta +\left( \pi /2\right) \right] }
\left\vert 0\right\rangle_1 \left\vert 2\gamma \right\rangle_2 +\int_{0}^{2\pi }\frac{d\theta }{2\pi }
e^{-i\left\vert \gamma \right\vert ^{2}\left[ 2\theta -\left( \pi /2\right) \right] }
\left\vert 2\gamma \right\rangle_1 \left\vert 0\right\rangle_2 \nonumber \\
\! &\propto& \!
\frac{\left\vert 0\right\rangle_1 \left\vert 2N\!-\!D\right\rangle_2 \,+\, \exp\left(i|\gamma|^2 \pi\right)\left\vert 2N\!-\!D\right\rangle_1
\left\vert 0\right\rangle_2 }{\sqrt{2}}\,,
\eea
where the last expression is a superposition of states with $2N-D$ photons in a single mode as required.
In practice it would be necessary to
limit the number of detections in the interference procedure
as much as possible, leading to components with visibility significantly less than $1$,
and use the states generated when the counts at the left and right detectors
are similar but not identical (when the difference of the localised relative phases is different from $\pi$)
so as to increase the yield.
The resulting states would be expected to
have a NOON state as a large amplitude component.  Future work could clarify the
usefulness of such approximate NOON states and optimise the suggested procedure.

\section{Relative optical phases and tests of superselection rules}
\label{sec:OPT2superselection}

A conservation law makes operational sense (or nonsense) only when related to the
procedures whereby the conserved physical quantities are measured. In particular, the
frame of reference against which the measurements are made plays a crucial role. Certain
frames of reference (e.g. position, atom number) are more in accord with everyday
experience than others (e.g. ``charge phase'', ``isospin phase''). This has perhaps more
to do with the ground state of the universe (the electromagnetic vacuum in particular)
which acts as a readily accessible reference frame,  than with any fundamental physical
restrictions.

A belief in absolute conservation laws leads to a belief in absolute superselection rules.
For example, in algebraic quantum field theory the existence of absolute conservation laws and
associated superselection rules, forbidding the creation of superpositions of states with
different values of the conserved quantities, is taken to be true axiomatically \cite{Wick52,Haag96}
(see also the discussion in section 2.7 of \cite{Weinberg95}).
To illustrate how the more relational approach works consider quantum
optics under the ``rotating wave approximation'' (refer for example \cite{Gerry04}),
taken to be equivalent to a strict superselection rule for energy under which the energy is additively
conserved, and in the absence of any absolute phase reference.
Under these assumptions superpositions of states of
different photon number (such as Glauber coherent states of light) are forbidden, as
are superpositions of non-degenerate atomic states ($|g\>,|e\>$).
Interaction Hamiltonians are explicitly
excitation-conserving so, for example, the familiar Jaynes-Cummings Hamiltonian
$\hat{H}=|e\>\!\<g|\otimes a+|g\>\!\<e|\otimes a^\dagger$ is allowed (where $a$ and $a^\dagger$ are
annihilation and creation operators for an optical mode).  For a discussion of Jaynes-Cummings dynamics
see \cite{Gerry04}.

If asked how to operationally create and verify the existence of a superposition
of atomic states of the form $|g\>+|e\>$, a simple response would be to send the atom,
initially in the ground state $|g\>$, through a cavity containing light in a large
amplitude coherent state\footnote{Sufficiently large to drive Rabi oscillations between $\ket{g}$ and $\ket{e}$.}
for an appropriate length of time,
where the interaction Hamiltonian is $\hat{H}$ above. Measurement of the atom after exiting the
cavity yields it in the ground state half the time. However this could be due to the
atom being in a mixed state.  To verify that a coherent superposition has been
obtained the atom could be sent through a second cavity, also in a optical coherent state with a large amplitude,
after which it would be found that the atom is always in the excited state $|e\>$ for example. This
demonstrates that the atom was indeed in the coherent superposition $|g\>+|e\>$ between the cavities.
Note that this notion of a superposition assumes that the phase reference provided
by the state of the first cavity is maintained.

What Aharonov and Susskind (AS) note in \cite{Aharonov67} is that the two cavities need not,
in fact, be prepared in coherent states. A state of the form \eq{OptOnepsiinfinity} discussed in
\sect{OptOneFockstates} of Chapter \ref{chap:OpticalOne} is
operationally just as good as initial coherent states for the purposes of demonstrating
a coherent superposition of atomic states using the procedure described above.\footnote{AS actually
considered the creation of a superposition of a proton (equivalent to $|g\>$) and a neutron
(equivalent to $|e\>$) using coherent states --- or otherwise --- of negatively charged
mesons (equivalent to photons), under a Hamiltonian identical to the Jaynes-Cummings one.}
As noted previously, this state has fixed total
energy, and thus there would be no violation of the conservation law globally.
An objection to this test having physical relevance can be made along the lines
that states of the form \eq{OptOnepsiinfinity} are not easy to come
by in nature, and in fact are highly entangled.  However the results of
\sect{OptOnePoissonian} and \sect{OptOnethermal} of Chapter \ref{chap:OpticalOne} demonstrate
that a mixed state of the form \eq{OptOnerhoinfinity} would do
just as well for the operational demonstration of coherent superposition envisaged by AS.
Such mixed states are much more easily preparable, and would seem to conform more closely
with the type of reference frame states that observers typically prepare.

There are some interesting ways this topic could be extended in future work.
One assumption made by AS in their calculations is that the cavity modes
are prepared with large intensities.  It would be of interest to check how effective
a phase locked state with only a small total number of photons would be.
Also it would be useful to check what happens when the interaction Hamiltonian is taken
to be $\hat{H}=(a+a^\dagger)(\ket{g}\!\bra{e}+\ket{e}\!\bra{g})$, which is derived
from the electric-dipole approximation without also making the rotating wave
approximation.  Perhaps the
easiest way to create the phase locked state \eq{OptOnepsiinfinity} is by splitting
the light from a single laser at a beam splitter, rather than by phase locking two
independent light sources.  However the process suggested by AS to demonstrate coherent superposition
could also be used to investigate features of the (partially) localised relative optical phase
induced in states prepared using the canonical interference procedure of \sect{OptOneFockstates}
of Chapter \ref{chap:OpticalOne}.

\section{Possible extensions}
\label{sec:OPT2opticalextensions}

It is clear that the programme of investigation in Chapters \ref{chap:OpticalOne} and \ref{chap:OpticalTwo}
can be extended to address further questions about localising relative optical phases.
For example, with reference to the canonical interference procedure,
further calculations could be done for a setup for which the ancilla modes are
combined at an unbalanced beam splitter, for which the reflection and transmission coefficients have different magnitudes.
The indistinguishability that the beam splitter provides is central to the process of localisation, and it might
be expected here that the localisation is slower, or only partial even after very many detections.
It would be interesting to clarify how the process localisation differs for the two cases of
asymmetric initial states (with different intensities in each mode) together with a $50:50$ beam splitter,
and symmetric initial states together with an unbalanced beam splitter.
Another question asks what happens when there is uncertainty in the path lengths of the apparatus during the interference procedure.
In this case the Kraus operators corresponding to photon detection take the modified form $\hat{a}\pm \left[\rm exp\left(i\hat{x}\right)\right]\hat{b}$ (where $\hat{a}$ and $\hat{b}$
are the annihilation operators for the two modes, and $\hat{x}$ is a position operator for the path length),
and the process of localisation acts on the sum of the path length and the relative optical phase.

%% file: LRDoF_v2_chapter_4.tex
\chapter{Interfering Independently Prepared Bose - Einstein Condensates and Localisation of the Relative Atomic Phase}
\label{chap:BEC}

This chapter looks at the interference of two Bose-Einstein condensates (BEC's) which have been prepared separately
and therefore have no prior phase correlation, a process in which localisation of the relative atomic phase plays an important role.
The focus is on spatial interference experiments in which the condensates are released from their traps and are imaged while falling,
as they expand and overlap.  High contrast interference patterns have been observed experimentally \cite{Andrews97,Hadzibabic04}.
Javanainen and Yoo were the first to provide a detailed numerical demonstration that such interference can be predicted
without assuming that each condensate has an a priori phase \cite{Javanainen96}.  Several studies have been done by other authors,
and \cite{Cirac96, Naraschewski96, Yoo97, Horak99} should be highlighted, though little was done analytically
(an exception being \cite{Castin97}).

This chapter presents and extends work published in \cite{Cable05}, investigating the key features of the localisation
of the relative atomic phase as a pattern of spatial interference is built up by a series of atomic detections.
The analysis has close connections with the study of localising relative optical phase in Chapter \ref{chap:OpticalOne}.
Most surprisingly it is shown that the localisation of the relative atomic phase takes place on the same rapid time scale
as in the optical example, and much faster than the emergence of a well-defined pattern of spatial interference
in the numerical simulations of Javanainen and Yoo and others.

The chapter proceeds as follows.  The conceptual difficulties associated with assigning a
BEC a coherent state macroscopic wavefunction with a definite a priori phase are reviewed in
\sect{BECstandardstory}.  The spatial interference of independently prepared BEC's is analysed in depth in \sect{BECspatialinterference}.
A simple measurement model is presented in \sect{BECdetectionmodelvisiblity} based on the approach adopted in \cite{Javanainen96}.
The visibility defined in \sect{OptOneQuantifyingLocalisation} of Chapter \ref{chap:OpticalOne},
which can be used to quantity the degree of localisation of the relative phase,
is found to have a simple interpretation here in terms of the probability density for detection of an atom at a particular position.
The case of initial Poissonian states is treated analytically in \sect{BECanalyticalPoissonian},
and the analytical predictions are compared with simulated results in \sect{BECnumerical}.
\sect{BECSomeOpenQuestions} discusses the issue of what information is available in principle about
the localisation of the relative atomic phase from a given pattern of spatial interference.  Finally
in \sect{BECOpticalRelPhaseMeasurement} a summary is given of a recent experiment and several different
theoretical proposals concerning methods which probe the relative atomic phase by optical means,
thereby performing a non-destructive measurement.

\section{The standard story and conservation of atom number}
\label{sec:BECstandardstory}

A common and useful description of BEC's makes use of a coherent state macroscopic
wavefunction for the condensates. Such a description is generally justified by invoking
standard stories about symmetry breaking (see for example Chapter 2 of \cite{PitaevskiiStringari}
and \cite{Javanainen86}).
However there are several reasons to be suspicious of this common treatment. The first is that it
requires a description of the state of a BEC in terms of a coherent superposition of states with different atom
numbers. If atom number is to be conserved then such a description is tricky to justify.
It is often argued that the BEC is surrounded by a thermal cloud with
which it is exchanging atoms, and thus the atom number is undetermined. However such a
process leads only to a mixed state for the BEC and not a pure coherent state.
Secondly, the symmetry breaking is generally invoked by the addition of auxiliary fields
with no clear physical relevance (see for example \cite{Naraschewski96} for a description and criticism
of an approach based on Bogoliubov auxiliary fields).  An assumption of an a priori symmetry breaking
also implies the philosophically precarious position of
writing down quantum mechanical states containing in principle unknowable parameters -
the absolute atomic phases\footnote{the absolute phase for a single mode is, roughly speaking,
the phase difference between the amplitudes for some number $n$ and $n+1$ particles.}.
Finally the most striking demonstrations of coherence
for BEC's come from interference experiments such as reported in
\cite{Andrews97,Hadzibabic04}, but such experiments do not require a
description in terms of atom-number-violating coherent states, as will be discussed below.

As an aside, there is a difference here with experiments in which a single
condensate is coherently ``cut'' into two parts and then allowed to re-interfere.
Such interference is obtainable without the use of coherent states.  As an optical analogue,
sending a photon Fock state --- or
even a thermal state --- through a Mach-Zehnder interferometer demonstrates perfect
interference!  Experiments in this category include \cite{Bloch00},
where interference was demonstrated between matter waves originating from spatially
separated parts of one trapped condensate, and \cite{Hall98} where the process of
splitting and recombining the condensate was done using two-photon pulses, with the two
condensates populating degenerate spin states.
Only the case that the BEC's are prepared separately is of interest in this chapter.

It is simple to imagine an experiment involving two BEC's that closely follows the
optical scenario described in Chapter \ref{chap:OpticalOne} for photons in cavities.
For instance, two condensates trapped in separate
potential wells may be allowed to slowly tunnel through a barrier. Atoms originating from
different wells can be rendered indistinguishable by mixing at an appropriate beam
splitter. While a standard description of the experiment would utilize interference
between coherent condensate fields $|\psi_1|e^{i\theta}$ and $|\psi_2|e^{i\phi}$, the
discussion of the Chapter \ref{chap:OpticalOne} can be carried over to conclude that such a
description is not necessary. In fact it is less desirable --- it violates atom number
conservation and invokes the use of the independent and unknowable phases
$\theta$ and $\phi$, which vary from run to run of the experiment and should therefore be
incorporated in a quantum mechanical framework by the use of mixed
states (leading to a description as in \sect{OptOnePoissonian} of Chapter \ref{chap:OpticalOne}).
In practice the most striking BEC interference patterns are not those which involve
leaking of single atoms onto a beam splitter and detection in one of only two channels,
but rather those in which spatial diffraction of the initially independent BEC's
occurs, and a spatial interference pattern is measured in the region of overlap.  In
what follows the discussion is extended to this type of experiment.

\section{Spatial interference of independently prepared Bose-Einstein condensates}
\label{sec:BECspatialinterference}

\subsection{Modelling the atomic detection process and quantifying the visibility of the interference patterns}
\label{sec:BECdetectionmodelvisiblity}

\begin{figure}[h]
\begin{center}
\includegraphics[height=7cm]{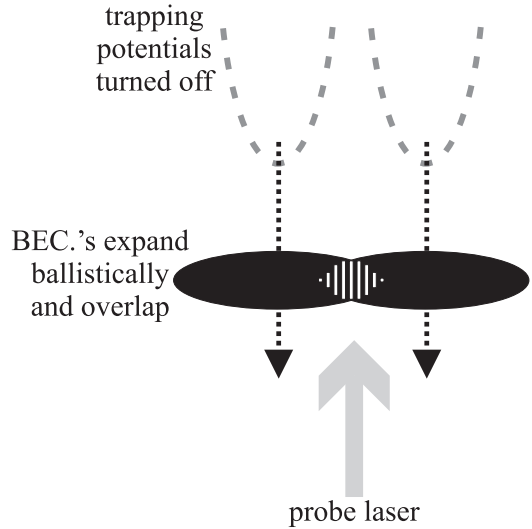}
\caption{\label{fig:BECspatialinterferencepic} A simplified illustration of a spatial
interference experiment for two initially uncorrelated Bose-Einstein condensates.  The condensates
are released from their traps, expand ballistically, and overlap.  The atomic density distribution
is imaged optically.  Interference fringes are observed in the region of overlap.}
\end{center}
\end{figure}

A simplified illustration of the class of experiment considered in this section is given in
\figu{BECspatialinterferencepic}.  Two condensates are prepared separately and are trapped
close together.  They are then released simultaneously from their traps and, as they fall under gravity,
the condensates expand ballistically and overlap.  A probe beam is shone on the region of overlap and the
atomic density profile is imaged.  High-contrast interference patterns have been observed in
experiments performed at MIT \cite{Andrews97} and at the ENS \cite{Hadzibabic04}.  The latter
involved the interference of a total of $30$ condensates, which were prepared in a one dimensional
optical lattice, and were released and imaged together.  Various steps were taken in this experiment
to ensure that the condensates did not have any prior phase correlation:
the lattice period was long and the potential barrier high
so as to eliminate tunneling between sites; in one version of the experiment the lattice was ramped up
during the cooling process so that the condensates were formed independently; and in another
the condensates were kept before the start for a ``holding time'' very much longer
than the observed decorrelation time.

It is not necessary to use a coherent state description to predict interference
between independently prepared BEC's, as was first shown in detail by
Javanainen and Yoo \cite{Javanainen96, Yoo97}.  They showed that interference patterns emerge
even if the atom number superselection rule is obeyed exactly, and the condensates are initially in
atomic Fock states with the same number of atoms.
The analysis which follows is based on the same simple model employed by Javanainen and Yoo,
but considers the case that the initial state of each condensate is mixed.
It is assumed that phase diffusion, the shape of the trapping potential, and edge effects can be ignored.
The condensates correspond to macroscopic occupation of
one-particle plane-wave states with opposite momenta $\pm k$.
They can therefore be described by second quantised fields
of the form $e^{\pm ikx} b_{\pm k}$, where the $b_{\pm k}$ are annihilation operators for each mode.
The initial states of the condensates are taken to be Poissonian, consistent with a superselection rule for atom number.
This choice of initial state is suggested by a maximum-entropy principle which can be applied here,
given the situation of continuous exchange of atoms between condensed atoms
and a surrounding thermal cloud where only the average size of the condensate is easily determined \cite{Jaynes03}.
Both condensates are assumed to have the same expected atom number $\bar{N}$.
To generate the pattern of spatial interference,
the condensates are assumed to merge over a linear array of atom detectors, with
the atoms detected one at a time.  The combined field operator, $\hat{\psi}=e^{ikx_j}b_k+e^{-ikx_j}b_{-k}$,
serves as the measurement operator for a detection at position $x_j$.

The situation here turns out to be dynamically equivalent to the optical problem discussed in Chapter
\ref{chap:OpticalOne} when the cavity modes are initially Poissonianly populated and the second cavity
undergoes random phase shifts between detections, for example
because of a frequency mismatch (see \sect{OptOnesymmetries}, third paragraph).
Inspecting the atomic measurement operator $\hat{\psi}$, it is seen
that atomic measurements $\pi/k$ apart are equivalent, and further,
that a detection in $\frac{\pi}{2k} \leq x_1 < \frac{\pi}{k}$ is equivalent to
one in $0 \leq x_1 < \frac{\pi}{2k}$ at $x_1-\frac{\pi}{2k}$ with operator $\hat{\psi}=e^{ikx_1}b_k-e^{-ikx_1}b_{-k}$.
For a mixed state
\[
\rho \propto \int d^2 \alpha d^2 \beta \, P(\alpha,\beta) \, | \alpha \>\!\< \alpha |_k \otimes | \beta \>\!\< \beta |_{-\!k},
\]
the probability density for measurement at $x_j$ with this periodic identification is proportional to
\begin{eqnarray*}
&\rm{tr}& \!\!\!\! \Big\{ \left( e^{ikx_j}b_k + e^{-ikx_j}b_{-k} \right) \rho \left( e^{-ikx_j}b^{\dagger}_k + e^{+ikx_j}b^{\dagger}_{-k} \right) \\
&& \,\, + \left( e^{ikx_j}b_k - e^{-ikx_j}b_{-k} \right) \rho \left( e^{-ikx_j}b^{\dagger}_k - e^{+ikx_j}b^{\dagger}_{-k} \right) \Big\} \\
&\propto& \!\!\!\!
\int d^2 \alpha d^2 \beta \, P(\alpha,\beta) \, (|\alpha|^2+|\beta|^2) \,.
\end{eqnarray*}
On this reduced range every $x_j$ is equally probable and the problem can be treated by assuming ``left''
and ``right'' Kraus operators, $K_{r,\tau} \propto b_k +e^{i\tau}b_{-k}$ and $K_{l,\tau} \propto b_k-e^{i\tau}b_{-k}$ with $\tau$ taking a
random value for each measurement.  In fact this equivalence between the optical
and BEC cases holds regardless of the states involved.

To formulate a measurement process which can be consistently interpreted in terms of probabilities it
is necessary to look at the normalisation of the measurement operators $\hat{\psi}(x)$.
For a general field operator $\varphi$, the operator $\varphi^{\dagger}(x)\varphi(x)$ is commonly
interpreted as the particle density at the position $x$.  The periodicity here of $\psi^{\dagger}(x)\psi(x)$
suggests restricting $x$ to a length of $\pi/k$.
It is easily seen that
$\int_0^{\pi/k} \hat{\psi}^{\dagger}(x)\hat{\psi}(x)
=\left(\frac{\pi}{k}\right)\left(\hat{N}_k+\hat{N}_{-k}\right)$
where $\hat{N}_{\pm k}$ are the number operators for the condensates with momentum $\pm k$.
Hence the operators,
\be
\label{eqn:normalisedmeasurementoperators}
\hat{K}(x):
\rho \rightarrow
\sqrt{\frac{k}{\pi}}
\left(e^{ikx}b_k+e^{-ikx}b_{-k}\right)\rho
\ee
define a complete set of measurement operators (for more on POVM's and quantum operations refer
for example sections 2.2.6 and 8.2 of \cite{NielsenChuang00}).
To allow a consistent probability interpretation account must be taken of the expected total
atom number for $\rho$, giving a final state
$\left\{ 1/{\rm tr} \! \left[ ( \hat{N}_k + \hat{N}_{-k}) \rho \right] \right\} \hat{K}(x) \rho \hat{K}^\dagger(x)$,
which is subnormalised with trace equal to the probability of the measurement outcome.
The action of the measurement procedure on the vacuum state with no atoms populating
either condensate mode is specified as $0$ to resolve the ambiguity in this case i.e. the probability is $0$ for
detecting an atom anywhere in this case.  Defining
\be
f(x_1,x_2,\cdots,x_r) = \mathcal{N}\,K(x_r) \cdots K(x_1) \rho K^{\dagger}(x_1) \cdots K^{\dagger}(x_r)
\ee
for a sequence of measurements at positions $x_1,...,x_r$,
where $\mathcal N$ is the normalisation factor correcting for the changing expected total atom number,
it is easy to check that
$\int_0^{\pi/k} f(x_1) dx_1 = 1$ when $r=1$.  For general $r$,
$\int_0^{\pi/k} f(x_1,\cdots,x_{r-1},x_r) dx_r = f(x_1,\cdots,x_{r-1})$, and hence $f(x_1,\cdots,x_r)$
is a well-defined probability density.

Consider now the effect of the measurement operator $\hat{K}{\left(x\right)}$ on an arbitrary state  $\rho$
for the two condensates (but not the vacuum!)
--- for example the state after a series of earlier atomic detections at different positions.
Then the probability for detection of an atomic at a position $x$ is given by,
\bea
P(x)
&=&
{\rm tr}\left\{ \left[ b_{k}^{\dagger }b_{k}+e^{i2kx}b_{-k}^{\dagger }b_{k}+e^{-i2kx}b_{k}^{\dagger }b_{-k}+b_{-k}^{\dagger }b_{-k}\right] \rho \right\} /{\rm tr}\left[ \left( b_{k}^{\dagger }b_{k}+b_{-k}^{\dagger }b_{-k}\right) \rho \right] \nonumber \\
&=&
1+\left[ e^{i2kx}tr\left( b_{k}\rho b_{-k}^{\dagger }\right) +e^{-i2kx}tr\left( b_{-k}\rho b_{k}^{\dagger }\right) \right] /{\rm tr}\left[ \left( b_{k}^{\dagger }b_{k}+b_{-k}^{\dagger }b_{-k}\right) \rho \right]. \nonumber
\eea
${\rm tr}\left( b_{k}\rho b_{-k}^{\dagger} \right)$ is a complex number and ${\rm tr}\left( b_{-k}\rho b_{k}^{\dagger}\right)$
its conjugate.  Defining
\be
\label{eqn:Vdefn}
{\rm tr}\left( b_{k}\rho b_{-k}^{\dagger} \right)
=
{\rm tr}\left[ \left( b_{k}^{\dagger }b_{k}+b_{-k}^{\dagger }b_{-k}\right) \rho \right]
\frac{V}{2}
\exp^{-i\varphi},
\ee
it follows that the probability density takes the simple form,
\be
\label{eqn:InterferenceProbDensity}
P(x) = 1+V \cos{\left( 2kx - \varphi \right)},
\ee
where $V$ is a positive number between $0$ and $1$, a fact which is guaranteed
by the fact that the density operator $\psi(x)^{\dagger}\psi(x)$ is positive.
The result \eq{InterferenceProbDensity}  was identified in \cite{Javanainen96, Yoo97}
but only for the case
when the initial states of the condensates are number states with the same number of atoms.

The form of \eq{InterferenceProbDensity} suggests $V$ as a measure of the degree of localisation of the relative atomic phase parameter of two condensates.  On the other hand, in \sect{OptOneQuantifyingLocalisation} of Chapter \ref{chap:OpticalOne}
a visibility was defined to quantify the localisation of the relative phase
for two modes as illustrated in \figu{OptOnevisibilitydef}, and this definition can be translated to the context
of atomic optics.  An experiment can be imagined in which one of the condensates is subject to a controlled phase shift $\tau$
and coherently combined with the other by a Josephson coupling in analogy with an optical $50:50$ beam splitter.
The atom number in one well could then be counted to provide an estimate of the intensity $I(\tau)$ defined in \sect{OptOneQuantifyingLocalisation}.
In principle the experiment would have to be repeated many times with identical preparations of the condensates to provide an accurate estimate
of $I(\tau)$, and repeated again for different values of $\tau$ so as to evaluate the visibility according to equation
\eq{OptOneVisFromInt}.  However there is another interpretation.  It is straightforward to show mathematically
that,
\be
I(\tau)=
{\rm tr}
\left[
\frac{1}{\sqrt{2}} \left( b^{\dagger}_{k}+e^{-i\tau}b^{\dagger}_{-k} \right)
\frac{1}{\sqrt{2}} \left( b_{k}+e^{i\tau}b_{-k} \right) \rho
\right].
\ee
Therefore $I(\tau)$ is proportional to the probability density $P\left(-\frac{\tau}{2k}\right)$
for detecting a particle at $x=-\frac{\tau}{2k}$.
Inspecting \eq{InterferenceProbDensity}, $V$ is seen to be the same as the visibility.

It is also worth noting the connection between the visibility and the method
used by M{\o}lmer for quantifying the degree of localisation of the relative optical phase
in his studies of the interference of independent light sources \cite{Molmer97,MolmerB97}.
The quantity considered by M{\o}lmer is
$\eta=\lambda 2 \Re \left\{ {\rm tr} \left[ a^{\dagger} b \, \rho \left( t \right) \right] \right\}$,
proportional to the real part of the trace quantity,
where $t$ is time,
$a$ and $b$ are annihilators for the optical modes, $\rho$ is the state of the cavities, and $\lambda$
is a normalisation factor scaling $\eta$ within $\pm 1$.  M{\o}lmer restricted
his attention to the case of initial number states, in which case $\lambda$ is the reciprocal of
the number of remaining photons, and he normalised the state of the two optical modes.
M{\o}lmer assumed a frequency difference between optical cavities, giving rise
to a rotating phase factor in the governing Hamiltonian
--- $\hbar  \left(\omega_b-\omega_a\right) b^{\dagger} b$ ---
which causes $\eta$ to oscillate.  When this oscillation is the principle evolution,
$\eta \simeq 2 \Re \left[ \frac{V}{2} e^{i\left(\varphi+(\omega_b-\omega_a) \tau\right)} \right]
= V \cos \left[\varphi+(\omega_b-\omega_a) \tau \right]$,
and the visibility is seen to govern the magnitude of oscillation of $\eta$.

\subsection{Analytical treatment for Poissonian initial states}
\label{sec:BECanalyticalPoissonian}

This section looks at how the localisation of the relative atomic phase of two condensates induced by a
series of measurements by operators defined in \eq{normalisedmeasurementoperators} can be treated analytically.
It is supposed that the condensates are initially in Poissonian states with the same mean number $\bar{N}$.
The problem can be treated in terms of ``left'' and ``right'' Kraus operators
$K_{l,\tau}\propto b_k-e^{i\tau}b_{-k}$ and $K_{r,\tau}=b_k+e^{i\tau}b_{-k}$.
To understand the characteristic localisation it is sufficient to
take half the detections at $\tau=0$ and the rest at $\tau=\pi/2$, the largest difference possible.
There is little advantage working as in \cite{Molmer97,MolmerB97,Castin97} with a probability density
for the full measurement record involving information about the precise spatial distribution of the atomic detections.
The commutativity of $K_{l,\tau}$ and $K_{r,\tau}$ allows the process to broken down as convenient. It is assumed that
there are $M$ measurements at each of $\tau=0$ and $\tau=\pi/2$. The numbers of ``left'' and ``right'' counts
are denoted by $l_1,r_1$ at $\tau=0$, and $l_2,r_2$ at $\tau=\pi/2$.

The $2M$ measurements cause the initial state with average atom number $\bar{N}$ for each condensate
\[
\rho_I = \int \tfrac{d\theta d\phi}{4\pi^2} | \alpha \>\< \alpha |_k \otimes | \beta \> \< \beta |_{-\!k},
\]
where $\alpha=\sqrt{\bar{N}}e^{i\theta}$ and $\beta=\sqrt{\bar{N}}e^{i\phi}$,
to evolve as
\begin{eqnarray*}
\rho_I \!\!\! &\rightarrow& \!\!\!
\frac{M!}{r_{2}!l_{2}!}\frac{M!}{r_{1}!l_{1}!}\hat{K}_{r,\frac{\pi}{2}}^{\, r_{2}}\hat{K}_{l,\frac{\pi }{2}}^{\, l_{2}}\hat{K}_{r,0}^{\, r_{1}}\hat{K}_{l,0}^{\, l_{1}}
\,\rho_I\,
\hat{K}_{l,0}^{\, l_{1}\dagger}\hat{K}_{r,0}^{\, r_{1}\dagger}
\hat{K}_{l,\frac{\pi }{2}}^{\, l_{2}\dagger}\hat{K}_{r,\frac{\pi }{2}}^{\, r_{2}\dagger} \\
\!\!\! &=& \!\!\!
\int \!\! \tfrac{d\theta d\phi}{4\pi^2} \;
|C_{l_{1},r_{1}}^{\tau=0}(\theta ,\phi ) \,
C_{l_{2},r_{2}}^{\tau=\! \frac{\pi}{2}}(\theta ,\phi )|^{2}
\left\vert \alpha \right\rangle \!\! \left\langle \alpha \right\vert_{k}
\!\! \otimes \! \left\vert \beta \right\rangle \!\! \left\langle \beta \right\vert_{-\!k},
\end{eqnarray*}
where
$\left|C_{l,r}^{\tau }(\theta ,\phi )\right|^{2}
\!=
\frac{ \left( r\!+ l \right) !}{r!l!} \,
\left|\cos \left(\! \tfrac{\Delta-\,\tau}{2} \!\right) \! \right|^{2r} \, \left|\sin \left(\! \tfrac{\Delta-\,\tau }{2} \!\right) \! \right|^{2l}\!$
and $\Delta \! \equiv \! \phi \!-\! \theta$. The peaked function $C_{l,r}$ is familiar from \sect{OptOneFockstates}
of Chapter \ref{chap:OpticalOne}, and the phase shift $\tau$ causes a translation.

The general features of the localisation are as in the optical analysis in \sect{OptOnePoissonian}
of Chapter \ref{chap:OpticalOne}.
However the effect of the phase shift is to ensure with high probability that exactly one value $\Delta_0$ for
the relative phase is picked out. This is so even when $M$ is small.
The phenomenon can be understood by careful inspection of the measurement record and with the aid of the asymptotic
expressions for $C_{l,r}(\theta,\phi)$, \eq{OptOnebothportsasympt} and \eq{OptOneoneportasympt} of Chapter \ref{chap:OpticalOne}.
Consider now the cases of $M=3$, $8$ and $15$, looking at the ``likely events'' --- defined as those with
probability greater than a equal fraction $1/(M+1)^2$. The probabilities of these events
total $0.9$, $0.8$ and $0.8$ respectively. In every case a unique value of $\Delta_0$ is picked out.
In very many cases --- all when $M=3$ and half when $M=15$ --- all the detections are of the same sort, all
$K_l$ (or all $K_r$), at $\tau=0$ or $\tau=\frac{\pi}{2}$, or both.
In other words at least one component of
$\left|C_{l_{1},r_{1}}^{\tau=0}(\theta ,\phi ) \, C_{l_{2},r_{2}}^{\tau=\! \frac{\pi}{2}}(\theta ,\phi )\right|^{2}$
is of the form $\left|C_{M,0}\right|^2$ (or $\left|C_{0,M}\right|^2$) which has only one peak and a larger spread than
otherwise. The product $\left|C_{l_{1},r_{1}}^{\tau=0}(\theta ,\phi ) \, C_{l_{2},r_{2}}^{\tau=\! \frac{\pi}{2}}(\theta ,\phi )\right|^{2}$
in turn has only one peak and is highly probable. Other probable events are such that
one peak of $\left|C_{l_{1},r_{1}}^{\tau=0}(\theta ,\phi )\right|^2$ strongly overlaps with one peak of
$\left|C_{l_{2},r_{2}}^{\tau=\frac{\pi}{2}}(\theta ,\phi )\right|^2$.
In short, the phase shift of $\pi/2$ makes it impossible for $\left|C_{l_{1},r_{1}}^{\tau=0}(\theta ,\phi )\right|^2$ and
$\left|C_{l_{2},r_{2}}^{\tau=\frac{\pi}{2}}(\theta ,\phi )\right|^2$
to strongly reinforce each other at more than one value for the relative atomic phase.

In the limit of a large number of detections,
\be
\rho_\infty =  \int\!\!\!d\theta \;
|\alpha\>\!\<\alpha|_k \! \otimes |\alpha e^{i\Delta_0}\>\!\<\alpha e^{i\Delta_0}|_{-\!k}.
\ee
When the relative phase is perfectly defined the atomic detections have a probability density
of $\cos^2 (kx_1 -\Delta_0/2)$ (where $x_1$ is the proper position). However, beginning from
initial states with no relative phase correlation, \emph{the value for the relative phase localises
much faster than it takes for the characteristic spatial interference pattern to become well established}
(a point also made in \cite{Horak99} for the case of initial number states).
The numerical studies of Javanainen and Yoo, for example, simulate interference patterns based on
$1000$ atomic measurements. However,
the dependence on the total number of detections $l+r$ in \eq{OptOnebothportsasympt} and \eq{OptOneoneportasympt}
of Chapter \ref{chap:OpticalOne} demonstrates that the underlying rate of localisation here
is similar to that at either of the two values of $\Delta_0$ which evolve when the phase
$\tau$ is fixed.

Beyond the very first few measurements,
the scalar function $\vert C_{l_{1},r_{1}}^{\tau=0} C_{l_{2},r_{2}}^{\tau=\! \frac{\pi}{2}} \vert  ^2$
is well estimated by a Gaussian with width between $\sqrt{2}/\sqrt{M}$ and $2/\sqrt{M}$.
The state of the condensates then takes the form,
\bea
\label{eqn:Gaussianstate}
\rho _{G}&=&\int_0^{2\pi} \!\! \int_0^{2\pi}
\frac{d\theta}{2\pi} \frac{d\phi}{2\pi}
\,\, 2 \pi P(\Delta) \,\, + \,\,
\left( \text{contributions for }2\pi \text{-periodicity in} \, \Delta \right) \nonumber \\
&& \times \,\,
\left\vert \sqrt{\bar{N}}e^{i\theta }\right\rangle \!\! \left\langle \sqrt{\bar{N}}e^{i\theta }\right\vert_k
\,\, \otimes \,\, \left\vert \sqrt{\bar{N}}e^{i\phi }\right\rangle \!\! \left\langle \sqrt{\bar{N}}e^{i\phi }\right\vert_{-k},
\eea
where $\Delta \equiv \phi-\theta$ and,
\be
P(\Delta )=\frac{1}{\sigma \sqrt{2\pi }}e^{-\left( \Delta -\Delta _{0}\right) ^{2}/\left( 2\sigma ^{2}\right) } \nonumber
\ee
is a normal distribution with mean $\Delta_0$ and variance $\sigma^2$.
The Gaussian width $2\sigma$ is small compared to $2\pi$ (the range of $\Delta$), and
\bea
\label{eqn:variancerange}
\sqrt{\frac{1}{2M}} \leq \sigma \leq \sqrt{\frac{1}{M}},
\eea
where a total of $2M$ measurements have been performed.
The visibility for a state of the form \eq{Gaussianstate} is derived in appendix \ref{chap:appendixB} and,
\be
\label{eqn:BECexpPredictedVsbly}
V = e^{-\frac{1}{2}\sigma ^{2}}.
\ee

\subsection{Numerical simulations for Poissonian initial states}
\label{sec:BECnumerical}

This section discusses how the emergence of a pattern of spatial interference can be simulated numerically,
extending the method described in \cite{Javanainen96,Yoo97} to general initial states, and looking specifically
at the case of initial Poissonian states considered in the previous section.

The approach used by Javanainen, Yoo and Ruostekoski to simulate the interference between
condensates initially in number states with the same number $N$ of atoms
makes key use of the simple form of the probability density for atomic detection in their measurement model
--- $P(x)=1+V\cos(2kx-\varphi)$ --- and is as follows.  $V$ is initially $0$.  $V$ and $\varphi$ are updated after
each atomic detection, and a simple method is used to generate the subsequent detection according to $P(x)$,
assuming a random number generator which supplies a number $u$ on $[0,1]$ according to the uniform probability density.
It is easy to show in general that given
a random variable $X$, with a probability density function $g(x')$ and a cumulative probability function
$G(x)=\int^x_{x_0} g(x')dx'$, the random variable $G(X)$ is uniformly distributed.
Hence to generate the next detection it is sufficient to solve the following equation in $x$:
\be
G(x)=\int_{0}^{x}1+V\cos \left( 2kx^{\prime}-\varphi \right)dx^{\prime}
=x+\frac{V}{2k}\left[ \sin \left( 2kx-\varphi \right)+\sin \left( \varphi \right) \right]=u,
\ee
which is easy to do numerically in practice.
To update $V$ and $\varphi$ after this detection it is sufficient to evaluate $P(x)$ at two points
by evaluating the full expression for $P(x)$ in terms of the positions of all the previous detections and $N$.
Since the simple form of $P(x)$ in \eq{InterferenceProbDensity} applies for arbitrary states,
the same method of stochastic simulation can also be applied to initial states other than number states.

In what follows the case that the BEC's are initially in Poissonian states
is considered in detail.  After measurements at positions $x_1,\cdots,x_r$ the initial state
\be
\rho _{I}=\int_{0}^{2\pi }\!\!\!\! \int_{0}^{2\pi }\frac{d\theta }{2\pi }\frac{d\phi }{2\pi }
\left\vert \alpha \right\rangle \!\! \left\langle \alpha \right\vert \otimes \left\vert \beta \right\rangle \!\! \left\langle \beta \right\vert,
\ee
where $\alpha =\sqrt{\bar{N}}e^{i\theta },\beta =\sqrt{\bar{M}}e^{i\phi }$
and, $\bar{N}$ and $\bar{M}$ are the initial expected atom numbers for each condensate,
evolves to
\be
\rho^{\prime}=\frac{1}{(\bar{N}+\bar{M})^r}K(x_r)...K(x_1)\rho _{I}K(x_1)^{\dagger }...K(x_r)^{\dagger }.
\ee
Here $K(x_j)$ is the measurement operator for a detection at position $x_j$
(refer \eq{normalisedmeasurementoperators}), and account has been taken of the total
atom number at each detection.  $\rho^\prime$ is subnormalised such that ${\rm tr}\left(\rho ^\prime\right)$
is a probability density.  It can be shown that,
\bea
\label{eqn:BECfullprobdensity}
{\rm tr}\left(\rho^{\prime }\right)
&=&\frac{1}{\left( \bar{N}+\bar{M}\right) ^{r}}\left( \frac{k}{\pi }\right) ^{r}\int \frac{d\theta }{2\pi }\frac{d\phi }{2\pi }\left[ \Sigma _{j=0}^{r}...c_j \alpha ^{j}\beta ^{r-j}...\right] \left[ \Sigma _{j=0}^{r}...c_j^{\ast }\alpha ^{\ast j}\beta ^{\ast \left( r-j\right) }...\right]
\nonumber \\
&=&\frac{1}{\left( \bar{N}+\bar{M}\right) ^{r}}\left( \frac{k}{\pi }\right) ^{r}\left[ \Sigma _{j=0}^{r}...\left\vert c_{j}\right\vert ^{2}\bar{N}^{j}\bar{M}^{r-j}...\right],
\eea
where $c_m$ is the coefficient of $\alpha^m \beta^{r-m}$ in the product $\Pi _{j=1}^{r}\left( e^{ikx_{j}}\alpha +e^{-ikx_{j}}\beta \right) $
and is easily computed.  This expression can be used to update the simulation parameters $V$ and $\varphi$ after each detection.

An interesting point concerning the measurement operators \eq{normalisedmeasurementoperators} emerges in the explicit
calculation of $\rho^\prime$ above ---
$\rm{tr}\left[ \left( b_k^{\dagger }b_k+b_{-k}^{\dagger }b_{-k}\right) \rho ^{\prime }\right] =\left( \bar{N}+\bar{M}\right) tr(\rho ^{\prime })$ ---
and it would appear that atom number is not conserved in this analysis.  In fact more extreme examples are possible.
If the two atomic modes are initially thermal and both with mean atom number $\langle N \rangle$, then the total atom
number after $r$ measurements can be shown to be $\langle N \rangle (r+2)$.  This difficulty can be understood
by reference to a quantum trajectory type approach which does conserve atom number (and as is used in \cite{Cirac96} for example).
In such an approach there is, in addition to jump operators corresponding to atomic detection, an additional non-Hermitian evolution
with Hamiltonian $H_\text{cond}=-i\frac{1}{2}\Gamma\left( b^{\dagger}_k b_k + b^{\dagger}_{-k} b_{-k} \right)$ where $\Gamma$ is some decay rate,
corresponding the effects of continuous observation of the combined condensate conditioned on no atoms being detected.
Hence the measurement operators \eq{normalisedmeasurementoperators} describe both a detection of an atom at a specific position
and a partial measurement of the total atom number.

It is necessary to check whether neglecting the evolution due to $H_\text{cond}$ affects the basic features of the localisation process.
For the case of initial number states $\ket{N}\ket{M}$ this is straightforward to see.  After $r$ atomic detections the state of
the condensates in a basis of number states is a superposition of states with total atom number $N+M-r$.
Therefore the evolution due to $H_\text{cond}$ merely affects the normalisation of the state.
However the case when the condensates begin in mixtures of states with different total atom number
--- and therefore different decay rates ---
is less clear.  That the evolution due to $H_\text{cond}$
is not important for the case of initial Poissonian states can be understood as follows.
A pair of coherent states evolves under $H_\text{cond}$ as
\be
\left\vert \sqrt{\bar{N}}e^{i\theta }\right\rangle \left\vert \sqrt{\bar{M}}e^{i\phi }\right\rangle \rightarrow e^{-\left( 1/2\right) \left( \bar{N}+\bar{M}\right) \left( 1-e^{-\Gamma t}\right) }{}^{{}}\left\vert \sqrt{e^{-\Gamma t}\bar{N}}e^{i\theta }\right\rangle ^{{}}\left\vert \sqrt{e^{-\Gamma t}\bar{M}}e^{i\phi }\right\rangle,
\nonumber
\ee
and the conditional evolution is seen not to change the ratio $\sqrt{\bar{N}/\bar{M}}$ of the amplitudes of the coherent states.
It is this ratio, rather than the absolute values of $\bar{N}$ and $\bar{M}$, which is the relevant factor in the probability density
after $r$ measurements \eq{BECfullprobdensity}
(and hence also the parameters $V$ and $\varphi$ in \eq{InterferenceProbDensity})
since
\be
\left( \bar{N}^{j}\bar{M}^{r-j}\right) /\left( \bar{N}+\bar{M}\right) ^{r}=1/\left\{ \left[ 1+\left( \bar{M}/\bar{N}\right) \right] ^{j}\left[ \left( \bar{N}/\bar{M}\right) +1\right] ^{r-j}\right\}.
\nonumber
\ee
The same is true the scalar factor $\left \vert c\left(\theta,\phi\right) \right \vert^2$ describing the localisation
in the relative phase $\Delta \equiv \phi-\theta$:
\be
\rho ^{\prime }/{\rm tr}\left( \rho ^{\prime }\right) =\int \frac{d\theta }{2\pi }\frac{d\phi }{2\pi }\left\vert c(\theta ,\phi )\right\vert ^{2}\left\vert \alpha \right\rangle \!\! \left\langle \alpha \right\vert \otimes \left\vert \beta \right\rangle
\!\! \left\langle \beta \right\vert
\nonumber
\ee
and,
\be
\left\vert c(\theta ,\phi )\right\vert ^{2}=\frac{1}{{\rm tr}\left( \rho ^{\prime }\right) }\frac{1}{\left( \bar{N}+\bar{M}\right) ^{r}}\left( \frac{k}{\pi }\right) ^{r}\left\vert \sum _{j=0}^{r}c_{j}\,\bar{N}^{\frac{j}{2}}\, \bar{M}^{\frac{\left( r-j\right) }{2}}\,e^{-ij\Delta }\right\vert ^{2}.
\ee

\begin{figure}[h]
\begin{center}
\includegraphics[height=8cm]{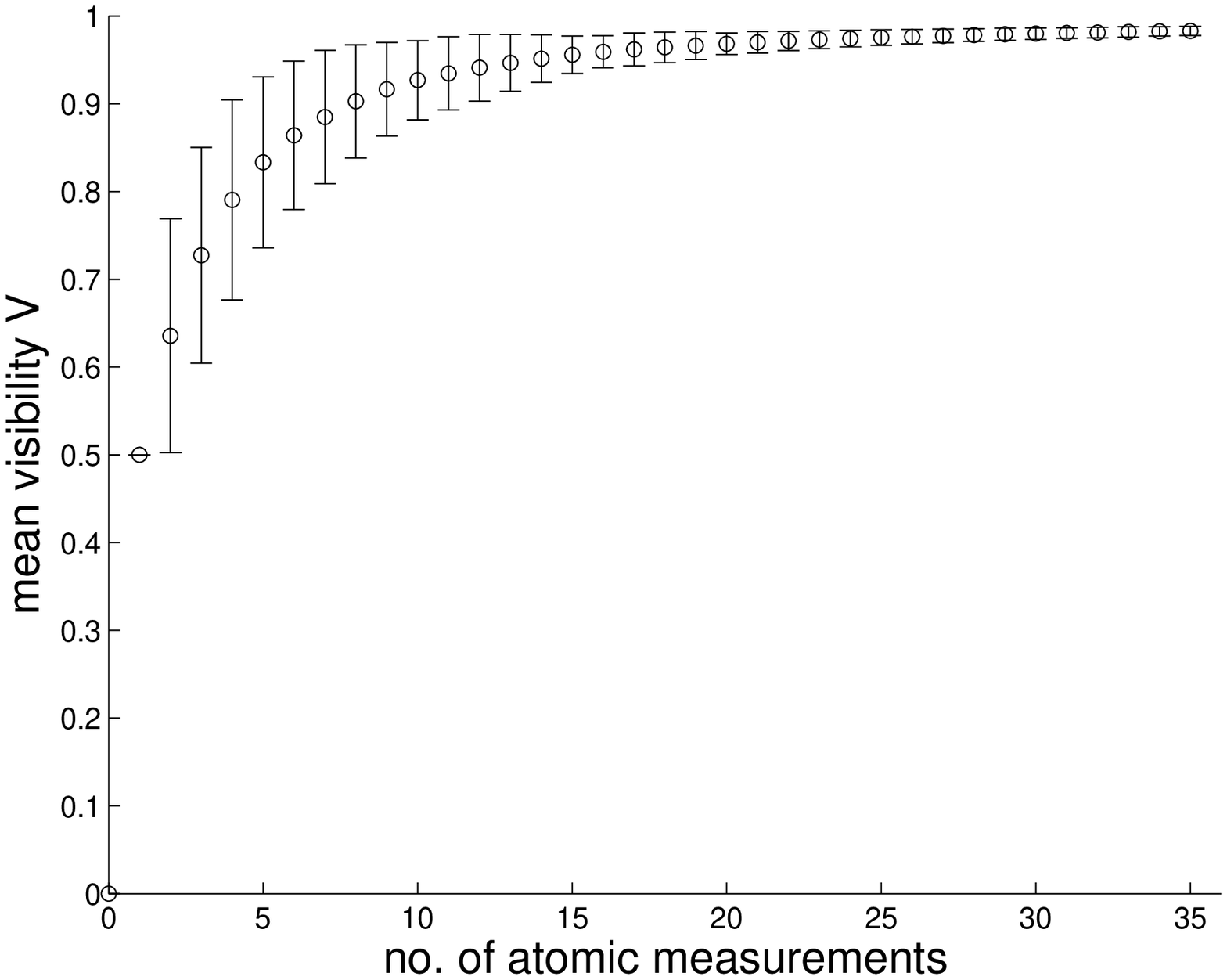}
\caption{\label{fig:BECVevolution}
Simulated evolution of the visibility $V$ for the case of initial Poissonian states
with the same mean number of atoms for both condensates.  $V$ is averaged over $5000$ runs.  The error
bars show $\pm$ one standard deviation.}
\vspace{0.6cm}
\includegraphics[height=7.5cm]{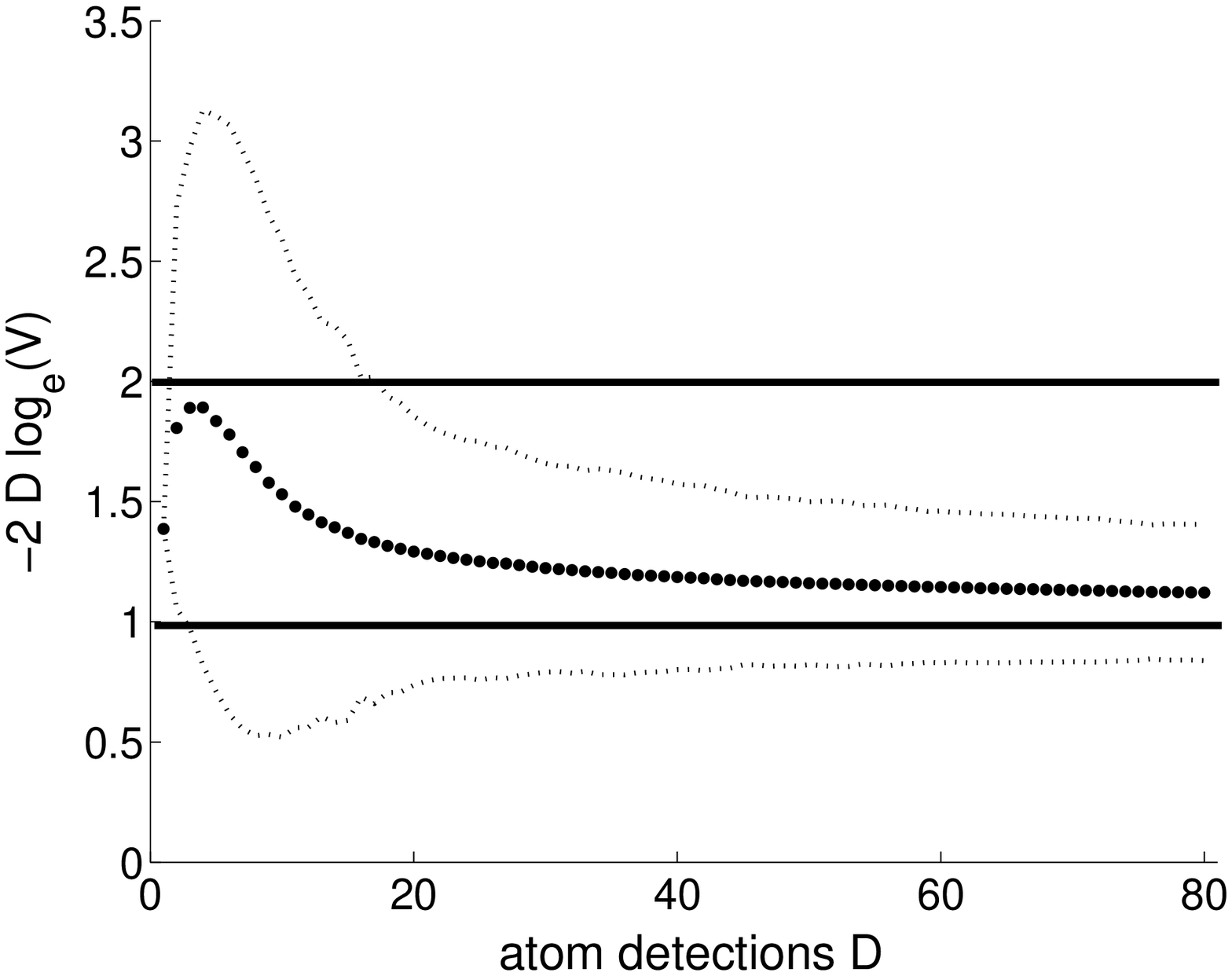}
\caption{\label{fig:BECrsigmasq}
The quantity $-2 D \log_e{V}$, where $V$ is a visibility,
is plotted against the number of atomic detections $D$.
The dotted curves are the results of numerical simulations.
For the centre one, $V$ is the average visibility after $D$ detections.
The lower and upper dotted curves
correspond to the average visibility plus and minus one standard deviatation
respectively.  The solid lines show the range predicted beyond the first
few detections by the analysis of \sect{BECanalyticalPoissonian}.}
\end{center}
\end{figure}
The results of repeated simulations for Poissonian initial states with the same mean atom number for both condensates
are shown in \figu{BECVevolution} and \figu{BECrsigmasq}.
The visibility $V$ for each run is always $0.5$ after one detection, but varies significantly from run-to-run beyond that,
and evolves rapidly to $1$ in all cases.
In \figu{BECVevolution} each point shows $V$ after a given number of detections, averaged over $5000$ runs, and the bars
mark $\pm1$ standard deviation.  \figu{BECrsigmasq} serves to verify the prediction made
at the end of \sect{BECanalyticalPoissonian} which specifies that
the scalar function $\vert C\left(\theta,\phi\right)\vert^2$ can
be well described by a single Gaussian with standard deviation in the range \eq{variancerange} beyond the first few detections.
From \eq{BECexpPredictedVsbly} the quantity $-2 D \log_e V$, where $D$ is the total number of atomic detections,
is expected to vary between $1$ and $2$ where the result applies.
The solid lines mark the predicted range of $-2 D \log_e V$, and the dotted curves
show the results of the numerical simulation.  The central dotted curve corresponds to the average value for $-2 D \log_e V$
and falls within the expected bounds.  The lower and upper
dotted curves are computed for the average visibility plus and minus one standard deviation.
It may be concluded that the analytical
prediction is substantially consistent with the simulated results, although the real
rate of localisation is slightly faster than estimated.

\subsection{Comments on the information available in principle from a pattern of spatial interference}
\label{sec:BECSomeOpenQuestions}

This section discusses two open questions concerning the information about the relative atomic phase available, at least in principle,
from a given pattern of spatial interference.  The first question is whether full knowledge about the relative phase is
contained in a precise record of the positions of the atomic measurements, or whether
some of the phase information is ``hidden'',
and can only be inferred using an additional experimental procedure.

Specifically the situation imagined here involves an experimentalist who is not concerned with the physics of the localisation
of the relative atomic phase, and who supposes that the BEC's have fixed but unknown phases prior to the experiment.
It is assumed that the experimentalist is able in principle to fully characterise all relevant features
of the experiment including: interferometric information - such as phase shifts due
to path length differences; the values of all control parameters determining, for example, the coupling of each condensate out of
its potential well.  The experimentalist is supposed to employ a highly efficient and accurate method of atomic detection,
and to keep a detailed account of how the condensates are prepared.

Adopting the standard approach of Bayesian statistical analysis,
the experimentalist would update the probability density $P(\Delta)$
summarising his or her ``knowledge'' of the relative phase distribution
after an atomic measurement at position $x_r$
according to the rule,
\be
P^\text{posterior} \left( \Delta \right)
=
\frac
{
f\left(x_j \vert \Delta\right) P^\text{prior} \left( \Delta \right)
}
{
\int_0^{2\pi}f(x_j \vert \Delta') P^\text{prior} \left( \Delta' \right) d \Delta'
}.
\ee
The question is then if there are circumstances where there are substantial differences
between $P\left(\Delta\right)$ and the actual quantum relative phase distribution.
Consider for example the situation considered in \sect{BECanalyticalPoissonian}
for which the condensates are known initially to be the same size and uncorrelated.
The prior probability density is flat.  As a pattern of spatial interference
is built up, a measurement at a position $x_j$ updates the Bayesian probability density
according to a factor proportional to $\cos^2\left(kx_j-\Delta/2\right)$.  After
a series of measurements at $x_1,\cdots,x_r$ the posterior distribution is proportional
to $\Pi_{j=1}^r \cos^2\left(kx_j-\Delta/2\right)$.
However, this is the same as the scalar factor $|c(\theta,\phi)|^2$ describing the
localisation of the relative phase parameter
in a basis of coherent states, when the condensates are initially Poissonian.
The fast localisation of the relative phase in this example is mirrored by the rapid narrowing of
the na\"{\i}ve Bayesian phase distribution.

As another example consider the case in which a pattern of spatial interference is obtained for a
state localised at two values for the relative phase,
\be
\rho_S=
\int_0^{2\pi} \frac{d\theta}{2\pi} \ket{\alpha}\bra{\alpha}
\otimes
\left[
\ket{\alpha e^{i\Delta_0}}\!\bra{\alpha e^{i\Delta_0}}
+
\ket{\alpha e^{-i\Delta_0}}\!\bra{\alpha e^{-i\Delta_0}}
\right],
\nonumber
\ee
where $\ket{\alpha}$ is a coherent state with parameter $\alpha=\sqrt{\bar{N}} e^{i\theta}$,
prepared in an atom-optical version of the experiment described in \sect{OptOnePoissonian} of Chapter \ref{chap:OpticalOne}.
If equal numbers of atoms were detected at the two ``output ports'' during the preparation procedure then
$\Delta_0=\pi/2$ and the visibility is $0$.
After the first detection of the spatial interference experiment at a position $x_1$, $\rho_S$ evolves to
\bea
\rho'_S&=&
\cos^2\left(kx-\pi/4\right)
\int_0^{2\pi} \frac{d\theta}{2\pi} \ket{\alpha}\!\bra{\alpha}
\otimes \ket{\alpha e^{i\pi/2}}\!\bra{\alpha e^{i\pi/2}}
\nonumber \\
&& +
\cos^2\left(kx+\pi/4\right)
\int_0^{2\pi} \frac{d\theta}{2\pi} \ket{\alpha}\!\bra{\alpha}
\otimes \ket{\alpha e^{-i\pi/2}}\!\bra{\alpha e^{-i\pi/2}},
\nonumber
\eea
where $\pm k$ is the momenta of the condensates as they overlap.
If $x_1$ is near either $\frac{\pi}{4k}$ or $\frac{3\pi}{4k}$ one or other of the components of $\rho_S$ is picked out and the
visibility jumps to approximately $1$, and the fact that the relative phase is perfectly localised at one value
is not contained in the interference pattern.
However it could be argued that the experimentalist who assumes a priori symmetry breaking would not be surprised by this,
as the na\"{i}ve Bayesian probability distribution, informed by the outcome of the preparation procedure, is again the same
as for the full quantum state of the condensates.

A second question is whether it is feasible to diagnose the characteristic localisation of
the relative atomic phase by looking at the variation in the points making up the pattern
of spatial interference.  One might expect that whenever the relative phase localises as the sequence of
atomic detections is made, the pattern is
less well defined after some number of detections compared to the case when the phases are fixed a priori.
This effect might be expected to be more pronounced when the localisation process is slow.
Note that the challenge here is different to the one usually considered in quantum parameter estimation,
where the aim is to efficiently characterise some quantum parameter of a given quantum state
by sampling a large number of identically prepared systems \cite{DAriano00}.

Yoo, Ruostekoski and Javanainen compared the case of initial number states with $N$ atoms in each mode
and that of initial coherent states in \cite{Yoo97}.
They looked at the statistics of the differences $x_2-x_1$, $x_4-x_3$, and so on,
for a given measurement record $x_1,x_2,...$, and performed a statistical analysis to determine the least number of runs that would be required
with initial number states, in order that at least half would be statistically significant against the null hypothesis of
initial coherent states (at the $5\%$ level).  The number of repetitions was found to increase rapidly with $N$ ---
for $N=1$, a total of $40$ repetitions would be needed, for $N=5$ the number is $700$ --- and too fast to form the basis
of a feasible experiment.

Another approach would be to investigate how the variation of estimates of the relative phase changes as increasing numbers of atoms are detected.
Given a record of atomic measurements at positions $x_1,...,x_r$ the relative phase could be estimated in various ways, for example
by maximum-likelihood estimation or by averaging single-shot estimates.  For the situation considered in \sect{BECanalyticalPoissonian}
this would mean extremising $\Pi_{j=1}^r \cos^2\left(kx_j-\Delta/2\right)$ in the variable $\Delta$,
and averaging the $\Delta_j \equiv 2 x_j /k$ respectively.
However the Cramer-Rao lower bound of classical statistics suggests
that such a method would not be particularly sensitive to rapid localisation in the relative phase.
Even for a classical probability distribution where
a random variable is phase locked with probability density function proportional to $\cos^2 \left(kx-\Delta_0/2\right)$ (with
$\Delta_0$ and $k$ fixed), the variance of any unbiased estimator of $\Delta_0$ can be shown to be at least $1/r$
for a number $r$ of detections.  In conclusion, experiments in which the relative phase parameter is expected to
localise at a slower rate may be better candidates for seeking statistical signatures of the localisation of the relative phase.

\section{Non-destructive measurement of the relative atomic phase by optical methods}
\label{sec:BECOpticalRelPhaseMeasurement}

An interesting direction for further work would explore experimentally feasible approaches to probing the localisation
of the relative phase between two BEC's, prepared separately, after an interference experiment involving only a small number of atomic detections.
It is easy to imagine an experiment in which only small fractions of two separated condensates are extracted, coherently combined and
subjected to a process of atomic detection, for which a pattern of interference would be obtained involving only a small number of points,
and leaving the remainders of the condensates in a partially localised state.
However demonstrating such partial localisation would go beyond any experiment reported to date.
The discussion below reviews in simple terms a body of work concerned with optical detection of the relative atomic phase between condensates.
This work treats the issue of the localisation of the relative phase briefly or not at all
(an exception being \cite{Ruostekoski97}), but it is of interest to identify to several
techniques which could be applied to the current problem.

A recent experiment \cite{Saba05} has demonstrated the non-destructive measurement of the relative phase for
two spatially separated and independently prepared BEC's, as follows.  It is straightforward to show that the density operator
$\hat{\psi}^\dagger (p) \psi{(p)}$, for a combined field operator $\hat\psi$ in momentum space, exhibits interference when the
component condensates are spatially separated but phase locked.  The method used in the experiment to probe this
interference is based on \cite{Pitaevskii99}.
Two counter-propagating lasers are used to continuously extract atoms from the
condensates which are trapped throughout the procedure.
The lasers interact with the condensates by a process of Bragg scattering (which continuously imparts recoil momentum
to a fraction of the atoms at high energy and momentum transfer), and outcouple them.  In addition
the Bragg beams overlap throughout a region covering both condensates,
so that the atoms exit in a single stream, with their origin from
either condensate indistinguishable.  The rate of outcoupling depends sinusoidally on the relative phase of the condensates.
An energy difference between the condensates causes the relative phase to change, and so
a fringe pattern is observed.  For each atom extracted
a photon is transferred between the Bragg beams (conserving momentum),
and hence the atomic fringe pattern is mirrored in variation of the beam intensities.

A number of theoretical papers propose schemes to measure the relative phase of two condensates spectroscopically, fully or partially.
Of particular note is\linebreak\cite{Imamoglu97} which considers a setup with condensates trapped in a two well
potential, and coupled via an excited trap state.  The ground state condensates are assumed sufficiently
far apart to allow them to be driven separately by lasers.  In one version of the scheme a laser acts as a
``local oscillator'' and illuminates one of the condensates.
It is predicted that light is scattered from the other condensate with a phase difference
(compared to the reference laser) proportional to the one between the condensates.
The optical phase difference could be measured by optical homodyning or heterodyning techniques.
This proposal is particularly interesting in as much as it suggests that it might be possible to map the relative phase
for the condensates onto an optical state which would be easier to probe experimentally.

Three further schemes aim to probe the relative phase of two condensates by exploiting features of the light which is scattered
when the two condensates occupy degenerate electronic ground states, are trapped together,
and are coupled via an (electronic) Raman system.  In each case it is supposed that the Raman system is driven
off-resonantly by phase-locked, copropagating lasers with different polarisations.  In \cite{JavanainenB96}
simultaneous amplification of one Raman beam and attenuation of the other is identified
as a signature of a well defined relative atomic phase.  \cite{Ruostekoski97}
also focus's on intensity variation in the scattered beams,
and looks at the localisation of the relative phase given initial number states by means of numerical simulation.
Finally in \cite{RuostekoskiB97} features of the spectrum of the scattered light are found to
depend on the value of the phase difference between in the condensates.
Another scheme \cite{Savage97} is closely related to these three and considers the situation
in which excited atoms, prepared independently, are launched towards two
condensates in overlapping ground states into which they can spontaneous decay.
Features of the emitted light are found to depend on the phase difference of the condensates.

Finally \cite{Corney98} considers theoretically the situation in which Josephson type oscillations between
two spatially separated BEC's in a double well potential
are measured by exploiting a dispersive interaction between one of the condensates and the electromagnetic mode of a cavity.
It is assumed that the cavity mode is off-resonantly coupled to an atomic transition, and that the cavity is strongly driven by an
optical field which leaks out of the opposing end mirror on the time scale of the condensate oscillations.
The optical signal field and the field which leaks out of the cavity are compared using a homodyne measurement scheme.
The intended interaction between the condensate and the cavity mode has a Hamiltonian operator proportional to
$b^{\dagger} b c^{\dagger} c$ (where $b$ and $c$ are the corresponding annihilation operators),
and so the optical phase is modulated by the varying atom number in the cavity in a manner depending on
the relative phase.

%% file: LRDoF_v2_chapter_5.tex
\chapter{Joint Scattering off Delocalised Particles and Localising Relative Positions}
\label{chap:ParticlesScatteringLight}

A recent article \cite{Rau03} by Rau, Dunningham and Burnett (RDB) examines
the localisation in relative position between two ``quantum mirrors'' in a Mach-Zehnder interferometer
(a ``rubber cavity''), or between two delocalised free particles, induced by the scattering of light.
Simple models of the scattering processes are studied numerically using a stochastic approach.
RDB emphasize the role of entanglement in the localisation process.  Furthermore they claim that
the localised relative positions have the properties of classical vector displacements
when three or more particles are involved, a statement which they substantiate in \cite{Dunningham04}.
This chapter presents work published in \cite{Cable05} which continues with the same two models
of scattering and advances the discussion considerably.

The two models of scattering in \cite{Rau03} are explained in \sect{PSLtwoscatteringmodels}.
For the first example --- that of a rubber cavity --- the localisation of the relative position between
two mirrors closely resembles that discussed in Chapter \ref{chap:OpticalOne}
for the relative phase between two optical modes.  Differently the localisation of the relative position is
many valued if the scattering photons have one fixed frequency (with the localised positions
having separation on the order of the wavelength of the light).

The second model is for scattering off free particles in two dimensions.  It is assumed
that each particle taken separately acts as a perfect point scatterer, deflecting the incident light in every direction
with equal probability, and with any internal excitation of the scatterer relaxing rapidly compared to the time for observation.
Detection of the photons in the far field renders scattering off two particles indistinguishable.
Differently from the rubber cavity model
the momentum kick on each photon detection is variable, causing the relative position of the particles to localise
at one value.  RDB assume in their analysis that the particles are delocalised in a one-dimensional region,
and it is pointed out here that there are difficulties extending the arguments to the case of
particles delocalised over a two dimensional region.

The technicalities of analysing the localisation processes in a basis of Gaussian states
are detailed in \sect{PSLActionOnGaussians}.  A basis of Gaussian states has some advantages compared to position eigenstates
(which were used in \cite{Rau03, Dunningham04}).  In particular position eigenstates disperse infinitely fast under free evolution, and
an analysis in terms of Gaussian states might be expected to run in close analogy with the studies of
localising relative phase in Chapters \ref{chap:OpticalOne}, \ref{chap:OpticalTwo} and \ref{chap:BEC} (which employ
coherent state representations).  However the situation here turns out to be more complicated since Gaussian states are not
eigenstates of the translation operators corresponding to momentum transfer.

The localisation of the relative position under the second, free particle, model of scattering
is considered in \sect{PSLlocalisingthermalparticles} for the case of particles with equal mass, each one
initially in a thermal state (with momentum given by the classical Maxwell-Boltzmann distribution).
In \cite{Rau03} the initial states for each particle are momentum eigenstates which
is not very realistic.  The quantum state of a thermal particle has a simple form as a diagonal mixture
of Gaussian states (over the position of the centre of mass), with spread given by the thermal de Broglie wavelength.
Results are presented for the cases when the incident light is monochromatic and thermal.
The situation considered is that of an observer viewing a distant light source ---
the incident light is either forward scattered by the particles into the field of view of the observer
or is deflected, in which case the light source is observed to dim.
Under these assumptions the localisation of the relative position between the particles is found to be partial,
in contrast to the sharp localisation reported in \cite{Rau03}.
For the monochromatic case the pattern of localisation is characterised by Bessel functions
of the first kind.

Finally possible future calculations are suggested in \sect{PSLfutureextensions}.
In particular it is observed that a substantial simplification of the calculations occurs for the important example
of scattering off a classical ideal gas at room temperature.  Entanglement does not play a
role in this example, and the free dynamics is straightforward to compute.

\section{Scattering in a rubber cavity and off delocalised free particles}
\label{sec:PSLtwoscatteringmodels}

\begin{figure}[h]
\begin{center}
\includegraphics[height=5.5cm]{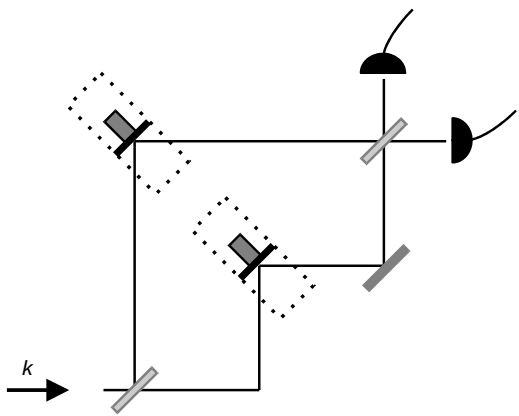}
\caption{\label{fig:PSLrubbercavity}
Photons with momentum $k$ pass through a  ``rubber cavity'' --- a Mach-Zehnder
interferometer in which two of the mirrors
are mounted on ``quantum springs'' and are initially delocalised along an axis.
Two photodetectors monitor the output channels.}
\end{center}
\end{figure}
The first model of scattering discussed in \cite{Rau03} --- that of  a ``rubber cavity'' ---
is illustrated in \figu{PSLrubbercavity} (Fig.~1 of \cite{Rau03}).
The relative position of two mirrors in a Mach-Zehnder
interferometer is localised by a series of single photons which pass through the device
and are detected by photodetectors monitoring the two output channels.
The Kraus operators corresponding to the two possible outcomes for each input photon can be
derived as follows.
Letting $\rho_{\rm mirrors}$ denote the initial state of the two mirrors,
labelling the two optical modes $a$ and $b$, and working in a Fock basis for the optical modes so that
the state of the input photon is $\ket{10}_{ab}$,
the action of the first beam splitter is given by:
\bea
&& \rho _{\rm mirrors} \! \otimes \! \left\vert 10\right\rangle \!\! \left\langle 10\right\vert _{ab} \nonumber \\
\,\longrightarrow\, &&
\rho_{\rm combined}\!=\!
\rho _{\rm mirrors} \! \otimes \!
\left( \frac{\left\vert 10\right\rangle _{ab}
\!+\!\left\vert 01\right\rangle _{ab}}{\sqrt{2}}\right) \!\!
\left( \frac{\left\langle 10\right\vert _{ab}
\!+\!\left\langle 01\right\vert _{ab}}{\sqrt{2}}\right). \nonumber
\eea
The action of the light in mode $a$ is to give a momentum kick $\sqrt{2}k\hat{N}_a$
to the delocalised mirror along its path (where $\hat{N}_a$ is the number operator and $k$ is the momentum of the input photon),
and is described by the operator $\exp\!\left(i\sqrt{2}k\hat{N}_a\hat{x}\right)$
(where $\hat{x}$ is the position operator for the mirror).
Similarly for mode $b$ the operator is $\exp\!\left(i\sqrt{2}k\hat{N}_b\hat{y}\right)$.
After the light in the interferometer interacts with the mirrors the state of the system is,
\[
\exp \left(i\sqrt{2}k\hat{N}_{b}\hat{y}\right)\exp\left(i\sqrt{2}k\hat{N}_{a}\hat{x}\right)
\rho_{\rm combined}
\exp \left(-i\sqrt{2}k\hat{N}_{a}\hat{x}\right)\exp\left(-i\sqrt{2}k\hat{N}_{b}\hat{y}\right).
\]
The second beam splitter, identical to the first, transforms the state of the system according to,
\bea
&& \!
\frac{
\exp\!\left(i\sqrt{2}k\hat{x}\right)
\left\vert 10\right\rangle _{ab}
+
\exp\!\left(i\sqrt{2}k\hat{y}\right)
\left\vert 01\right\rangle _{ab}}{\sqrt{2}}
 \nonumber \\
&\longrightarrow& \!
\left[\frac{\exp\!\left(i\sqrt{2}k\hat{x}\right)-\exp\!\left(i\sqrt{2}k\hat{y}\right)}{2}\right] \!\! \left\vert 10\right\rangle_{ab}
+
\left[\frac{\exp\!\left(i\sqrt{2}k\hat{x}\right)+\exp\!\left(i\sqrt{2}k\hat{y}\right)}{2}\right] \!\! \left\vert 01\right\rangle_{ab}.
\nonumber
\eea
The photon is detected at one of the two output channels, and the Kraus operators are proportional to,
\be
\label{eqn:PSLrubbercavityoperators}
\exp \! \left( i\sqrt{2}k \hat{x} \right) \pm \exp \! \left( i\sqrt{2}k \hat{y} \right).
\ee

The localisation of the relative position in this example resembles that of relative optical phase in
Chapter \ref{chap:OpticalOne}.  In fact the (numerically produced) Fig.~2 of \cite{Rau03} is
essentially identical to the (analytic) \figu{OptOneRLoptical} in \sect{OptOnelocalisingscalarfunction}
of Chapter \ref{chap:OpticalOne}.
The action of the Kraus operators in a basis of position eigenstates is analogous to that of the
optical operators $K_l$ and $K_r$ on optical coherent states, although
in the latter case only the basis is overcomplete. Differently from the optical case the
pattern of relative spatial localisation which emerges has a periodicity of $\pi\sqrt{2}/k$ and
extends throughout the region where the mirrors are at the start.

\begin{figure}[h]
\begin{center}
\includegraphics[height=2.8cm]{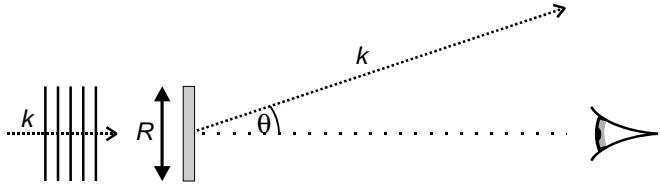}
\caption{\label{fig:PSLfreeparticlescattering}
Plane wave photons with momentum $k$ scatter off two free particles, delocalised in a region of length $R$,
and are either deflected at an angle $\theta$ or continue in the forward direction. The observer can
``see'' photons which forward scatter or which are deflected only by a small amount.}
\end{center}
\end{figure}
The second example considered by RDB is the localisation in relative position
of two free particles boxed in a one dimensional region
due to the scattering of plane wave photons.
\figu{PSLfreeparticlescattering} illustrates the situation.
The Kraus operators are derived as follows.
It is supposed that each particle
taken by itself acts as a perfect point scatterer, scattering as S-waves with certainty.
The scattered photons are detected in the far field at some angle $\theta$ of deflection.
This simple scattering cannot yield information about the position of the particle
(see \cite{Heisenberg49}).  Rather each event imparts a variable momentum kick
$k \sin \theta$ with operator $\exp \! \left( i k \sin \theta \hat{x} \right)$, where $k$ is the momentum of
the incident photon, assumed to approach perpendicularly.
For two particles, and making the assumption that the scattering off each particle is indistinguishable,
the Kraus operators take the form
\be
\label{eqn:PSLfreeparticleoperators}
\exp \left( i k \sin \theta \hat{x} \right) + \exp \left( i \phi \right) \exp \left(i k \sin \theta \hat{y} \right),
\ee
where $\phi$ is some phase factor to be determined.  These operators are seen to form a density.
The condition that the two particles should act as a single point scatterer when together sets $\phi=0$.
In addition, unitarity implies the possibility for forward scattering and the corresponding Kraus operator
is
\be
\label{eqn:PSLfreeparticleoperatortwo}
\sqrt{ \int_0^{2\pi} \tfrac{d\theta}{2\pi} \sin^2 \left[ \tfrac{k \sin \theta}{2} (\hat{y}-\hat{x}) \right]}\,,
\ee
(there is just one).
These forward scattering events do contribute to the localising process and cannot be ignored.

It could be asked why the two particles are not considered to be delocalised in a two dimensional
region.  In fact there is a complication in this case.
Following the same arguments as above it is easy to write down the relevant Kraus operators.
Resolving vectors in Cartesian components with the ``x-direction'' aligned along the direction
of propagation of the incident plane wave:
\begin{eqnarray*}
\hat{K}_{S} &=& \sqrt{\tfrac{1}{2\pi}}\exp \left( i\Delta k\cdot\hat{m}\right) \cos \left( \tfrac{\Delta k}{2}\cdot\hat{r}\right) \\
\hat{K}_{N} &=& \sqrt{\int_{0}^{2\pi }\tfrac{d\theta}{2\pi} \sin ^{2}\left( \tfrac{\Delta k}{2}\cdot\hat{r}\right) }\,\,,
\end{eqnarray*}
where the vector position operator for the $j^{\rm th}$ particle is $(\hat{x}_j,\hat{y}_j)$,
$\hat{r}=\left( \hat{x}_{2}\!-\!\hat{x}_{1},\hat{y}_{2}\!-\!\hat{y}_{1}\right) $,
$\hat{m}=\left( \frac{\hat{x}_{1}+\hat{x}_{2}}{2},\frac{\hat{y}_{1}+\hat{y}_{2}}{2}\right) $
and the momentum kick is
$\Delta k=k\left( 1-\cos \theta ,-\sin \theta \right) $.
The problem is this --- the operators depend on the vector $\hat{r}$ rather than just its magnitude
$|\hat{r}|$.
The localisation process potentially resolves an absolute orientation as well as a relative position, with the fixed
direction of the incident photons providing a reference.

The key difference between the rubber-cavity and free particle scattering models, as demonstrated in the numerical
studies of \cite{Rau03}, is that the changeable momentum kick
of the latter localises a single value for the relative position rather than a periodic array.
Localisation at any value within the initial region is possible.
Notice however that the symmetry about $0$ remains --- the scattering process localises the
absolute value of the relative position, giving two values for the relative displacement whenever
the initial conditions allow for it.  Comparison should be made with Chapter \ref{chap:BEC}
where translations rather than frequency shifts prevent multiple values for the localisation of the relative parameter,
and the multiplicity is eliminated entirely.

\section{Action of the scattering processes in a basis of Gaussian states}
\label{sec:PSLActionOnGaussians}

In what follows the models introduced by RDB are investigated further.
For this a basis of Gaussian states has some advantages over position eigenstates ---
position eigenstates are dynamically fragile in as much as they disperse infinitely rapidly under free
evolution, and using Gaussian states facilitates analogy with the localisation of relative phase discussed
in earlier chapters. However the case of localising relative position turns out to be more technically
complex since Gaussian states are not eigenstates of translation operators.  This section
takes a diversion to set out the technicalities of working with Gaussian states.

The notation
$\left\vert \psi _{k,a,d}\right\rangle$ is adopted for a Gaussian state with mean momentum $k$, mean position $a$
and spatial spread parameter $d$,
\be
\label{eqn:PSLGaussianexpanded}
\left\vert \psi _{k,a,d}\right\rangle \propto \sqrt{d} \int_{-\infty }^{\infty }dxe^{ik(x-a)}G_{a,d}(x)\left\vert x\right\rangle,
\ee
where $G_{a,d}$ denotes a Gaussian probability distribution with mean $a$ and spread
$d$,\footnote{To simplify the notation the standard deviation $d$ is suppressed when it is constant in consecutive steps.}
\[
G_{a,d}(x)=\frac{1}{d \sqrt{2\pi }}e^{-\frac{1}{2}\left( \frac{x-a }{d}\right) ^{2}}.
\]
Consider now the effect of a sequence of localising Kraus operators,
$K(\hat{x}_{1},\hat{x}_{2})=K_{i_{N}}(\hat{x}_{1},\hat{x}_{2})...K_{i_{1}}(\hat{x}_{1},\hat{x}_{2})$,
acting on an arbitrary basis state for the two particles. Rather than
explicit localisation in the relative mean position parameter $a_2\!-\!a_1$,
each basis state evolves to a superposition as follows,
\begin{eqnarray}
\left\vert \psi _{k_{1},a_{1},d}\right\rangle \! \otimes \! \left\vert \psi _{k_{2},a_{2},d}\right\rangle
\!\!\! &\rightarrow& \!\!\!
\left( \tfrac{1}{2\pi }\right) ^{2} \!\! \int \! dx_{1}dx_{2}dp_{1}dp_{2} \nonumber \\
\!\!\! &\times& \!\!\!
K(x_{1},x_{2})e^{ip_{1}(a_{1}-x_{1})}e^{ip_{2}(a_{2}-x_{2})} \nonumber \\
\!\!\! &\times& \!\!\!
\left\vert \psi _{p_{1}+k_{1},a_{1},d}\right\rangle \! \otimes \! \left\vert \psi _{p_{2}+k_{2},a_{2},d}\right\rangle
\end{eqnarray}
where $K(x_{1},x_{2})$ is evaluated at number values and, unless otherwise specified, the integral is
$\int_{-\infty}^{\infty}...\int_{-\infty}^{\infty}\!.$
This can be understood more simply.
Writing $K(x_{1},x_{2})=e^{+i\zeta \frac{(x_{1}+x_{2})}{2}}C_{\Theta }(\frac{x_{2}-x_{1}}{2})$,
a product of translation to the centre of mass (where $\frac{\zeta}{2}$ is the cumulative momentum kick)
and the localising function $C_{\Theta}(\frac{x_{2}-x_{1}}{2})$, and assuming
that the Fourier transform $\widetilde{C}_{\Theta }(p)=\int \! dze^{-ipz}C_{\Theta }(z)$
may be defined, the final state of the particles is proportional to,
\be
\label{eqn:PSLrelpositionFT}
e^{i\zeta \frac{a_{1}+a_{2}}{2}}
\!\! \int \!\! dp
e^{i\frac{p}{2}(a_{2}-a_{1})}\widetilde{C}_{\Theta }(p)
\left\vert \psi _{\frac{\zeta }{2}-\frac{p}{2}+k_{1},a_{1}}\right\rangle
\otimes \left\vert \psi _{\frac{\zeta }{2}+\frac{p}{2}+k_{2},a_{2}}\right\rangle.
\ee
It is seen that the basis states evolve to an increasingly flat superposition over
relative mean momentum with the phase terms defining the location of the relative spatial maxima.
In particular $\widetilde{C}_{\Theta }(p)\simeq e^{-i\frac{p}{2}\Delta _{0}}$
when $K(\hat{x}_1,\hat{x}_2)$ enforces sharp localisation of the relative position
at a single value $\Delta_0$.

The picture can be further clarified by changing basis,
regarding the Hilbert space as a tensor product of spaces
for the centre of mass and the relative position rather than spaces for each
particle,
\[
\left\vert x \right\>\! \otimes \!\left\vert y \right\>
\longleftrightarrow
\left\vert \tfrac{x+y}{2} \right\>_{\mathrm{COM}} \! \otimes \! \left\vert \tfrac{y-x}{2} \right\>_{\mathrm{Rel}} \!.
\]
The basis states considered above can be rewritten as another product of Gaussian states,
\be
\left\vert \psi _{k_{1},a_{1},d}\right\rangle
\! \otimes \!
\left\vert \psi _{k_{2},a_{2},d}\right\rangle
\propto
\left\vert \psi _{k_{1}+k_{2},\frac{a_{1}+a_{2}}{2},\frac{d}{\sqrt{2}}}\right\rangle_{\!\!\mathrm{COM}}
\!\!\! \otimes \!
\left\vert \psi _{k_{2}-k_{1},\frac{a_{2}-a_{1}}{2},\frac{d}{\sqrt{2}}}\right\rangle_{\!\!\mathrm{Rel}}\!\!\!\!. \nonumber
\ee
In the new notation the final state, after $K(\hat{x}_1,\hat{x}_2)$ has acted enforcing sharp localisation at $\Delta_0$,
is proportional to,
\be
\label{eqn:PSLRLcomrellimit}
e^{i\zeta \frac{a_{1}+a_{2}}{2}}\left\vert \psi _{k_{1}+k_{2}+\zeta ,\frac{a_{1}+a_{2}}{2},\frac{d}{\sqrt{2}}}\right\rangle_{\! \mathrm{COM}}
\otimes \!\! \int \!\! dp \, e^{i\frac{p}{2}(a_{2}-a_{1}-\Delta _{0})}\left\vert \psi _{k_{2}-k_{1}+p,\frac{a_{2}-a_{1}}{2},\frac{d}{\sqrt{2}}}\right\rangle _{\! \mathrm{Rel}}\!\!\!\!.
\ee
The centre of mass component is merely translated. The localisation in the relative component
is best seen by comparison with the following identity, expanding an arbitrary position
eigenstate in terms of Gaussians:
\[
\int dp \, e^{ip(a-X)}\left\vert \psi _{k+p,a,d}\right\rangle
\propto e^{ik(X-a)}G_{a,d}(X)\left\vert X\right\rangle.
\]
If $X$ is far from $a$ the norm vanishes.

\section{Localising thermal particles with monochromatic and thermal light}
\label{sec:PSLlocalisingthermalparticles}

This section looks at the localisation of relative position
for two particles as might occur in nature. Rather than the
pure momentum states chosen by RDB for initial states,
the localisation between two thermal particles is considered here.
It is supposed that the two particles have the same mass $m$ and temperature $T$.
At first it will be assumed that the localising process can pick out one
value $\Delta_0$ for the relative position.

A thermal state for a single particle is given by a mixture of
momentum eigenstates weighted according to the classical Maxwell-Boltzmann distribution,
and can be expressed in terms of Gaussian states in a simple diagonal form,
\be
\label{eqn:PSLMaxwellBoltzmann}
\int \!\! dp \,
\sqrt{\tfrac{1}{2\pi m k_{B}T}} \exp \left( -\tfrac{p^{2}}{2mk_{B}T}\right)
\! \left\vert p\right\rangle \!\!
\left\langle p\right\vert
\,=\,
\int \!\! \tfrac{da}{2\pi}
\left\vert \psi_{0,a} \right\rangle \!\!
\left\langle \psi_{0,a} \right\vert.
\ee
The spatial spread parameter of the Gaussian states is given by $d=\sqrt{\tfrac{1}{2mk_{B}T}}$
(where $k_B$ is Boltzmann's constant) and may be identified as the thermal de Broglie wavelength
(the average de Broglie wavelength for an ideal gas at temperature $T$).
When the particle is heavy and hot the Gaussian states approximate position eigenstates.
To attain normalisable states the infinite limits in \eq{PSLMaxwellBoltzmann} are dropped
and the initial states are assumed to be delocalised over a finite region $R$. For two particles together,
\be
\rho_I \propto \int_{R}\int_{R}da_{1}da_{2}\left\vert \psi _{0,a_{1}}\right\rangle \!\!
\left\langle \psi _{0,a_{1}}\right\vert \otimes \left\vert \psi _{0,a_{2}}\right\rangle \!\! \left\langle \psi _{0,a_{2}}\right\vert.
\ee

Under the action of the localising process $K(\hat{x}_1,\hat{x}_2)$ the initial state $\rho_I$ is transformed
as follows, staying with the notation introduced in \sect{PSLActionOnGaussians},
\begin{eqnarray}
\rho_I \!\!\! &\rightarrow& \!\! K_{i_N}...K_{i_2} K_{i_1} \rho_I K_{i_1}^\dagger K_{i_2}^\dagger ... K_{i_N}^\dagger \nonumber \\
&\propto& \int_{R}\int_{R}\int \! da_{1}da_{2}d^{2}x_{1}^{(\prime )}d^{2}x_{2}^{(\prime )}d^{2}p_{1}^{(\prime )}d^{2}p_{2}^{(\prime )} \nonumber \\
&\times&
e^{-i\left\{ p_{1}(x_{1}-a_{1})+p_{2}(x_{2}-a_{2})\right\} }
e^{i\left\{ p_{1}^{\prime }(x_{1}^{\prime }-a_{1}^{\prime })+p_{2}^{\prime }(x_{2}^{\prime }-a_{2}^{\prime })\right\} }
\nonumber \\
&\times& K(x_{1},x_{2})K(x_{1}^{\prime },x_{2}^{\prime })^{\ast } \nonumber \\
&\times&
\left\vert \psi _{p_{1},a_{1},d}\right\rangle \!\! \left\langle \psi _{p_{1}^{\prime },a_{1},d}\right\vert \otimes \left\vert \psi _{p_{2},a_{2},d}\right\rangle \!\! \left\langle \psi _{p_{2}^{\prime },a_{2},d}\right\vert,
\end{eqnarray}
where $d^2x_{1}^{(\prime)}$ abbreviates $dx_{1}dx^{\prime}_{1}$ etc.
When the $K(\hat{x}_1,\hat{x}_2)$ operators enforce sharp localisation --- at $\Delta_0$ say --- the final state takes the simple form,
\begin{eqnarray}
\label{eqn:PSLsharplylocalizedparticlesA}
\rho _{f}
\!\!\! &\propto& \!\!\!
\int_{R} \!\! \int_{R} \! \int \!
da_{1}da_{2}d^2p^{(\prime)}
e^{\frac{i}{2}(p-p^{\prime })(a_{2}-a_{1}-\Delta _{0})} \nonumber \\
\!\!\! &\times& \!\!\!
\left\vert \psi _{\frac{\zeta }{2}-\frac{p}{2},a_{1}}\right\rangle \!\! \left\langle \! \psi _{\frac{\zeta }{2}-\frac{p^{\prime }}{2},a_{1}}\right\vert
 \otimes
\left\vert \psi _{\frac{\zeta }{2}+\frac{p}{2},a_{2}}\right\rangle \!\! \left\langle \! \psi _{\frac{\zeta }{2}+\frac{p^{\prime }}{2},a_{2}}\right\vert\!\!.
\end{eqnarray}
The centre of mass and relative positions remain
unentangled throughout the localising process as is clear for example from \eq{PSLRLcomrellimit}, and
as would certainly be expected. The two particles however
evolve from being separable to being highly entangled. This contrasts to the localisation of relative optical
phase for two initially Poissonian or thermal states, which remain separable despite
the emergence of strong correlation between them (as discussed in \sect{OptOneMixed} of Chapter \ref{chap:OpticalOne}).

The localisation in relative position induced by the
scattering operators \eq{PSLfreeparticleoperators}
and \eq{PSLfreeparticleoperatortwo}, describing the general case
of light scattering off two free particles, will now be treated in detail.
Differently from \cite{Rau03} it is supposed here
that there is no access to a detailed record of every scattering event.
Rather it is assumed that there are two types of measurement outcome:
``forward scattering'' where the incident photon continues
without scattering or is scattered into a small angle between $-\epsilon$ and $\epsilon$; and ``deflection''
where the photon is scattered outside of this range and the light source dims.
It is necessary then to mix over all possible events constituting each of these measurement outcomes.
This is a reasonable model for a real observer monitoring light from a distant source
scattering off two particles, who only has a limited field of view and cannot
measure the angle of deflection.

As previously the initial states for the particles are supposed to be thermal and
to be delocalised in a region $R$.
Changing basis to separate out the centre of mass and relative components,
the initial state of the particles is,
\[
\rho _{I} \propto \int d\tfrac{a_{1}+a_{2}}{2}
\left\vert \psi _{0,\frac{a_{1}+a_{2}}{2},\frac{d}{\sqrt{2}}}\right\rangle \!\!
\left\langle \psi _{0,\frac{a_{1}+a_{2}}{2},\frac{d}{\sqrt{2}}}\right\vert _{\mathrm{COM}}
\!\!\!\! \otimes \rho_{\,\mathrm{Rel}},
\]
where,
\[
\rho_{\,\mathrm{Rel}}
\! \propto \!
\int_{L_{\rm lower}}^{L_{\rm upper}}
\!\! d\tfrac{a_{2}-a_{1}}{2}
\left\vert \psi _{0,\frac{a_{2}-a_{1}}{2},\frac{d}{\sqrt{2}}}\right\rangle \!\!
\left\langle \psi _{0,\frac{a_{2}-a_{1}}{2},\frac{d}{\sqrt{2}}}\right\vert_{\mathrm{Rel}}\!\!\!.
\]
Having changed integration variables from $a_1$ and $a_2$ to $\frac{a_1+a_2}{2}$ and $\frac{a_2-a_1}{2}$, $L_{\rm lower}$ and $L_{\rm upper}$
denote the lower and upper limits of the inner integral which corresponds to the relative component of the
two particle state.  $L_{\rm lower}$ and $L_{\rm upper}$ depend on the outer integration variable $\frac{a_1+a_2}{2}$, which in effect ensures
that the particles remain within the original region $R$.

\begin{figure}[h]
\begin{center}
\includegraphics[height=11.5cm]{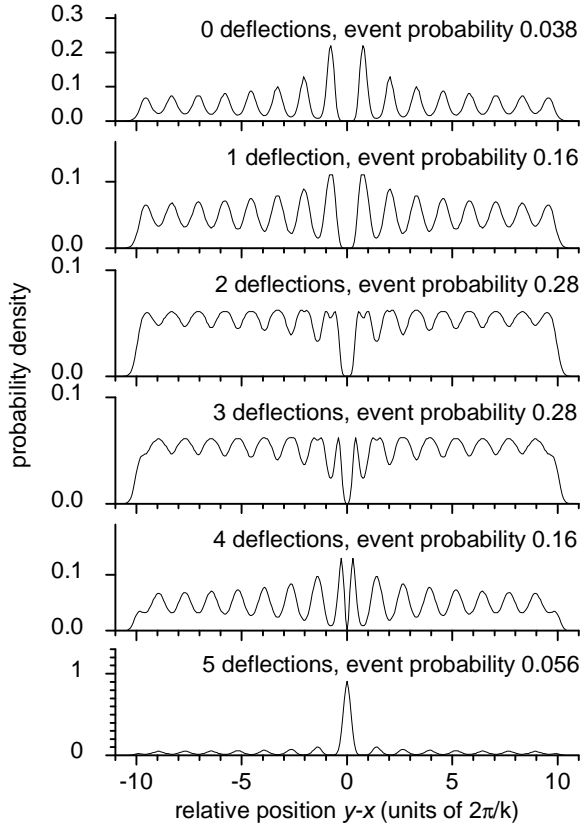}
\caption{\label{fig:PSLmonochromaticlightfreeparticles}
Probability densities between $L_{\rm lower}$ and $L_{\rm upper}$ for the relative separation of two free thermal particles after
$5$ photons, each with momentum $k=5$, have scattered off them,
either being deflected into some large angle or continuing in the forward direction.
The spatial spread parameter
$d=\sqrt{\tfrac{1}{2mk_{B}T}}$ is set to $0.2$ (units $2\pi/k$).}
\end{center}
\end{figure}

Consider first the case that the incident light takes the form of single photons at
one frequency.  With $S$
deflection and $F$ forward-scattering events $\rho_{\,\mathrm{Rel}}$ evolves to $\rho^{\prime}_{\,\mathrm{Rel}}$ as follows,
tracing out the centre of mass component at the end,\footnote{
The momentum kick imparted to the centre of mass depends on the angle of scattering
but the linearity of the partial trace procedure ensures that this causes no additional complication.}
\begin{eqnarray}
\rho^{\prime}_{\rm Rel} \!\!\!\! &=& \!\!\!
\int \!\! \int \!\!
\int_{L_{\rm lower}\left( \frac{a_{1}+a_{2}}{2}\right) }^{L_{\rm upper}\left( \frac{a_{1}+a_{2}}{2}\right) }
d\tfrac{r}{2} \;\; d\tfrac{r^{\prime }}{2} \;\; d\tfrac{a_{2}-a_{1}}{2}
\,\,\,G_{\frac{a_{2}-a_{1}}{2},\frac{d}{\sqrt{2}}}(\tfrac{r}{2})
\,\,G_{\frac{a_{2}-a_{1}}{2},\frac{d}{\sqrt{2}}}(\tfrac{r^{\prime }}{2})
\nonumber \\
\!\!\!\! &\times& \!\!\!
\Bigg[ \sqrt{\int_{0}^{2\pi }\!\!\! d\theta \sin ^{2}\left( \tfrac{k\sin \theta r}{2}\right)
\int_{0}^{2\pi }\!\!\!d\theta \sin ^{2} \left( \tfrac{k\sin \theta r^{\prime }}{2}\right) }
\,+\int_{-\epsilon }^{\epsilon }
\!\!d\theta \cos \left( \tfrac{k\sin \theta r}{2}\right) \cos \left( \tfrac{k\sin \theta r^{\prime }}{2}\right) \Bigg] ^{F}
\nonumber \\
\!\!\!\! &\times& \!\!\!
\left[ \int_{\epsilon }^{2\pi -\epsilon }
\!\!\!\!\!\!\!\!
d\theta \cos \left( \tfrac{k\sin \theta r}{2}\right) \cos \left( \tfrac{k\sin \theta r^{\prime }}{2}\right) \right] ^{S}
\!\! \left\vert \tfrac{r}{2}\right\rangle \!\! \left\langle \tfrac{r^{\prime }}{2}\right\vert _{\mathrm{Rel}}\!.
\end{eqnarray}
A typical pattern of localisation is shown in \figu{PSLmonochromaticlightfreeparticles}
which plots the probability density
$P(y-x)\propto\langle \frac{y-x}{2} \vert \rho^{\prime}_{\rm Rel} \vert \frac{y-x}{2} \rangle$ for different
ratios of ``deflection'' and ``forward-scattering'' events, where $x$ and $y$ are the (precise) positions
of each particle; prior to the scattering process $P(y-x)$ is uniformly distributed.
In contrast to the sharp localisation reported in \cite{Rau03},
limited knowledge of the scattering record
means that the localisation of the relative position is only partial
even after many photons have been scattered.  The localisation takes the form of
complex interference patterns rather than sharp peaks.

As would be expected the degree of localisation
for different outcomes is found to be insensitive to the precise value of the small parameter
$\epsilon$ describing the narrow range of angles visible to the observer
(as are the associated probabilities). Taking $\epsilon=0$
the interference patterns are characterized by Bessel functions
of the first kind $P_{\mathrm{Rel}}(y-x)\sim \left[ 1-J_{0}(k(y-x))\right] ^{F}\left[ 1+J_{0}(k(y-x))\right] ^{S}$.
The localisation is symmetric about the origin.
Sharp localisation at one specific value --- $y=x$ --- is possible with small probability
and occurs when every photon is deflected.

\begin{figure}
\begin{center}
\includegraphics[height=11.5cm]{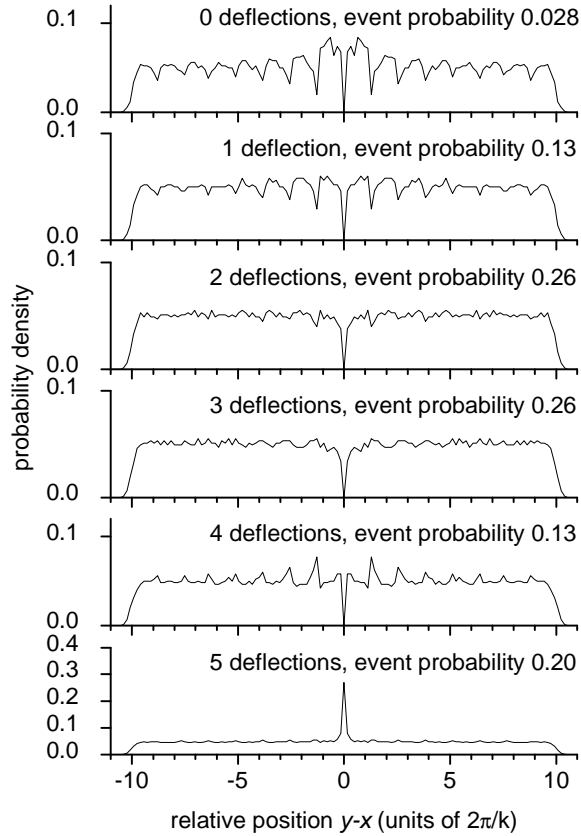}
\caption{\label{fig:PSLthermallightfreeparticles}
Probability densities for the relative separation of two free thermal particles after
$5$ thermal wave packets have scattered off them,
either being deflected into some large angle or continuing in the forward direction.
The momentum parameter $k=5$ and the spatial spread parameter $d=0.2$ (units $2\pi/k$).
The thermal wave packets have mean photon number $\bar{n}=5$.}
\end{center}
\end{figure}

Consider finally the case that the particles are initially in thermal states and that
the scattering light is also thermal.
Each incident wave packet is then described by the mixture
$\rho=\sum_n \left[ \bar{n}^n / (1+\bar{n})^{n+1} \right]\vert n\>\!\<n \vert$
(where $\vert n\>$ denotes an $n$ photon Fock state).
The scattering operators \eq{PSLfreeparticleoperators} and
\eq{PSLfreeparticleoperatortwo} must be modified here --- the fixed momentum kick
$k\sin\theta$ is replaced by the operator $\hat{N} k\sin\theta$ where $\hat{N}$ is the number operator
for the optical mode. Scattering of a single thermal wave packet leads to a
variable photon count, all detected at a single angle $\theta$ of deflection.
A typical pattern of localisation is shown in \figu{PSLthermallightfreeparticles}
which plots the probability density $P(y-x)$ after five thermal wavepackets have scattered;
the results are not sensitive to the precise value of the small parameter
$\epsilon$.

\section{Possible extensions}
\label{sec:PSLfutureextensions}

This section discusses some possible extensions of the calculations in this chapter.
The scattering of light off the particles of an ideal gas at everyday temperatures is of particular interest, and
it turns out that a substantial simplification of the previous calculations is possible for this situation.
Consider first the definition \eq{PSLGaussianexpanded} of
\sect{PSLActionOnGaussians} where $\ket{\psi_{k,a,d}}$ denotes a minimum uncertainty Gaussian basis state
with expected momentum $k$, expected position $a$ and spread $d$.  The action of the momentum kick operator
${\rm exp} \! \left(i p \hat{x} \right)$ (transferring momentum $p$) on these basis states
has a simple form whenever the length scales $1/p$ and $1/k$ are large compared to $d$,
${\rm exp}\left( i p \hat{x} \right) \ket{\psi_{k,a,d}} \rightarrow e^{i p a} \ket{\psi_{k,a,d}}$.
Now for a particle of a classical Maxwell-Boltzmann gas, the state of the particle is mixed and may be represented
diagonally in terms of Gaussian basis states as in \eq{PSLMaxwellBoltzmann}, with zero mean momentum and
a spread given by the thermal de Broglie wavelength.  The thermal de Broglie wavelength is very much smaller
than the (classically derived) average distance between particles and the wavelength of light at optical
frequencies (the thermal de Broglie wavelength is for example $0.02$nm for molecular nitrogen at $300K$).
Hence the simple form of the action of the momentum kick operator above would be expected to apply for the particles
of such an ideal gas for parameter values relevant to the scattering of light from an everyday light source.

Two initially uncorrelated particles of an ideal gas would be expected to evolve
under the action of the scattering processes given by
\eq{PSLrubbercavityoperators}, or \eq{PSLfreeparticleoperators} and \eq{PSLfreeparticleoperatortwo},
to good approximation as follows,
\bea
\label{eqn:PSLinanologywithopticalPoissonian}
&& \int_{R}\int_{R}da_{1}da_{2}
\left\vert \psi _{0,a_{1}}\right\rangle \! \left\langle \psi _{0,a_{1}}\right\vert \otimes
\left\vert \psi _{0,a_{2}}\right\rangle \! \left\langle \psi _{0,a_{2}}\right\vert \nonumber \\
&\rightarrow&
\int_{R}\int_{R}da_{1}da_{2}
\, \vert K\left(a_1,a_2\right)\vert^2 \,
\left\vert \psi _{0,a_{1}} \right\rangle \! \left\langle \psi _{0,a_{1}}\right\vert \otimes
\left\vert \psi _{0,a_{2}}\right\rangle \! \left\langle \psi _{0,a_{2}}\right\vert,
\eea
(continuing with the notation of \sect{PSLlocalisingthermalparticles}).
Here the localisation of the relative position is contained in the scalar function
$\vert K\left(a_1,a_2\right)\vert^2$, and the analysis has close analogy with that
in \sect{OptOnePoissonian} of Chapter \ref{chap:OpticalOne}
concerning the localisation of relative optical phase for the case of initial Poissonian states.
A full analysis would need to check carefully the regimes
in which the result \eq{PSLinanologywithopticalPoissonian} holds.

Two simple features of the localisation in \eq{PSLinanologywithopticalPoissonian} should be noted.  First
the particles stay unentangled throughout the localising process.  However in \cite{Rau03,Dunningham04}
RDB look at examples with pure initial states for which the particles do evolve to highly entangled states,
and they suggest that entanglement is central to scattering-induced localisation of relative positions in general.
This is seen here not to be true in an important example.  Second the free evolution under the
Schr\"{o}dinger equation of the localised particles has a simple form,
shedding some light on the robustness of the localised states.  For the mixed diagonal state
\eq{PSLinanologywithopticalPoissonian} it is easy to see that the free unitary evolution is simply given by replacing the
Gaussian basis states with the states to which they would evolve separately i.e.
$\ket{\psi_{k,a,d}} \rightarrow \exp \left(-i \hat{H}_0 t\right) \ket{\psi_{k,a,d}}$, where $\hat{H}_0$
is the kinetic energy operator.
The free evolution of the one dimensional Gaussian state is well known \cite{Tannoudji77}.
A Gaussian state of the form $\ket{\psi_{k,a,d}}$ remains Gaussian at all times with fixed momentum
dispersion and steadily increasing spatial dispersion.  The evolved state is not a minimum uncertainty
state and is not of the specific form of the Gaussian basis states as defined in \eq{PSLGaussianexpanded}.
Two particles, initially in localised states of the form \eq{PSLinanologywithopticalPoissonian},
will each have a spatial dispersion of
$2\sqrt{1/\left(2mk_{B}T\right)}\sqrt{1+4k_{B}^{2}T^{2}t^{2}}$
at a time $t>0$ after the start of the free evolution
(where $m$ is the particle mass, $T$ is the temperature and $k_B$ is the Boltzmann constant).

In another direction, the model of free particle scattering treated by RDB and in this chapter is rather simple ---
it is assumed that for a single scatterer incident photons are scattered with certainty and with equal
probability in every direction --- and it is of interest to study more realistic models relevant to
the scattering of light from an everyday source off air or dust.  For example,
two types of elastic scattering dominate the scattering of sunlight off the atmosphere  ---
Rayleigh scattering off molecules and Mie scattering off particles larger than a wavelength of the light.
For Rayleigh scattering the scattered intensity is proportional to $(1+\cos^2\theta)/\lambda^4$, where $\theta$
is the angle with the forward direction and is $\lambda$ the wavelength of the light.
The dependence of the total cross section on $\lambda$ might have implications for the scattering of thermal light
which has components with different energies.  Mie scattering is not strongly dependent on the wavelength of light,
but the scattered intensity has an irregular distribution with strong scattering in the forward direction.
Interestingly the patterns of localisation discussed in \cite{Rau03} and in this chapter
are characterised by the wavelength of the light but for Mie scattering the scatterers are already of this order.

%% file: LRDoF_v2_chapter_6.tex
\chapter{Outlook}
\label{chap:Outlook}

This final chapter suggests several ways in which the ideas and methods developed in this thesis could be
applied to new problems.  It is clear that the ``modus operandi'' for Chapters \ref{chap:OpticalOne} through to
\ref{chap:ParticlesScatteringLight} can be extended, and also applied to many more physical systems, and this is discussed
first.  Many of the wider issues arising in this thesis in the contexts of quantum optical processes and Bose-Einstein
condensates, are also relevant to circuits with Bardeen-Cooper-Schrieffer superconductor components.  The coherent coupling of bulk
superconductors in close proximity by a mechanically oscillating superconducting grain has been discussed theoretically,
and provides a concrete system where localisation of the relative superconducting order parameter could be investigated.
There are a number of ``big debates'' with connections to the thesis topic.  One of these concerns different mathematical
descriptions of quantum phase measurements, and compares the features of the various approaches.
Another is why and how time should be treated as a quantum variable.  Here processes of localisation are relevant to
synchronising ``quantum clocks''.  Finally, there is the question of the extent to which the theory of decoherence
fully accounts for the emergence of classicality in quantum systems.  States with a well-defined relative correlation,
of the type whose preparation is discussed in this thesis, have potential application here as pointer states --- states
which are comparatively long lived with respect to coupling to an environment.

\begin{center}
\textbf{A ``modus operandi'' for analysing processes which localise some relationally defined degree of freedom in different physical systems}
\end{center}

It is clear that the methodology set out in Chapters \ref{chap:OpticalOne} through to
\ref{chap:ParticlesScatteringLight} can be easily extended.
The key ideas and methods discussed in this thesis are also relevant to many more physical systems.
Consider, for example, the localisation of a relative angle degree of freedom defined by two spin systems.  This problem is relevant
to proposals and experiments concerned with using measurements on spin systems as the elemental operations in a quantum
computer. Experimental detection of the magnetic force between a ferromagnetic ``tip'' and a single electron spin has
recently been reported, using the technique of magnetic resonance force microscopy \cite{Rugar04}.
For this system the ferromagnet might be treated theoretically as a ``quantum gyroscope'', serving
as a reference for the spin-1/2 particle. Angular momentum coherent states have several properties
analagous to those of optical Glauber coherent states (as explained in \cite{Peres95} for example),
and might provide a mathematically convenient representation along the lines of the analysis
in this thesis.

It should be pointed out that there are also many systems where it is of interest
to study some dynamical process localising a relationally defined parameter, which are quite unlike
the measurement-based processes studied in this thesis.  To pick just one example consider an electron and a
proton, initially well separated and ``free'', evolving under the Coulomb potential to a stationary state of a
hydrogen atom with a well-defined particle separation.  Taking a simple model the stationary states here could just be
taken to be those of the time-independent Schr\"{o}dinger equation for a ``simple'' hydrogen atom
(as detailed in introductory texts).

\vspace{0.3cm}
\begin{center}
\textbf{Localising relative superconductor phase}
\end{center}

The Bose condensation of Cooper pairs (weakly bound pairs of electrons) to a macroscopically occupied ground state
plays a key role in the phenomenon of superconductivity.  A superconductor is then typically assigned an order parameter
corresponding to a definite, absolute, phase.  However, a line of thinking similar to that in Chapter \ref{chap:BEC}
concerning atomic Bose-Einstein condensates asks the question of whether this description has possible shortcomings,
in particular with respect to careful considerations of charge conservation.
The coherent coupling of two bulk superconductors in close proximity by a superconducting grain shuttling
mechanically between them has been discussed theoretically \cite{Gorelik01,Isacsson02},
and provides a concrete example where localisation of the relative superconducting phase could be investigated in detail.
In this system a moveable nanometer sized superconducting grain acts as a Cooper pair box.  The grain behaves like a two level system
defined by states different by twice the fundamental electron charge (i.e. by occupation by a single Cooper pair).
It is able to interact with the bulk superconductors when touching them via a suitable gate mechanism.
In addition, the bulk superconductors can be connected electronically by a Josephson junction.
The magnitude of the current through the junction depends sinusoidally on the value for the relative phase.
Future investigations might aim to clarify the roles of the mechanical and electronic couplings in establishing well-defined relative
phases between the bulk superconductors, identify the most appropriate description of the initial states of the superconductors
(including the possibility of mixed states), and explore analogies with localising optical and atomic relative phases.

\vspace{0.3cm}
\begin{center}
\textbf{Quantum phase operators and distribution functions}
\end{center}

A variety of methods exist for treating phase variables in quantum mechanics, such as Pegg-Barnett phase
operators and Wigner functions.  There has been much discussion of the features of these different approaches
(see for example \cite{Buzek92,Garraway92}).
The POVM (positive operator-valued measure) operators $\hat{a}\pm\hat{b}$
(where $\hat{a}$ and $\hat{b}$ are annihilation operators for two simple harmonic oscillator modes),
or $\hat{a}+e^{i\xi}\hat{b}$ (where the phase shift $\xi$ takes values on a continuous range),
summarise the measurement processes in Chapters \ref{chap:OpticalOne} through to \ref{chap:BEC},
and constitute natural measurement operators for the relative phase between two modes.
It would be valuable to study how the ``relative number'' variable defined by the oscillator modes
changes as the corresponding relative phase becomes well-defined under the action of these measurement operators.
The situation becomes especially interesting when the initial state has a larger intensity in one mode.

\vspace{0.3cm}
\begin{center}
\textbf{Synchronising quantum clocks}
\end{center}

Another topic which is related to this thesis is that of ``quantum clocks''.
Time in quantum mechanics is commonly and implicitly taken to be a classical variable ---
``coordinate time''.  This is the case in the Schr\"{o}dinger equation for example.
However, an extensive literature exists which attempts to treat time in terms of measurements on systems
of quantum resources --- spin systems, or systems of simple harmonic oscillators for example.
The motivation here is to deal with various conceptual difficulties associated with using coordinate
time in quantum mechanics, such are manifested in the well-known Wheeler-DeWitt equation.
If time is treated quantumly, dynamics that would traditionally be described as a unitary process
must now be treated probabilistically (as is done in the conditional probability interpretation
of Page and Wootters \cite{Page83} for example).  The issue of the dynamical localisation
of relative degrees of freedom is relevant here for problems concerning the synchronisation of quantum clocks.
In fact the localisation of relative phase between optical modes initially in Poissonian
states, as analysed in Chapters \ref{chap:OpticalOne} and \ref{chap:OpticalTwo}, is immediately applicable ---
for a laser constitutes a simple, concrete, example of a quantum clock
(a matter treated at length in \cite{Wiseman04}).

\vspace{0.3cm}
\begin{center}
\textbf{Pointer states for decohering systems}
\end{center}

Another possible direction for this programme of research is an investigation of the relational aspects of the theory of
decoherence.  The situation here is of a focus
system interacting with a reservoir (an environment).  The two are assumed not to be correlated at the start.
After some time the focus system and the reservoir evolve to become highly correlated.
In general, it is not possible to keep track of the exact state of an environment,
and the state of reservoir should be traced out.  This yields the focus system in a highly mixed state.
Overall, the key process of decoherence is the damping of the off-diagonal coherences of the focus system
in a special basis of ``einselected pointer states''.  An initial state of the form,
$\rho _{\rm focus}^{\rm initial}\!=\!\sum_{i,j}c_{i}\,c_{j}^{\ast} \left\vert \phi _{i}\right\rangle \!\! \left\langle \phi _{j}\right\vert,$
where the $\ket{\phi_i}$ are pointer states, evolves to the diagonal form
$\rho _{\rm focus}^{\rm decohered}\!=\!\sum_{i} \vert c_{i} \vert^2 \left\vert \phi _{i}\right\rangle \!\! \left\langle \phi _{i}\right\vert$.
Einselected pointer states have special properties --- they are particularly stable against interaction
with the environment, and they lose their purity only slowly.  However,
for a given system it is difficult in general to determine a basis of pointer states.
The existence and completeness of such states is not guaranteed, and they can be hard to
identify.

States with a well localised value of a relative parameter, analagous to those whose
preparation is explained in the thesis, are proposed here as natural candidates for pointer states
in a more relational approach to decoherence theory.  As demonstrated in this thesis,
sharp and rapid localisation of a relationally defined degree of freedom is possible in some systems even when the
initial states are mixed, and happens as readily for an apparatus with small instabilities or asymmetries as
in the ideal case.  This suggests that such dynamical localisation is not a fragile phenomenon that can
only be observed in highly controlled experiments, but is something of relevance to processes occurring in nature.
Examples of processes of localisation acting on highly asymmetric initial states are particularly relevant here.

However even after adopting a more relational approach to the study of decoherence processes,
the problem often raised of why definite events are actually observed in the real world
(see for example the debate in \cite{PhysicsToday91,PhysicsToday93})
is likely to remain. In particular,
the correct interpretation of the final mixed state $\rho_{\rm focus}^{\rm decohered}$,
diagonal in a basis of pointer states, remains to be addressed.  $\rho_{\rm focus}^{\rm decohered}$
is an improper mixture and cannot be interpreted as describing a situation where one of the outcomes ``actualises''
according to the probability distribution $\vert c_i \vert^2$.
It remains an open question as to whether or not careful consideration
of relationism in quantum mechanics has anything to say on this issue.

%% file: LRDoF_v2_appendix_opt_one.tex
\chapter{Derivation of the visibilities for Poissonian and thermal initial states}
\label{chap:appendixA}

\section{Derivation of the visibility for Poissonian initial states}
\label{appendix:PoissonianVisibilityDerivation}

The initial state of the cavity fields, a product of two Poissonian states
both with average photon number $\bar{N}$,
\[
\rho=
\int \frac{d\theta d\phi}{4\pi ^{2}} \left\vert \alpha \right\rangle \left\langle \alpha \right\vert \otimes \left\vert \beta \right\rangle \left\langle \beta \right\vert
\]
where $\alpha=\sqrt{\bar{N}}e^{i\theta}$ and $\beta=\sqrt{\bar{N}}e^{i\phi}$,
acquires a factor
\be
C_{l,r}(\theta ,\phi )=
\left\< r \, {\Big\vert} \frac{\sqrt{\epsilon }\alpha +\sqrt{\epsilon }\beta }{\sqrt{2}}\right\rangle
\left\< l \, {\Big\vert} \frac{-\sqrt{\epsilon }\alpha +\sqrt{\epsilon }\beta }{\sqrt{2}}\right\rangle
\ee
extracting the $l$ and $r$ photon components of the coherent states
under the canonical localising process, in which a fraction $\epsilon \bar{N}$
leaks out of
each cavity and, $l$ and $r$ photons are detected at the left and right detectors respectively.
$\vert C_{l,r} \vert $ is a function of $\Delta \equiv \phi-\theta$ and is peaked at $\pm \Delta_0$ given by $2 \arccos \sqrt{r/r+l}$.
The final state is then,
\[
\rho ^{\prime } \!\!=\!\! \frac{\epsilon ^{r+l}}{4\pi ^{2}r!l!}e^{-2\epsilon \bar{N}}
\!\!\! \int \!\!\!
d\theta d\phi
\!
\left\vert \frac{\alpha +\beta }{\sqrt{2}}\right\vert ^{2r}
\!\!
\left\vert \frac{-\alpha +\beta }{\sqrt{2}}\right\vert ^{2l}
\!\!\!\!
\left\vert \alpha ^{\prime }\right\rangle \!\! \left\langle \alpha ^{\prime }\right\vert \! \otimes \!
\left\vert \beta ^{\prime }\right\rangle \!\! \left\langle \beta ^{\prime }\right\vert
\]
where $\alpha ^{\prime}=\sqrt{1-\epsilon}\alpha$ and $\beta ^{\prime}=\sqrt{1-\epsilon}\beta$.

\subsection{Calculation of the probabilities of different measurement records}
\label{appendix:Poissonianprobabililities}

A probability can be calculated using,
\[
\int d\theta d\phi \cos ^{2r}\frac{\phi-\theta}{2}\sin ^{2l}\frac{\phi-\theta}{2}
= 4 \pi
\frac{\Gamma (r+0.5)\Gamma (l+0.5)}{\Gamma (r+l+1)}
\]
\begin{eqnarray*}
P_{l,r}(\epsilon ,\bar{N}) &=& {\rm tr}\rho ^{\prime } \\
&=& \frac{(2\epsilon \bar{N})^{r+l}}{r!l!}e^{-2\epsilon \bar{N}}\frac{\Gamma (r+0.5)\Gamma (l+0.5)}{\pi \Gamma (r+l+1)}
\end{eqnarray*}

\subsection{Calculation of the visibility after a sequence of detections}
\label{appendix:PoissonianvisibilitiesSCS}

The visibility of $\rho ^{\prime}$ is computed as follows. The second mode undergoes a
variable phase shift of $\tau$ and both modes are then combined at a $50:50$
beamsplitter. $\rho ^{\prime}$ goes to $\rho ^{\prime\prime}$ according to
\[
\left\vert \alpha ^{\prime }\right\rangle \left\langle \alpha ^{\prime }\right\vert \otimes \left\vert \beta ^{\prime }\right\rangle \left\langle \beta ^{\prime }\right\vert \rightarrow
\]
\[
\left\vert \frac{\alpha ^{\prime }\!+\!\beta ^{\prime }e^{i\tau }}{\sqrt{2}}\right\rangle \left\langle \frac{\alpha ^{\prime }\!+\!\beta ^{\prime }e^{i\tau }}{\sqrt{2}}\right\vert \otimes \left\vert \frac{-\alpha ^{\prime }\!+\!\beta ^{\prime }e^{i\tau }}{\sqrt{2}}\right\rangle \left\langle \frac{-\alpha ^{\prime }\!+\!\beta ^{\prime }e^{i\tau }}{\sqrt{2}}\right\vert.
\]
And an intensity at the left detector is then given by
\begin{eqnarray*}
I(\tau )&=& {\rm tr}(a^{\dagger }a\rho ^{\prime \prime }) \\
&\propto& \int d\theta d\phi \left\vert \frac{\alpha +\beta }{\sqrt{2}}\right\vert ^{2r}\left\vert \frac{-\alpha +\beta }{\sqrt{2}}\right\vert ^{2l}\left\vert \frac{\alpha ^{\prime }+\beta ^{\prime }e^{i\tau }}{\sqrt{2}}\right\vert ^{2},
\end{eqnarray*}
where the constant of proportionality is of no interest (it divides out when computing the visibility).
Expanding the last term of the integrand,
\[
\left\vert \frac{\alpha ^{\prime }+\beta ^{\prime }e^{i\tau }}{\sqrt{2}}\right\vert ^{2}=2(1-\epsilon )\bar{N}\cos ^{2}(\frac{\Delta +\tau }{2}),
\]
where $\Delta\equiv\phi-\theta$.
The expression for $I(\tau)$ may be simplified.
\begin{eqnarray*}
I(\tau ) &\propto& \int d\theta d\phi \cos ^{2r}(\frac{\Delta }{2})\sin ^{2l}(\frac{\Delta }{2}) \\
&& \Big[ \cos ^{2}(\frac{\Delta }{2})\cos ^{2}\frac{\tau }{2}\\
&& -2\cos (\frac{\Delta }{2})\sin (\frac{\Delta }{2})\cos \frac{\tau }{2}\sin \frac{\tau }{2}\\
&& +\sin ^{2}(\frac{\Delta }{2})\sin ^{2}\frac{\tau }{2} \Big]
\end{eqnarray*}
The first and last contributions can be resolved in terms of Gamma functions, as for the probability
above, and the second term evaluates to $0$. So,
\[
I(\tau )\propto r\cos ^{2}\frac{\tau }{2}+l\sin ^{2}\frac{\tau }{2}+\frac{1}{2}
\]
and extremising at $\tau=0$ and $\tau=\pi$,
\begin{eqnarray*}
V&=&\left( I_{\rm max}-I_{\rm min} \right) / \left( I_{\rm max} +I_{\rm min} \right) \\
&=& \frac{|r-l|}{r+l+1}.
\end{eqnarray*}

\subsection{Revised calculation of the visibilities for constituent components localised at one value of the relative phase}
\label{appendix:SplitPoissonianVisibilities}

The localising scalar $C_{l,r}(\theta ,\phi )$ is the same as for the case as initial Fock states with the same number,
illustrated in \figu{OptOneRLoptical}.  It is clear that $\rho'$ may be considered a sum of two separate components, one
localised at $+\Delta_0$ with the relative phase parameter $\Delta/2=\frac{\phi-\theta}{2}$ varying on $0$ to $\pi/2$,
and another at $-\Delta_0$ with $\Delta/2$ restricted to $-\pi/2$ to $0$:
\bea
\rho ^{\prime }
\!\!&=&\!\! \frac{\epsilon ^{r+l}}{\pi r!l!}e^{-2\epsilon \bar{N}}
\!\!\! \int ^\frac{\pi}{2} _0 \!\!\!
d\left(\frac{\phi-\theta}{2}\right)
\!
\left\vert \frac{\alpha +\beta }{\sqrt{2}}\right\vert ^{2r}
\!\!
\left\vert \frac{-\alpha +\beta }{\sqrt{2}}\right\vert ^{2l}
\!\!\!\!
\left\vert \alpha ^{\prime }\right\rangle \!\! \left\langle \alpha ^{\prime }\right\vert \! \otimes \!
\left\vert \beta ^{\prime }\right\rangle \!\! \left\langle \beta ^{\prime }\right\vert \nonumber \\
\!\!&+&\!\! \frac{\epsilon ^{r+l}}{\pi r!l!}e^{-2\epsilon \bar{N}}
\!\!\! \int _{-\frac{\pi}{2}} ^0 \!\!\!
d\left(\frac{\phi-\theta}{2}\right)
\!
\left\vert \frac{\alpha +\beta }{\sqrt{2}}\right\vert ^{2r}
\!\!
\left\vert \frac{-\alpha +\beta }{\sqrt{2}}\right\vert ^{2l}
\!\!\!\!
\left\vert \alpha ^{\prime }\right\rangle \!\! \left\langle \alpha ^{\prime }\right\vert \! \otimes \!
\left\vert \beta ^{\prime }\right\rangle \!\! \left\langle \beta ^{\prime }\right\vert. \nonumber
\eea
In what follows the visibility is computed for the $+\Delta_0$ component separately.  The computation
proceeds as in Sec.~\ref{appendix:PoissonianvisibilitiesSCS}.  Differently from this previous calculation,
the reduced range of the integration variable $\Delta/2$ means that the intensity $I(\tau)$ has an additional
contribution (the ``cross term'' no longer drops out).  In detail,
\bea
I(\tau )
&\propto& \cos ^{2}\left( \frac{\tau }{2}\right) \int_{0}^{\pi /2}dx\cos ^{2r+2}\left( x\right) \sin ^{2l}\left( x\right) \nonumber \\
&-& 2\cos \left( \frac{\tau }{2}\right) \sin \left( \frac{\tau }{2}\right) \int_{0}^{\pi /2}dx\cos ^{2r+1}\left( x\right) \sin ^{2l+1}\left( x\right)
\nonumber \\
&+& \sin ^{2}\left( \frac{\tau }{2}\right) \int_{0}^{\pi /2}dx\cos ^{2r}\left( x\right) \sin ^{\left( 2l+2\right) }\left( x\right), \nonumber
\eea
where the integrands are all positive on the $0$ to $\pi/2$ quadrant.
Evaluating the integrals in terms of Gamma functions:
\bea
I(\tau )
&\propto&
\cos ^{2}\left( \frac{\tau }{2}\right) \Gamma \left( r+\frac{3}{2}\right) \Gamma \left( l+\frac{1}{2}\right) \nonumber \\
&-& 2\cos \left( \frac{\tau }{2}\right) \sin \left( \frac{\tau }{2}\right) \Gamma \left( r+1\right) \Gamma \left( l+1\right) \nonumber \\
&+& \sin ^{2}\left( \frac{\tau }{2}\right) \Gamma \left( r+\frac{1}{2}\right) \Gamma \left( l+\frac{3}{2}\right) \nonumber
\eea
Computing the maximum intensity, $I_{\max}$, by setting $\tau=-\Delta_0$, and the
minimum intensity, $I_{\min}$, by setting $\tau=-\Delta_0-\pi$, the visibility
can be computed:
\begin{eqnarray*}
V&=&\left( I_{\rm max}-I_{\rm min} \right) / \left( I_{\rm max} +I_{\rm min} \right) \\
&=&
\frac{\left( r-l\right) ^{2}+\left\{ 4\sqrt{rl}\Gamma \left( r+1\right) \Gamma \left( l+1\right) /\left[ \Gamma \left( r+\frac{1}{2}\right) \Gamma \left( l+\frac{1}{2}\right) \right] \right\} }{\left( r+l\right) \left( r+l+1\right) }.
\end{eqnarray*}

\section{Derivation of the visibility for thermal initial states}
\label{appendix:ThermalVisibilityDerivation}

These calculations follow a similar line to the Poissonian case above.
The initial state of the two cavity fields, a product of two thermal states
with the same average photon number $\bar{N}$,
\[
\rho = \left( \frac{1}{\bar{N}\pi }\right) ^{2}
\!\!\! \int \!\! d^{2}\alpha d^{2}\beta \exp \!-\!\left( \frac{|\alpha |^{2}
\!+\!
|\beta |^{2}}{\bar{N}}\right) \left\vert \alpha \right\rangle \!\! \left\langle \alpha \right\vert
\! \otimes \! \left\vert \beta \right\rangle \!\! \left\langle \beta \right\vert,
\]
where $\alpha \!=\! \sqrt{\bar{n}}e^{i\theta }$ and $\beta \!=\! \sqrt{\bar{m}}e^{i\phi }$,
acquires a factor
\[
\left\< r \, {\Big\vert} \frac{\sqrt{\epsilon }\alpha +\sqrt{\epsilon }\beta }{\sqrt{2}}\right\rangle
\left\< l \, {\Big\vert} \frac{-\sqrt{\epsilon }\alpha +\sqrt{\epsilon }\beta }{\sqrt{2}}\right\rangle,
\]
extracting the $l$ and $r$ photon components of the coherent states
under the canonical localising process.

The final state is
\begin{eqnarray*}
\rho ^{\prime } \!\! &=& \!\!
\frac{\epsilon ^{r+l}}{\bar{N}^{2}\pi ^{2}r!l!}\int d^{2}\alpha d^{2}\beta
\exp -\left( \frac{|\alpha |^{2}\!+\!|\beta |^{2}}{\bar{N}}\right) \\
&\times& \!\! \left\{\! \exp \!\!-\!\!\epsilon \! \left( \left\vert \frac{\alpha \!+\! \beta }{\sqrt{2}}\right\vert ^{2} \!+\! \left\vert \frac{-\alpha \!+\! \beta }{\sqrt{2}}\right\vert ^{2}\right) \!\right\} \left\vert \frac{\alpha \!+\!\beta }{\sqrt{2}}\right\vert ^{2r}
\!\left\vert \frac{-\alpha \!+\! \beta }{\sqrt{2}}\right\vert ^{2l} \\
&\times& \!\! \left\vert \alpha ^{\prime }\right\rangle \! \left\langle \alpha ^{\prime }\right\vert
\otimes
\left\vert \beta ^{\prime }\right\rangle \! \left\langle \beta ^{\prime }\right\vert,
\end{eqnarray*}
where
$\alpha ^{\prime }=\sqrt{1-\epsilon }\alpha $ and $\beta ^{\prime }=\sqrt{1-\epsilon }\beta $.
This expression may be simplified using the parallelogram rule,
\[\left| \frac{\alpha +\beta }{\sqrt{2}} \right|^{2}+\left|\frac{-\alpha +\beta }{\sqrt{2}}\right|^{2}
=|\alpha |^{2}+|\beta |^{2},\]
giving
\begin{eqnarray*}
\rho ^{\prime } \!\! &=& \!\!
\frac{\epsilon ^{r+l}}{\bar{N}^{2}\pi ^{2}r!l!}\int d^{2}\alpha d^{2}\beta \\
&\times&
\left\{ \exp -\left( \epsilon +\frac{1}{\bar{N}}\right) \left( \left\vert \frac{\alpha +\beta }{\sqrt{2}}\right\vert ^{2}+\left\vert \frac{-\alpha +\beta }{\sqrt{2}}\right\vert ^{2}\right) \right\} \\
&\times&
\left\vert \frac{\alpha +\beta }{\sqrt{2}}\right\vert ^{2r}\left\vert \frac{-\alpha +\beta }{\sqrt{2}}\right\vert ^{2l}\left\vert \alpha ^{\prime }\right\rangle \! \left\langle \alpha ^{\prime} \right\vert
\otimes
\left\vert \beta ^{\prime }\right\rangle
\!\left\langle \beta ^{\prime }\right\vert.
\end{eqnarray*}

A probability can be calculated, changing variables of integration such that
$\frac{\alpha+\beta}{\sqrt{2}} \rightarrow \alpha$ and $\frac{-\alpha+\beta}{\sqrt{2}} \rightarrow \beta$,
and evaluating with
$\frac{d^{2}\alpha }{\pi } \!=\! d\bar{n} \frac{d\theta }{2\pi }$ and
$\frac{d^{2}\beta }{\pi } \!=\! d\bar{m} \frac{d\phi }{2\pi }$,
\begin{eqnarray*}
P_{l,r}(\epsilon,\bar{N}) \!\! &=& \!\!  {\rm tr}\rho ^{\prime } \\
&=& \!\! \frac{\epsilon ^{r+l}}{\bar{N}^{2}\pi ^{2}r!l!} \!\! \int \!\! d^{2}\alpha d^{2}\beta \\
&& \times \!
\left\{ \exp \! - \! \left( \epsilon \!+\! \frac{1}{\bar{N}}\right) \left( |\alpha |^{2} \!+\! |\beta |^{2}\right) \right\}
\left\vert \alpha \right\vert ^{2r} \!\! \left\vert \beta \right\vert ^{2l} \\
&=& \!\! \frac{(\epsilon \bar{N})^{r+l}}{\left( 1+\epsilon \bar{N}\right) ^{r+l+2}}.
\end{eqnarray*}

The calculation for the intensity $I(\tau)$ for $\rho ^{\prime}$ proceeds as follows,
\begin{eqnarray*}
I(\tau )&=& {\rm tr}(a^{\dagger }a\rho ^{\prime \prime }) \\
&\propto& \int \frac{d^{2}\alpha }{\pi }\frac{d^{2}\beta }{\pi }\left\{ \exp -\left( \epsilon +\frac{1}{\bar{N}}\right) \left( |\alpha |^{2}+|\beta |^{2}\right) \right\} \\
&& \times \left\vert \alpha \right\vert ^{2r}\left\vert \beta \right\vert ^{2l}
\left\vert \alpha ^{\prime }+\beta ^{\prime }+(-\alpha ^{\prime }+\beta ^{\prime })e^{i\tau }\right\vert ^{2} \\
&\propto& \int d\bar{n}\frac{d\theta }{2\pi }d\bar{m}\frac{d\phi }{2\pi}\left\{ \exp -\left( \epsilon +\frac{1}{\bar{N}}\right) \left( \bar{n}+\bar{m}\right) \right\} \\
&& \times \bar{n}^{r}\bar{m}^{l}\left\vert \sqrt{\bar{n}}e^{i\theta }(1-e^{i\tau })+\sqrt{\bar{m}}e^{i\phi }(1+e^{i\tau })\right\vert ^{2}.
\end{eqnarray*}
Now,
\begin{eqnarray*}
&& \left\vert \sqrt{\bar{n}}e^{i\theta }(1-e^{i\tau })+\sqrt{\bar{m}}e^{i\phi }(1+e^{i\tau })\right\vert \\
&=&
\bar{n}|1-e^{i\tau }|^{2}+\bar{m}|1+e^{i\tau }|^{2}+(..)e^{i\theta }e^{-i\phi }+(..)e^{-i\theta }e^{i\phi },
\end{eqnarray*}
and the latter two contributions integrate to $0$. Hence
\begin{eqnarray*}
I(\tau) \!\! &\propto& \!\! |1-e^{i\tau }|^{2} \!\! \int \!\! d\bar{n}d\bar{m}\bar{n}^{r+1}\bar{m}^{l} \! \exp \!- \left\{ \! \left( \epsilon \!+\! \frac{1}{\bar{N}}\right) \left( \bar{n}\!+\!\bar{m}\right) \! \right\} \\
\!\! &+& \!\! |1+e^{i\tau }|^{2} \!\! \int \!\! d\bar{n}d\bar{m}\bar{n}^{r}\bar{m}^{l+1} \! \exp \!- \left\{ \! \left( \epsilon \!+\!\frac{1}{\bar{N}}\right) \left( \bar{n}\!+\!\bar{m}\right) \! \right\}.
\end{eqnarray*}
Evaluating the integrals is as for the probability calculation above.
\begin{eqnarray*}
I(\tau ) &\propto& (|1-e^{i\tau }|^{2}(r+1)+|1+e^{i\tau }|^{2}(l+1) \\
&\propto& l\cos ^{2}\frac{\tau }{2}+r\sin ^{2}\frac{\tau }{2}+1
\end{eqnarray*}
and the visibility is given by
\[ V=\frac{|r-l|}{r+l+2} \]
extremising at $\tau=0$ and $\tau=\pi$.

%% file: LRDoF_v2_appendix_GaussianVis.tex
\chapter{Derivation of the visibility for a Gaussian distribution of the relative phase}
\label{chap:appendixB}

For a normal distribution with mean $\Delta_0$ and variance $\sigma^2$, the probability density function takes the form:
\be
P(\Delta )=\frac{1}{\sigma \sqrt{2\pi }}e^{-\left( \Delta -\Delta _{0}\right) ^{2}/\left( 2\sigma ^{2}\right) } \nonumber
\ee
In what follows the visibility $V$ is computed for a state $\rho_G$ of the form,
\bea
\rho _{G}&=&\int_0^{2\pi} \!\! \int_0^{2\pi}
\frac{d\theta}{2\pi} \frac{d\phi}{2\pi}
\,\, 2 \pi P(\Delta) \,\, + \,\,
\left( \text{contributions for }2\pi \text{-periodicity in} \, \Delta \right) \nonumber \\
&& \times \,\,
\left\vert \sqrt{\bar{N}}e^{i\theta }\right\rangle \left\langle \sqrt{\bar{N}}e^{i\theta }\right\vert
\,\, \otimes \,\, \left\vert \sqrt{\bar{N}}e^{i\phi }\right\rangle \left\langle \sqrt{\bar{N}}e^{i\phi }\right\vert,
\eea
where $\Delta \equiv \phi-\theta$ and $\sqrt{\bar{N}}e^{i\theta }$ denotes a Glauber coherent state with mean atom (or photon) number $\bar{N}$
and phase $\theta$.  It is assumed that the Gaussian width is small compared to $2\pi$.

The following identity can be used following the $2\pi$-periodicity of the variables $\Delta$, $\phi$ and $\theta$ and the lack of dependence
on the mean phase variable $\frac{\theta+\phi}{2}$:
\be
\int_0^{2\pi} \!\! \int_0^{2\pi}
\frac{d\theta}{2\pi} \frac{d\phi}{2\pi}
\left( \cdot \right) = \int_{\Delta_\text{offset}-\pi}^{\Delta_\text{offset}+\pi} \frac{d\Delta}{2\pi} \left( \cdot \right).
\ee
If $\sigma$ is less than $\pi/2.58\sim1.2$ then
${\rm tr} \rho_G = \int_{\Delta_{0}-\pi}^{\Delta_{0}+\pi} d\Delta \, P\left(\Delta\right) > 0.99$ and $\rho_G$ is seen to be well normalised.

$V$ can be computed by comparing the probability for measurement at $x$, where the measurement is given by
the Kraus operator $\hat{M}\left( x\right) =\sqrt{1/2\bar{N}} \left( e^{i\pi x}a+e^{-i\pi x}b \right) $,
with the standard form $1+V\cos\left(2\pi x - \varphi \right)$ (the momentum parameter $k$ has be set to $\pi$ units)
- refer \sect{BECspatialinterference}.

This probability is computed as follows:
\bea
{\rm tr}\left[ M(x)\rho_{G}M(x)^{\dagger} \right]
\!\! &=& \!\!
\int_{\Delta _{0}^{{}}-\pi }^{\Delta _{0}+\pi } \!\!\! d\Delta
\, \frac{1}{\sigma \sqrt{2\pi }} \,\, e^{-\left( \Delta -\Delta _{0}\right) ^{2}/\left( 2\sigma ^{2}\right) }
\, \left[ 1+\cos \left( 2\pi x-\Delta \right) \right] \nonumber \\
\!\! &=& \!\!
1 + \cos \left( 2\pi x-\Delta _{0}\right) \int_{\Delta _{0}^{{}}-\pi }^{\Delta _{0}+\pi} \!\!\! \frac{d\Delta}{\sigma \sqrt{2\pi }} \, e^{-\left( \Delta -\Delta _{0}\right) ^{2}/\left( 2\sigma ^{2}\right) }\cos \left( \Delta _{0}\!-\!\Delta \right) \nonumber \\
\!\! && \!\!
- \sin \left( 2\pi x-\Delta _{0}\right) \int_{\Delta _{0}^{{}}-\pi }^{\Delta _{0}+\pi } \!\!\! \frac{d\Delta}{\sigma \sqrt{2\pi }} \, e^{-\left( \Delta -\Delta _{0}\right) ^{2}/\left( 2\sigma ^{2}\right) }\sin \left( \Delta _{0}\!-\!\Delta \right) \nonumber
\eea
The second integral evaluates to $0$ because of change of sign about $\Delta_0$.
$V$ may be identified with the coefficient of $\cos\left(2\pi x -\Delta_0\right)$.  Extending the range of integration to $\pm \infty$,
\bea
V &\simeq& \int_{-\infty }^{\infty} d\Delta \frac{1}{\sigma \sqrt{2\pi }} \, e^{-\Delta ^{2}/\left( 2\sigma ^{2}\right) } \cos \Delta \nonumber \\
&=& e^{-\frac{1}{2}\sigma ^{2}}.
\eea
This expression for $V$ may be taken as exact when $\sigma \lesssim 1$.

As an example consider the optical interference experiment analysed in \sect{OptOnePoissonian} of chapter \ref{chap:OpticalOne}
for which two optical states, initially Poissonian with equal
mean photon numbers, are combined at a beamsplitter.  In the case that $r$ photons are counted at the right output channel,
the relative phase localisation is described by a scalar function
$c\left(\theta,\phi\right) \propto \cos^{2r}\frac{\Delta}{2}$
and the visibility $V$ is calculated exactly as $r/\left(r+1\right)$.  For larger values of $r$ $c\left(\theta,\phi\right)$ is well approximated
by a Gaussian with standard deviation $\sigma=\sqrt{2/r}$ and the calculation above suggests a value for $V$ as $\exp\left(1/r\right)$.
The fractional error in the latter estimate does indeed decrease for increasing $r$ and is 0.26 for $r=1$, $0.09$ for $r=2$, $0.04$ for $r=3$, less than $0.01$ by $r=7$ and less than $0.001$ for $r=23$.

% Local Variables:
% TeX-master: "../thesis"
% End:

%% file: LRDoF_v2_backmatter.tex
\bibliographystyle{thesbib}
\bibliography{LRDoF_v2_thesis}

%%% Local Variables:
%%% mode: latex
%%% TeX-master: thesis.tex
%%% End: